\pdfoutput=1

\documentclass[a4paper,11pt]{article}
\usepackage{jheppub}

\usepackage{bm}

\usepackage{color}
\usepackage[usenames,dvipsnames,svgnames,table]{xcolor}

\usepackage{amsmath}
\usepackage{amssymb}
\usepackage{graphicx}
\usepackage{slashed}
\usepackage{soul}
\usepackage{lscape}
\usepackage{graphicx}
\usepackage{xcolor}
\usepackage{multirow}
\usepackage{placeins,hyperref}
\usepackage[separate-uncertainty=true]{siunitx}
\usepackage{caption}
\usepackage{subcaption}
\usepackage{mathrsfs}
\usepackage[version=3]{mhchem}
\usepackage{amsmath, amssymb, amscd, amsthm, amsfonts}
\usepackage{wrapfig}

\counterwithin{figure}{section}

\usepackage{lineno}

\begin{document}

\title{\centering Mineral Detection of Neutrinos and Dark Matter 2024 \\ Proceedings}

\author[1]{Sebastian~Baum,}
\affiliation[1]{Institute for Theoretical Particle Physics and Cosmology, RWTH Aachen University, 52056 Aachen, Germany}

\author[2]{Patrick~Huber,}
\affiliation[2]{Center for Neutrino Physics, Department of Physics, Virginia Tech, Blacksburg, VA 24061, USA}

\author[3]{Patrick~Stengel;}
\affiliation[3]{Istituto Nazionale di Fisica Nucleare, Sezione di Ferrara, via Giuseppe Saragat 1, I-44122 Ferrara, Italy}

\author[4]{Natsue~Abe,}
\affiliation[4]{Center for Mathematical Science and Advanced Technology (MAT), Japan Agency for Marine-Earth Science and Technology (JAMSTEC), Yokohama, Kanagawa 236-0001, Japan}

\author[5]{Daniel~G.~Ang,}
\affiliation[5]{Quantum Technology Center, University of Maryland, College Park, MD 20742, USA}

\author[6]{Lorenzo~Apollonio,}
\affiliation[6]{Dipartimento di Fisica, Università degli Studi di Milano, via Celoria 16 Milano, Italy}

\author[7]{Gabriela~R.~Araujo,}
\affiliation[7]{Physics Department, University of Zurich, Switzerland}

\author[8]{Levente~Balogh,}
\affiliation[8]{Department of Mechanical and Materials Engineering, Queen's University, 130 Stuart Street, Kingston, Ontario K7L 3N6, Canada}

\author[9]{Pranshu~Bhaumik}
\affiliation[9]{National Security Institute, Virginia Tech, 1311 Research Center Drive, Blacksburg, VA 24060, USA}

\author[10]{Yilda~Boukhtouchen,}
\affiliation[10]{Department of Physics, Engineering Physics, and Astronomy, Queen’s University, Kingston, Ontario, K7L 2S8, Canada}

\author[10]{Joseph~Bramante,}

\author[11]{Lorenzo~Caccianiga,}
\affiliation[11]{Istituto Nazionale di Fisica Nucleare, Sezione di Milano, Via Celoria 16, Milano, Italy}

\author[12]{Andrew~Calabrese-Day,}
\affiliation[12]{Department of Physics, University of Michigan, Ann Arbor, Michigan 48109, USA}

\author[13]{Qing~Chang,}
\affiliation[13]{Volcanoes and Earth’s Interior Research Center, Japan Agency for Marine-Earth Science and Technology (JAMSTEC), Yokosuka, Kanagawa 237-0061, Japan}

\author[14]{Juan~I.~Collar,}
\affiliation[14]{Enrico Fermi Institute, University of Chicago, Chicago, IL 60637, USA}

\author[5]{Reza~Ebadi,}

\author[15]{Alexey~Elykov,}
\affiliation[15]{Institute for Astroparticle Physics, Karlsruhe Institute of Technology, 76021 Karlsruhe, Germany}

\author[16,17,18]{Katherine~Freese,}
\affiliation[16]{Department of Physics, The University of Texas at Austin, Austin, TX 78712, USA}
\affiliation[17]{The Oskar Klein Centre, Department of Physics, Stockholm University, AlbaNova, SE-106 91 Stockholm, Sweden}
\affiliation[18]{Nordita, Stockholm University and KTH Royal Institute of Technology, Hannes Alfvéns väg 12, SE-106 91 Stockholm, Sweden}

\author[10]{Audrey~Fung,}

\author[19]{Claudio~Galelli,}
\affiliation[19]{LUTh, Observatoire de Paris, 5 Pl. Jules Janssen, 92190 Meudon, France}

\author[20]{Arianna~E.~Gleason,}
\affiliation[20]{SLAC National Accelerator Laboratory, Stanford University, 2575 Sand Hill Road, Menlo Park, CA 94025, USA}

\author[9]{Mariano~Guerrero~Perez,}

\author[21]{Janina~Hakenm{\"u}ller,}
\affiliation[21]{Department of Physics, Duke University, Durham, NC 27708, USA}

\author[13]{Takeshi~Hanyu,}

\author[22]{Noriko~Hasebe,}
\affiliation[22]{Institute of Nature and Environmental Technology, Kanazawa University}

\author[4]{Shigenobu~Hirose,}

\author[2]{Shunsaku~Horiuchi,}

\author[23]{Yasushi~Hoshino,}
\affiliation[23]{Department of Physics, Kanagawa University, Yokohama, Japan}

\author[24]{Yuki~Ido,}
\affiliation[24]{Department of Earth and Environmental Sciences, University of Nagoya, Japan}

\author[9,25,26]{Vsevolod~Ivanov,}
\affiliation[25]{Department of Physics, Virginia Tech, Blacksburg, VA 24061, USA}
\affiliation[26]{Center for Quantum Information Science and Engineering, Virginia Tech, Blacksburg, VA 24061, USA}

\author[27]{Takashi~Kamiyama,}
\affiliation[27]{Faculty of Engineering, Hokkaido University, Kita 13 Nishi 8, Kita-ku, Sapporo, Hokkaido, Japan}

\author[28]{Takenori~Kato,}
\affiliation[28]{Institute for Space-Earth Environmental Research, Nagoya University, Furo-cho, Chikusa-ku, Nagoya, 464-8601, Japan}

\author[4]{Yoji~Kawamura,}

\author[29]{Chris~Kelso,}
\affiliation[29]{Department of Physics, University of North Florida, 1 UNF Dr, Jacksonville, FL 32224, USA}

\author[25]{Giti~A.~Khodaparast,}

\author[29]{Emilie~M.~LaVoie-Ingram,}

\author[30]{Matthew~Leybourne,}
\affiliation[30]{Department of Geological Sciences \& Geological Engineering, Queen’s University, 36 Union street Kingston, ON K7L 3N6, Canada}

\author[5]{Xingxin~Liu,}

\author[8]{Thalles~Lucas,}

\author[25]{Brenden~A.~Magill}

\author[6]{Federico~M.~Mariani,}

\author[30]{Sharlotte~Mkhonto,}

\author[31]{Hans~Pieter~Mumm,}
\affiliation[31]{National Institute of Standards and Technology, 100 Bureau Dr., MS 8461, Gaithersburg, MD 20899, USA}

\author[32]{Kohta~Murase,}
\affiliation[32]{The Pennsylvania State University, University Park, PA 16802, USA}

\author[33]{Tatsuhiro~Naka,}
\affiliation[33]{Department of Physics, Toho University, 2-2-1 Miyama, Funabashi, Japan}

\author[4]{Kenji~Oguni,}

\author[12]{Kathryn~Ream,}

\author[21]{Kate~Scholberg,}

\author[5]{Maximilian~Shen,}

\author[12]{Joshua~Spitz,}

\author[34]{Katsuhiko~Suzuki,}
\affiliation[34]{Submarine Resources Research Center, Japan Agency for Marine-Earth Science and Technology (JAMSTEC), Yokosuka, Kanagawa 237-0061, Japan}

\author[12]{Alexander~Takla,}

\author[5]{Jiashen~Tang,}

\author[35]{Natalia~Tapia-Arellano,}
\affiliation[35]{Department of Physics and Astronomy, University of Utah, Salt Lake City, UT 84112, USA}

\author[36]{Pieter~Vermeesch,}
\affiliation[36]{Department of Earth Sciences, University College London, London WC1E 6BT, UK}

\author[10]{Aaron~C.~Vincent,}

\author[37]{Nikita~Vladimirov,}
\affiliation[37]{University Research Priority Program (URPP), Adaptive Brain Circuits in Development and Learning, University of Zurich, Zurich, Switzerland}

\author[5]{Ronald~Walsworth,}

\author[38]{David~Waters,}
\affiliation[38]{Department of Physics \& Astronomy, University College London, London WC1E 6BT, UK}

\author[29]{Greg~Wurtz,}

\author[39]{Seiko~Yamasaki}
\affiliation[39]{Geological Survey of Japan, AIST, 1-1-1 Higashi Tsukuba Ibaraki, Japan}

\author[40]{and Xianyi~Zhang}
\affiliation[40]{Lawrence Livermore National Laboratory, 7000 East Ave, Livermore, CA 94536, USA}

\abstract{The second ``Mineral Detection of Neutrinos and Dark Matter'' (MD$\nu$DM'24) meeting was held January 8-11, 2024 in Arlington, VA, USA, hosted by Virginia Tech's Center for Neutrino Physics. This document collects contributions from this workshop, providing an overview of activities in the field. MD$\nu$DM'24 was the second topical workshop dedicated to the emerging field of mineral detection of neutrinos and dark matter, following a meeting hosted by IFPU in Trieste, Italy in October 2022. Mineral detectors have been proposed for a wide variety of applications, including searching for dark matter, measuring various fluxes of astrophysical neutrinos over gigayear timescales, monitoring nuclear reactors, and nuclear disarmament protocols; both as {\it paleo-detectors} using natural minerals that could have recorded the traces of nuclear recoils for timescales as long as a billion years and as detectors recording nuclear recoil events on laboratory timescales using natural or artificial minerals. Contributions to this proceedings discuss the vast physics potential, the progress in experimental studies, and the numerous challenges lying ahead on the path towards mineral detection. These include a better understanding of the formation and annealing of recoil defects in crystals; identifying the best classes of minerals and, for paleo-detectors, understanding their geology; modeling and control of the relevant backgrounds; developing, combining, and scaling up imaging and data analysis techniques; and many others. During the last years, MD$\nu$DM has grown rapidly and gained attention. Small-scale experimental efforts focused on establishing various microscopic readout techniques are underway at institutions in North America, Europe and Asia. We are looking ahead to an exciting future full of challenges to overcome, surprises to be encountered, and discoveries lying ahead of us.}

\maketitle
\flushbottom

\section*{Preface}
\addcontentsline{toc}{section}{Preface}

Minerals record persistent damage features from ions -- often called tracks -- that can be read out using a variety of microscopy techniques. The damage features from heavy ions with tens of MeV of energy produced by spontaneous fission of uranium, thorium, and other heavy radioactive trace elements are routinely imaged in minerals including mica, apatite, zircon and many others using microscopy techniques ranging from transmission electron microscopy over X-ray techniques to conventional optical microscopy, often after previously enlarging the damage features by chemical etching~\cite{Fleischer:1964,Fleischer383,Fleischer:1965yv,GUO2012233}. While crystal damage does anneal with time, the timescales are enormous in many minerals. For example, the annealing time in diopside at room temperature is $t_{\rm ann} \sim 10^{59}\,$yr~\cite{Fleischer:1965yv}. By counting the density of nuclear fission tracks in a natural sample and independently measuring the concentration of uranium and thorium in the sample, one can obtain a {\it fission track age} of a sample, a standard dating technique in geoscience~\cite{Wagner:1992,Malusa:2018}. Fission track ages as large as $\sim 0.8\,$Gyr in  apatite~\cite{Murrell:2003,Murrell:2004,Hendriks:2007} and $\sim 2\,$Gyr in zircon~\cite{Montario:2009} have been reported in the literature, presenting a proof-of-existence for geological environments on Earth that are sufficiently cold and stable for fission tracks to be preserved over gigayear timescales. 

Persistent damage features in minerals have also been observed from particles with energies and/or stopping powers much lower than those of the $\mathcal{O}(10)\,$MeV, $Z \sim 50$ fission fragments. The damage features produced by the nuclear remnants of $\alpha$-decays of $^{238}$U and similar heavy radioactive isotopes that receive typical nuclear recoil energies of some tens of keV are used for alpha-recoil track (ART) dating similar to fission track dating~\cite{Goegen:2000,Glasmacher:2003}. In the 1980s, Price and Salamon observed natural ``tracks'' from Al and Si ions with a few hundred keV of nuclear recoil energy in 0.5\,Gyr old mica by optical microscopy following chemical etching~\cite{Price:1986}. In the 1990s, Snowden-Ifft and collaborators made a first attempt at searching for damage features in muscovite mica produced by dark matter induced nuclear recoils~\cite{Snowden-Ifft:1995zgn}. In calibration studies, Snowden-Ifft and collaborators were able to image damage features from few-keV nuclear recoils of $Z \sim 10$ ions produced directly by ion implantation as well as indirectly by fast neutron irradiation using chemical etching and atomic force microscopy~\cite{SnowdenIfft:1993,Snowden-Ifft:1995rip,Snowden-Ifft:1995zgn}. During the last decade, various groups have demonstrated efficient readout of damage ``tracks'' produced by few-MeV $\alpha$-particles and even $\mathcal{O}(100)\,$MeV protons (having even smaller energy losses than MeV $\alpha$-particles) in doped sapphire (Al$_2$O$_3$:C,Mg)~\cite{Akselrod:2006,Bartz:2013,Kouwenberg:2018,Akselrod:2018,Kusumoto:2022} as well as (undoped) lithium fluoride (LiF)~\cite{Bilski:2017,Bilski:2019a,Bilski:2019b,Bilski:2020} using fluorescence microscopy of {\it color centers} created along the ion trajectories. These results demonstrate that, at least in some materials, persistent lattice damage that can be read out with existing microscopy techniques is produced by ions with stopping powers $dE/dx \sim 100\,{\rm keV}/\mu{\rm m}$ for chemical/plasma etching and atomic force/optical microscopy (see Refs.~\cite{SnowdenIfft:1993,Snowden-Ifft:1995rip,Snowden-Ifft:1995zgn} as well as contributions to this proceedings) and stopping powers as small as $dE/dx \sim 1\,{\rm keV}/\mu{\rm m}$ for fluorescence microscopy sensitive to color centers. However, while substantial experience exists in the physics and geoscience communities about lattice damage produced by ions with stopping powers in the traditional fission track and alpha-recoil track regime ($dE/dx \sim 10^4\,{\rm keV}/\mu{\rm m}$) in many materials (including annealing behavior, imaging methods, and an understanding of the microphysical nature of the latent damage tracks -- amorphization of the crystal structure and mechanical stress), much less is known about persistent damage features produced in the lower stopping-power regimes.

During the last years, natural as well as laboratory produced minerals have been proposed as passive nuclear recoil detectors to measure the fluxes of man-made (e.g., from nuclear reactors~\cite{Cogswell:2021qlq,Alfonso:2022meh}) and astrophysical neutrinos (e.g., from our Sun~\cite{Tapia-Arellano:2021cml}, supernovae~\cite{Baum:2019fqm,Baum:2022wfc}, and cosmic rays interacting with Earth's atmosphere~\cite{Jordan:2020gxx}), search for dark matter~\cite{Rajendran:2017ynw,Baum:2018tfw,Drukier:2018pdy,Edwards:2018hcf,Sidhu:2019qoa,Marshall:2020azl,Ebadi:2021cte,Acevedo:2021tbl,Baum:2021jak,Bramante:2021dyx,Ebadi:2022axg} and more exotic hypothetical objects such as magnetic monopoles or charged black holes (remnants)~\cite{Lehmann:2019zgt}, as well as for more practical applications such as neutron-detectors for nuclear safety and non-proliferation~\cite{Cogswell:2021qlq,Alfonso:2022meh} or various geoscience applications; see Ref.~\cite{Baum:2023cct} for an overview. In many aspects, these ideas revive previous attempts to use minerals as passive detectors for sources of low-energy and/or rare nuclear recoils, including Refs.~\cite{Goto:1958,Goto:1963zz,Fleischer:1969mj,Fleischer:1970vm,Fleischer:1970zy,Alvarez:1970zu,Kolm:1971xb,Eberhard:1971re,Ross:1973it,Price:1983ax,Kovalik:1986zz,Price:1986ky,Guo:1988,Ghosh:1990ki,Snowden-Ifft:1995zgn,Collar:1994mj,Engel:1995gw,Collar:1995aw,Snowden-Ifft:1996dug,Jeon:1995rf,Snowden-Ifft:1997vmx,Baltz:1997dw,Collar:1999md}. The main driver behind the revival of these ideas in the last few years are the advances over the past decades in microscopy technology as well as data analysis techniques. Regarding microscopy techniques, technologies such as conventional optical and fluorescence microscopy in diffraction-limited or super-resolution mode, laboratory-based X-ray techniques such as micro- and nanoCT (computer tomography), hard X-ray imaging techniques at synchrotron and free-electron laser light sources, scanning probe microscopes, electron and focused ion (including He-ion) beam microscopes, and atom probe tomography allow one to image samples with ever increasing resolution, lower detection thresholds, rapidly increasing throughput, and can increasingly produce three-dimensional images. Regarding data analysis techniques, most envisaged applications of mineral detectors will require processing enormous amounts of data: for example, the damage features produced by the few-keV nuclear recoils that would be produced by reactor, solar or supernova neutrinos or conventional dark matter candidates have characteristic sizes of $\mathcal{O}(1)\textit{--}\mathcal{O}(100)\,$nm. Scanning $\mathcal{O}(1)\,$kg of material (corresponding to a volume with linear dimensions of $\mathcal{O}(10)\,$cm) would correspond to some $10^{18}-10^{21}$ voxels. While a suitable combination of lower- and higher-resolution imaging techniques will allow for a massive reduction in data size (and make the imaging of the required amount of material in microscopes feasible), automated data analysis techniques for the online identification of the relevant patterns in image(-like) data will be a crucial component for virtually any application of mineral detectors. Artificial intelligence/machine learning (AI/ML) techniques are ideally suited to this task. 

To illustrate the potential of mineral detectors, consider the search for dark matter. One of the leading hypothesized candidates for the dark matter of our Universe, the missing 85\,\% of its matter budget, are so-called ``Weakly Interacting Massive Particle'' (WIMPs), elementary particles with mass roughly comparable (within an order of magnitude or two) to that of ordinary atomic nuclei and feeble interactions with ordinary matter. While the dynamics of our galaxy give us a fairly good idea of what to expect for the number density ($n_{\rm DM} \sim 10^{-3}\,{\rm cm}^{-3} \times 100\,{\rm GeV}/m_{\rm DM} = 1\,{\rm \ell}^{-3} \times 100\,{\rm GeV}/m_{\rm DM}$, with $m_{\rm DM}$ the particle mass) and speed relative to Earth ($v \sim 300\,$km/s), we know much less about the interaction strength of these particles with ordinary matter, parameterized via the scattering cross section. Motivated by the typical mass scale and relative speeds of WIMPs, one particular type of interaction often employed to search for them is elastic scattering off atomic nuclei in so-called direct detection experiments~\cite{Drukier:1984vhf,Goodman:1984dc,Drukier:1986tm}, giving rise to keV-scale nuclear recoils. The conventional direct detection approach is to assemble a large amount of clean and well-controlled material in a low background environment and to actively instrument this target volume with detectors that can measure phonons, scintillation photons, or the ionization charge produced by the dark matter induced nuclear recoils. From an technological viewpoint, the direct detection program has been an enormous success. Starting from few-hundred gram germanium detectors in the 1980s~\cite{Ahlen:1987mn}, detector technology has evolved to ever-larger detectors, lower recoil energy thresholds, and perhaps most importantly, backgrounds have been reduced in step with the increasing size of the detectors to levels that were hard to imagine at the start of the program more than four decades ago; see, for example, Refs.~\cite{Schumann:2019eaa, Billard:2021uyg} for reviews. For most of the classical WIMP mass region, $5\,{\rm GeV} \lesssim m_{\rm DM} \lesssim 10\,$TeV, time-projection chambers filled with liquid noble xenon or argon have emerged as the leading technology. The current generation of these detectors have reached the few-ton scale~\cite{DarkSide:2018kuk,DEAP:2019yzn,PandaX-4T:2021bab,LZ:2022lsv,XENON:2023cxc}. Alas, to date these detectors have not detected dark matter but instead given us increasingly stringent upper limits on the dark matter--nucleon interaction strength, evolving from $\sigma_n^{\rm SI}(m_\chi = 60\,{\rm GeV}) \lesssim 10^{-41}\,{\rm cm}^2$ in 1987~\cite{Ahlen:1987mn} to $\sigma_n^{\rm SI}(m_\chi = 60\,{\rm GeV}) \lesssim 10^{-47}\,{\rm cm}^2$ to date~\cite{LZ:2022lsv}, where $\sigma_n^{\rm SI}$ is the spin-independent dark matter--nucleon scattering cross section. The next generation of these detectors is envisaged to reach the tens-of-tons target mass scale~\cite{DarkSide-20k:2017zyg,Aalbers:2022dzr,PandaX:2024oxq}. While the track record of the community in constructing and operating these detectors is excellent, suggesting that the next generation will also be a technological success, these detectors are becoming ever more challenging and expensive to construct and operate. To exemplify the scale of the effort, note that the next generation of xenon-based detectors is envisaged to use $50 - 100$\,tons of xenon -- this is comparable to the annual worldwide production. While argon is much more widely available, the cosmogenic activation of argon extracted from Earth's atmosphere leads to prohibitively large background levels for next-generation direct detection experiments. Thus, a dedicated argon extraction chain is currently under construction, involving a deep-underground argon well (the Urania project in Colorado) and a 350\,m underground cryogenic distillation column (the ARIA/SERUCI project on Sardinia). 

Mineral detection offers an alternative path for the direct detection of dark matter. The concept of {\it paleo-detectors}~\cite{Snowden-Ifft:1995zgn,Baum:2018tfw,Drukier:2018pdy} is to use natural minerals formed hundreds of millions of years ago as natural recorders of dark matter interactions. These recording detectors can be read out by microscopically imaging the damage features produced by dark matter induced nuclear recoils in the sample. Paleo-detectors achieve a large exposure, the product of detector size and integration time, by maximizing the time over which the ``detector'' records signals rather than by maximizing the size of the target. Natural minerals found in geologically stable regions on Earth are as old as ${\sim}\,2\,$Gyr. In order to match the exposure the current generation of liquid noble gas detectors is aiming for, ${\sim}\,10\,$ton\,yr, with a mineral detector that has recorded the traces of dark matter interactions over $1\,$Gyr, one would only have to image $10\,$mg of target material, or a volume of ${\sim}\,1\,$mm linear dimensions. Conversely, by imaging a larger volume, one could realize exposures orders of magnitude larger than what next-generation direct detection experiments envision. The price to pay for this enormous exposure is that rather than operating a well-understood experiment in a controlled laboratory environment, paleo-detectors rely on naturally produced minerals with history that can only be known as well as the geological environment it resides in is understood. A number of theoretical studies have tried to characterize the backgrounds in such geological samples (see, for example, Refs.~\cite{Snowden-Ifft:1995zgn,Collar:1994mj,Engel:1995gw,Collar:1995aw,Snowden-Ifft:1996dug,Baum:2018tfw,Drukier:2018pdy,Edwards:2018hcf,Baum:2019fqm,Baum:2021jak}), giving reasons to be optimistic about the potential of paleo-detectors. In particular, the large exposure may offer a path to controlling backgrounds different from conventional direct detection experiments, where the traditional strategy is to design a signal region that is (close to) background free: in paleo-detectors, one would instead attempt to predict spectral distributions (e.g., in the size of damage features) of the various background components as well as of the hypothetical signal and then compare a background-only to a background+signal model fit of the data similar to the conceptual approach when, for example, searching for new particles at colliders. Reducing backgrounds as much as possible is of course nonetheless crucial for any paleo-detectors; some of the requirements discussed in the literature are that any geological mineral sample to be used as a dark matter paleo-detector must be taken from deep underground (in order to shield it from cosmic rays) and must stem from a relatively radiopure environment (for example, from, ultramafic deposits).

The enormous exposure times could allow mineral detectors to compete with conventional direct detection experiments in terms of the raw dark matter discovery potential. Moreover, integration times reaching up to $\mathcal{O}(1)\,$Gyr may allow one to use paleo-detectors to search for answers to questions inaccessible to conventional experiments. The number of events recorded in a direct detection experiment is proportional to the product of the dark matter interaction cross section and the flux of dark matter particles through the detector. A conventional experiment is thus sensitive to the current density and velocity distribution of dark matter in the vicinity of Earth, while a paleo-detector would be sensitive to the averaged dark matter distribution Earth encountered over its integration time. The solar system oscillates vertically through the Milky Way's stellar disk with amplitude $\Delta z \sim 100\,$pc and period $T_z \sim 90\,$Myr, and orbits the Milky Way with ${\sim}\,250\,$Myr period in the galactic plane in an approximately circular orbit (with variation in galactocentric distance $\Delta R \sim 300\,$pc). By comparing the signal in multiple mineral samples of different damage retention ages, one could thus measure the changes in the rate at which any dark matter induced signal is produced over timescales of tens of millions of years to a billion years, potentially providing information about the distribution of dark matter in our Galaxy~\cite{Baum:2021chx}. Another possible application of mineral detectors for dark matter searches is to use them as large passive detectors that can record dark matter induced events on laboratory timescales. The damage features produced in crystals by nuclear recoil carry information about the direction of the nuclear recoil and are thus correlated with the direction of the dark matter particle scattering off the nucleus. Thus, mineral detectors could be used as {\it directional} dark matter detectors~\cite{Spergel:1987kx}. Such a detector could, for example, be realized by actively instrumenting a large segmented volume of suitable natural or artificial crystals~\cite{Rajendran:2017ynw,Marshall:2020azl,Ebadi:2022axg}. The active instrumentation, e.g. via phonon sensors, would serve to obtain a time-stamp of any potential dark matter induced nuclear recoil event as in a conventional (solid state) direct detection experiment. By microscopically imaging the detector segment where the event occurred, one could then reconstruct the direction of the nuclear recoil. Such a detector might be a more compact alternative to using time projection chambers filled with low-pressure gas as directional dark matter detectors~\cite{Vahsen:2020pzb,Vahsen:2021gnb}.

Mineral detectors have also been discussed to measure sources of nuclear recoils other than dark matter. In particular, a wide variety of possible applications as neutrino detectors has been explored. Similar to WIMP Dark Matter, neutrinos in the MeV to GeV energy range interact with nuclei primarily via elastic scattering, mostly via so-called coherent elastic neutrino nucleus scattering (CE$\nu$NS), giving rise to nuclear recoils with typical energies in the keV range. Natural minerals could be used as paleo-detectors to measure the change of the solar neutrino fluxes over timescales of hundreds of millions of years, possibly shedding light on the solar composition problem~\cite{Tapia-Arellano:2021cml}. Mineral detectors could also measure the flux of neutrinos from core collapse supernovae in our Galaxy, shedding light on changes in the galactic supernova rate and hence the Milky Way's star formation history~\cite{Baum:2019fqm} and the flavor content of supernova neutrino bursts~\cite{Baum:2022wfc}. Neutrinos with larger energies, for example, atmospheric neutrinos produced by cosmic ray interactions with Earth's atmosphere, would interact with atomic nuclei predominantly via quasi-elastic as well as deep inelastic scattering. Searching for the damage features produced from the corresponding nuclear recoil cascades with paleo-detectors may allow one to probe the cosmic ray history of the Milky Way over gigayear timescales~\cite{Jordan:2020gxx}. Mineral detectors have also been discussed as passive (directional) neutrino detectors to measure neutrino fluxes on laboratory timescales. For example, mineral detectors could be used to monitor nuclear reactors~\cite{Cogswell:2021qlq,Alfonso:2022meh}. Another possibility discussed in this proceedings is to use a large array of mineral detectors similar to what would be required for the directional detection of dark matter as a directional detector for the neutrinos from the next nearby core collapse supernova.

Proposed applications of mineral detectors include also much more exotic as well as more practical cases. On the exotic end, paleo-detector applications of mineral detectors could be used to search for exotica that could leave spectacular signals in minerals and have only escaped detection in any type of detector to date because of their low flux. For example, if dark matter is composed of heavy composite objects or charged black hole (remnants), the flux of such particles on Earth would be $\Phi \sim f_{\rm DM} \times 1\,{\rm m}^{-2}\,{\rm yr}^{-1}\times(m_{\rm DM}/10^{19}\,{\rm GeV})$, where $f_{\rm DM}$ is the fraction of the dark matter comprised of these objects. For $m_{\rm DM} \gtrsim 10^{19}\,$GeV such particles could have escaped detection in conventional direct detection experiments which have ${\lesssim}\,1\,{\rm m^2}$ cross sections, and for $m_{\rm DM} \gtrsim 10^{23}\,$GeV, they could have easily escaped detection in low-threshold neutrino detectors such as Borexino, NO$\nu$A, or Super-Kamiokande which are ${\lesssim}\,10^2\,{\rm m^2}$ in size. Exotica such as heavy composite dark matter candidates~\cite{Sidhu:2019qoa,Ebadi:2021cte,Acevedo:2021tbl}, magnetic monopoles, or charged black hole remnants~\cite{Lehmann:2019zgt} could leave long ($\gtrsim$\,cm) damage tracks with macroscopic cross sections ($\gtrsim$\,$\mu$m$^2$) in materials that would be relatively easy to search for using, for example, fast scanning electron microscopy techniques~\cite{Ebadi:2021cte}. If one imagines scanning a (stack of) m$^2$-sized surface(s) of Gyr old materials for such damage features, one would obtain an exposure equivalent to a conventional detector with $\sim$\,yr exposure times and a ${\sim}\,10^3$\,km$^2$ cross section. 

On the perhaps more practical end, mineral detectors could, for example, be used as passive recording detectors for fast neutrons. Fast neutrons traveling through a crystal produce keV-scale nuclear recoils via elastic, quasi-elastic, and inelastic interactions with nuclei. Thus, a mineral detector could measure the neutron flux integrated over the time it is exposed to a neutron source. Furthermore, mineral detectors would be imaging neutron detectors since they would record the location and direction of the neutron-induced nuclear recoils, although mineral detectors would not have any time-resolution beyond microscopically imaging the sample prior to and post irradiation. As neutron detectors, mineral detectors have potential applications relevant to nuclear security, e.g., they could play a key role in certain systems of nuclear arms control~\cite{Cogswell:2021qlq,Alfonso:2022meh}. In geoscience, mineral detectors have of course long been used for dating as well as to obtain time-temperature profiles of geological deposits by measuring (partially annealed) fission and/or alpha-recoil tracks. However, the standard imaging techniques employed to date in geoscience rely on chemically enlarging the tracks by applying an etching agent to a (prepared) surface of a sample and then imaging the etched features. This approach is inherently two dimensional since the chemical etching will only attack crystal damage intersecting the surface the etchant is applied to, limiting these techniques to samples with a relatively high density of fission or alpha recoil tracks. Practically all of the applications of mineral detectors above would required a volumetric readout of damage features -- such a technique could revolutionize applications of mineral detectors in geoscience, for example, allowing one to obtain the thermal history of a single mineral grain~\cite{Baum:2023cct}.

The potential applications of mineral detectors for neutrinos, dark matter, and a number of other applications are vast, as the above discussion and a number of theoretical studies over the last years have demonstrated. Sparked by this motivation, experimental efforts to explore the feasibility of mineral detectors have begun at institutions in North America, Europe, and Asia during the past years. The challenges are, without doubt, enormous: we are lacking a detailed understanding of the nature of damage features formed by keV-scale nuclear recoils, their annealing behavior, how to best image them, how to analyze the data, which geological environments and classes of materials are best suited for paleo-detector applications, and have no experimental understanding of the backgrounds that will be encountered when imaging a sample of the required size with the necessary spatial resolution, to name only some of the known unknowns. This is a highly cross-disciplinary endeavor, combining fundamental and applied physics, materials science, geoscience, AI/ML techniques and quantum information. In October 2022, the first ``Mineral Detection of Neutrinos and Dark Matter'' (MD$\nu$DM) workshop took place at the Institute for Fundamental Physics of the Universe (IFPU) in Trieste, Italy\footnote{\href{https://agenda.infn.it/event/32181/}{\url{https://agenda.infn.it/event/32181/}}}. Following that workshop, the community published a whitepaper summarizing the state and prospects of this emerging field~\cite{Baum:2023cct}. This proceedings collects contributions from the second MD$\nu$DM workshop which took place in January 2024 in Arlington, VA, USA, hosted by Virginia Tech's Center for Neutrino Physics\footnote{\href{https://indico.phys.vt.edu/event/62/}{\url{https://indico.phys.vt.edu/event/62/}}}. As displayed by the contributions below, MD$\nu$DM is a vibrant field bringing together researchers with a range of backgrounds, stimulating interesting discussions and ideas. A number of experimental groups have presented exciting results at MD$\nu$DM'24, motivating the continued exploration of mineral detectors as well as sharpening our understanding of the challenges ahead. Bringing mineral detection to a stage where it can have sensitivity rivaling that of existing conventional detectors will require years of concerted effort by different groups bringing together the required expertise. However given the scale conventional detectors have reached, as discussed above for direct dark matter detection experiments, such an effort seems well-motivated to us. Furthermore, mineral detection could provide answers to questions inaccessible for conventional detectors such as the variation of signal rates over Myr to Gyr timescales.  We are setting ourselves ambitious goals. The next MD$\nu$DM meeting is scheduled for May~2025 hosted by JAMSTEC in Yokohama, Japan -- some of the topics on which the community aims to make progress until then are:
\begin{itemize}
    \item better understanding of damage feature formation, nature, and retention in a variety of materials by controlled irradiation measurements and imaging with a variety of microscopy techniques,

    \item development of low-level data analysis techniques to reconstruct nuclear recoils from microscopy data,

    \item improvement of modeling tools of crystal damage bridging the gap between (quantum) molecular dynamics simulations and parameterized particle transport codes such as SRIM/TRIM, FLUKA or GEANT4,

    \item characterize and model sources of backgrounds in natural samples,

    \item identify promising (classes of) artificial materials for laboratory-timescale exposure applications as well as natural minerals and promising geological environments for paleo-detector applications -- ideal mineral detectors are available in sufficient volumes, susceptible to substantial and well-understood damage effects of nuclear recoils in the energy range of interest, highly crystalline, and stable during sample collection and analysis,
    
    \item demonstrate the detection of any type of damage feature from astroparticle interactions.

\subsection*{Acknowledgements}
We thank the Center for Neutrino Physics at Virginia Tech for hosting the MD$\nu$DM'24 workshop in January 2024.

\end{itemize}

\FloatBarrier
\newpage
\section{Understanding control and modulation of color center defects and their use for dark matter detection}

Authors: {\it Mariano~Guerrero~Perez$^{1}$, Pranshu~Baumik$^{1}$, Vsevolod~Ivanov$^{1,2,3}$}
\vspace{0.1cm}\\
$^{1}$National Security Institute, Virginia Tech, $^{2}$Department of Physics, Virginia Tech,\\
$^{3}$Center for Quantum Information Science and Engineering, Virginia Tech
\vspace{0.3cm}\\
Color centers are few-atom defects in optically transparent materials that can host electron spins. Transitions between the localized energy levels of these defects can emit photons, making color centers a promising spin-photon interface that could form the basis of a numerous quantum technologies, including quantum networking, positioning, and sensing. Such applications require color centers with specific properties, which has led to various high-throughput searches for defect candidates~\cite{Ivanov:2023cqn, Bertoldo2022, Xiong:2023svj} and a flurry of experimental efforts in synthesizing them~\cite{Higginbottom:2022day, Udvarhelyi:2022qew, Redjem:2022nty, Zhiyenbayev23, Jhuria:2023wkm, Wolfowicz20}. 

Due to the sensitive nature of their electronic and optical properties, color center defects are now being explored as an avenue for detecting nuclear recoil events caused by high-energy particles such as neutrons, neutrinos, and dark matter. As a result of their promise for quantum applications, a variety of computational workflows have been developed to precisely predict defect properties, which we can take advantage of to design defects for dark matter detection. These properties are highly sensitive to external influence, and in fact significant work has been done to understand how these properties can be modulated or controlled by temperature, strain, and local disorder. For the purpose of sensing nuclear recoils, these studies remain highly relevant, but instead of controlling defect properties, particular external influences can be detected by measuring the changes in these properties. 
 
There are several possible ways in which high energy particles might be detected using color center defects - 1. direct interaction with a color center, leading to electronic excitation or structural transformation of the defect, 2. damaging the lattice, affecting the optical properties of nearby defects, or 3. creating optically active color center defects directly from the nuclear recoils. Here we explore the potential of these approaches, using prior simulations of color center defects in silicon for context and to assess the feasibility. 

For the particles we are considering, namely neutrinos and dark matter, collision cross sections with atomic nuclei in materials are already quite low, and using a detection scheme predicated on direct interaction with a rare defect in a crystal lattice further lowers observation probabilities by several order of magnitude. Nevertheless, this approach has the advantage of being able to detect much lower energy recoils, as defect structural transformations can be on the order of 10s of meV~\cite{Song1990}. In certain defects, a structural transformation is accompanied by a dramatic change in brightness~\cite{Ivanov:2022wfa, Jhuria:2023wkm}, making the transformation convenient to detect optically.  

At much higher energies particles will form large regions of damage and long tracks in crystalline materials. Such damage can be detected both through the high local temperatures it generates and the local lattice strain from the local damage. Color center defects are highly sensitive to lattice distortions - we have previously shown in silicon how local lattice temperatures can broaden emission peaks~\cite{Redjem:2022nty}, and also cause an overall shift in the emission frequency~\cite{Zhiyenbayev23}. The underlying mechanism for the emission frequency shift is due to local temperature-driven lattice strain, which can also arise directly when the lattice is damaged~\cite{Liu:2023flm}. Since in general the shift response to strain is anisotropic, this type of detection is also directional.

\begin{figure}
   \centering
   \includegraphics[width=0.5\textwidth]{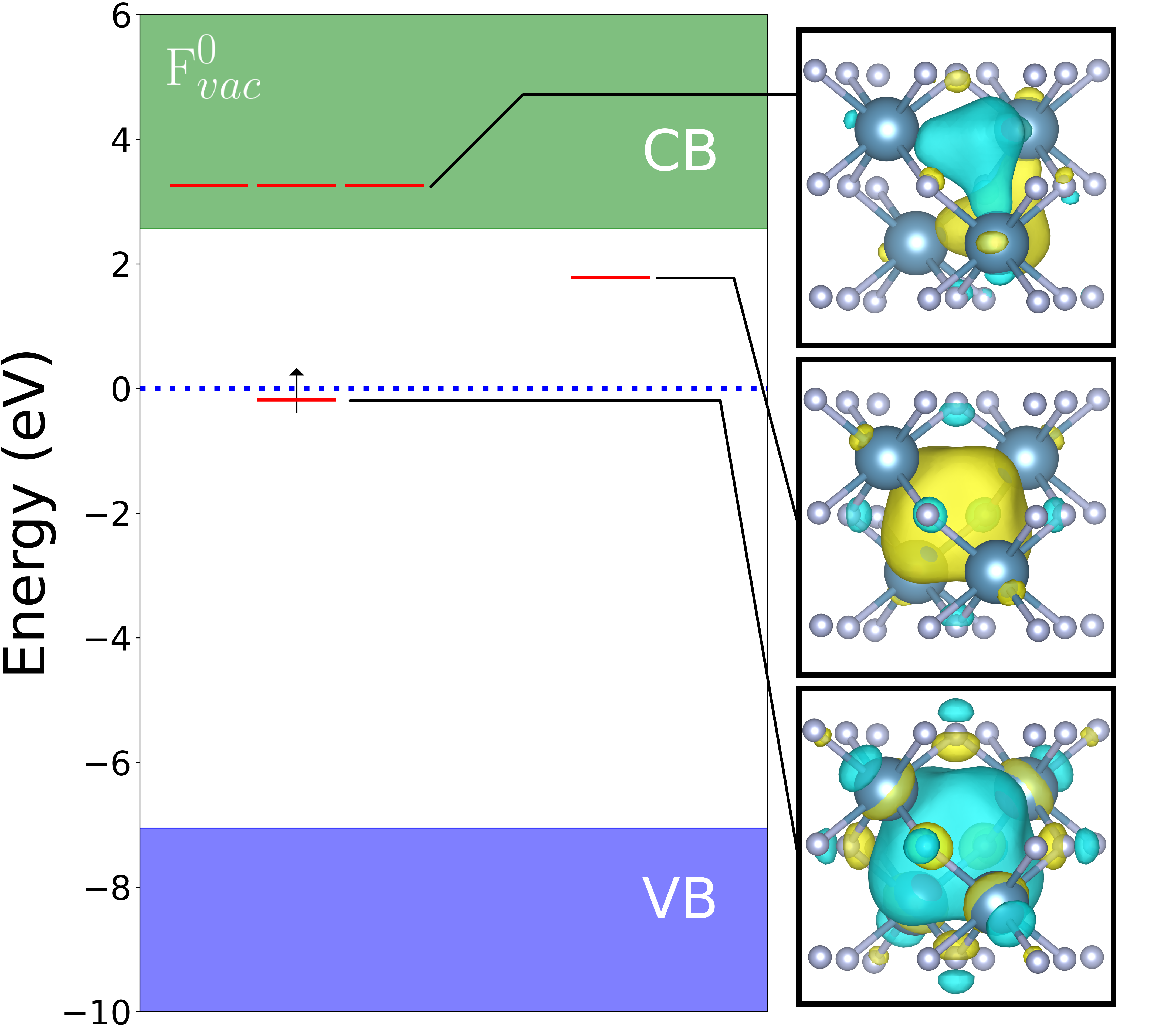}
   \caption{Energy level diagram for the neutral F vacancy defect, showing the valence band (VB, blue), conduction band (CB, green), and localized defect levels (red lines). Wavefunctions corresponding to the local levels are shown on the right}
   \label{caf2_defect}
\end{figure}

By far the most promising approach to detect dark matter using color center defects is observing their direct formation. For this, a suitable detector material needs to be selected. The detector material must be have a wide band gap (optically transparent), have optically active color center defects, and most importantly, be cheap to produce with extremely high purity. The cost and band gap requirements are relatively straightforward to satisfy, leading to a large number of candidate materials which are optimal to screen for optically active defects using first-principles methods. We demonstrate our defect property workflows by computing the formation energies and zero-phonon line (ZPL) emission of self-interstitial and vacancy defects in two such materials, LiF and CaF$_2$. We find that the fluorine vacancy defect in both materials is optically bright, with a ZPL emission of 3.82 eV (324 nm) in LiF and 1.52 eV (815 nm) in CaF$_2$. Figure~\ref{caf2_defect} shows the positions of the localized defect levels within the gap for the fluorine vacancy in CaF$_2$, along with the corresponding wavefunctions. 

An optimal detector material would have color center defects with photoemission within the visible band (400-700nm), so that they can be readily observed using off-the-shelf photodetectors, with LiF and CaF$_2$ lying just outside this range. The formation energies of these defects can also be computed, which can then be used in combination with optical measurements of detector materials to establish energy constraints in dark matter detection experiments~\cite{Alfonso:2022meh}.

\subsection*{Acknowledgements}
This work is supported by startup funding from Virginia Tech, and by the Inclusive Excellence project at Virginia Tech (\href{https://ie.vt.edu/}{\url{https://ie.vt.edu/}}), based on a grant from the Howard Hughes Medical Institute.

\FloatBarrier
\newpage
\section{DMICA: exploring Dark Matter in natural MICA}

Authors: {\it Shigenobu~Hirose$^1$, Natsue~Abe$^1$, Qing~Chang$^1$, Takeshi~Hanyu$^1$, Noriko~Hasebe$^2$, Yasushi~Hoshino$^3$, Takashi~Kamiyama$^4$, Yoji~Kawamura$^1$, Kohta~Murase$^5$, Tatsuhiro~Naka$^6$, Kenji~Oguni$^1$, Katsuhiko~Suzuki$^1$, and Seiko~Yamasaki$^7$}
\vspace{0.1cm} \\
$^1$Japan~Agency~for~Marine-Earth~Science~and~Technology, $^2$Kanazawa~University, \\
$^3$Kanagawa~University, $^4$Hokkaido~University, $^5$The~Pennsylvania~State~University,\\ \
$^6$Toho~University, $^7$National~Institute~of~Advanced~Industrial~Science~and~Technology
\vspace{0.3cm}

\subsection{Mica as Dark Matter Detector}
Muscovite mica, referred to here as mica, is an efficient solid-state track detector. Utilizing this property, a mica-based dark matter (DM) detection experiment was conducted by \cite{Snowden-Ifft:1995zgn}. Prior to these experiments, DM-scattered atomic nuclei with energies in the keV/amu range had initiated cascades of nuclear scatterings within the crystal, which created series of atomic vacancies and formed latent tracks in the mica. These tracks, once etched as part of the experimental procedure, appeared as microscopic pits, marking the scattering events for analysis.

The etching process only reveals a portion of the latent track, preventing a direct determination of the nuclear recoil energy associated with the track's full extent. To utilize mica as a detector with energy resolution capabilities, a model linking the depth of nanometric pits with the recoil energy was employed in \cite{Snowden-Ifft:1995zgn}. Developed by \cite{Snowden-Ifft:1995rip} based on ion irradiation experiments, this model facilitates the conversion of the recoil energy spectrum into a pit-depth histogram, an observable quantity derived from etching. After examining 80,720 $\mu$m$^2$ of mica, corresponding to an exposure of approximately $10^{-6}$ ton-year, and not detecting the anticipated features in the histogram's region of interest (see Sec. \ref{sec:backgrounds_and_sensitivity_projection_for_dmica}), constraints were set on the interaction cross-section of dark matter \cite{Snowden-Ifft:1995zgn} (cf. Fig.\ref{fig:DMICA_sensitivity}).

\subsection{Goal and Current Status of Project DMICA}
\begin{figure}
   \centering
   \includegraphics[width=0.7\textwidth]{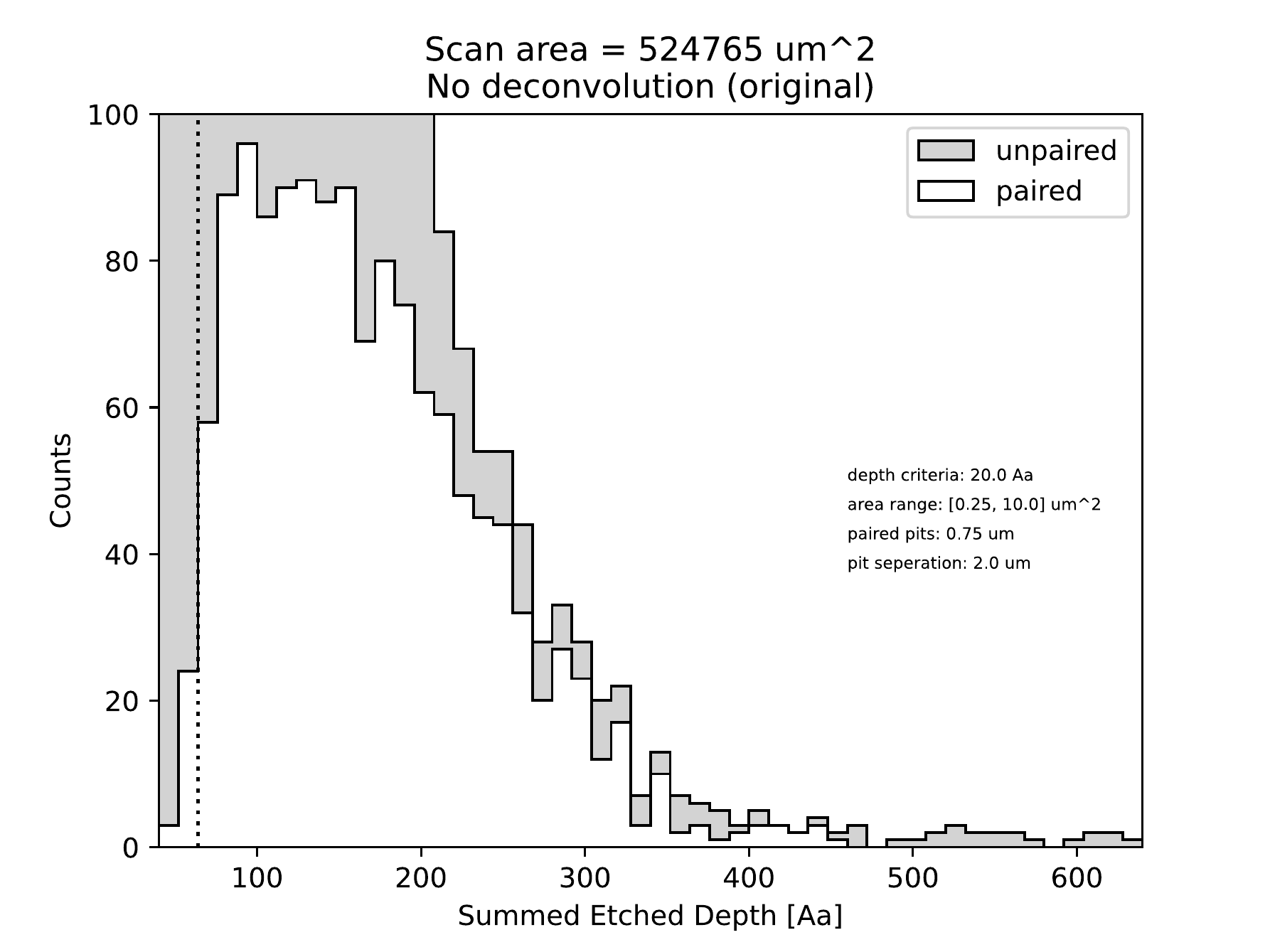}
   \caption{Pit-depth histogram from the processed mica data covering 524,765 $\mu$m$^2$. Currently, the experimental setup has not secured any ARTs-free bins (see Sec.\ref{sec:backgrounds_and_sensitivity_projection_for_dmica}).}
   \label{fig:DMICA_histogram}
   \vspace{1cm}
   \includegraphics[width=0.7\textwidth]{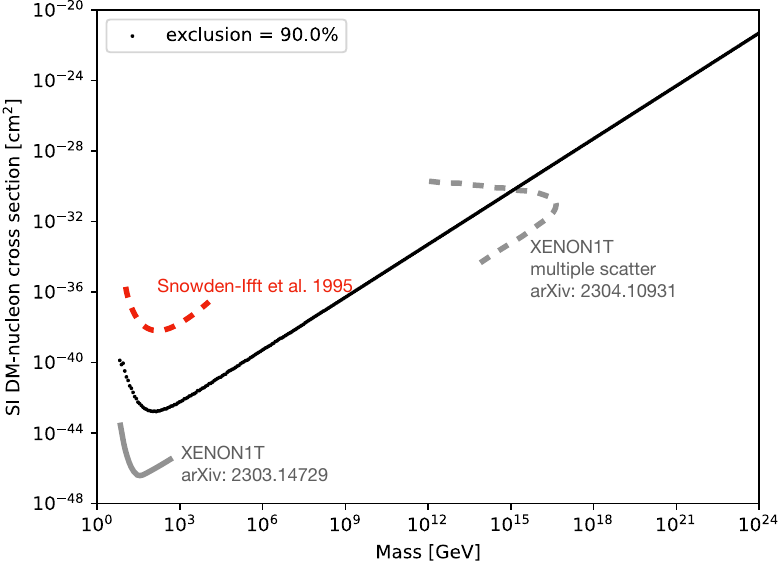}
   \caption{Projected 90\%-confidence exclusion curve for spin-independent DM-nucleon cross-section for DMICA with an exposure of 1 ton-year (black), compared with previous work \cite{Snowden-Ifft:1995zgn} (red dashed) and the XENON1T experiments \cite{XENON:2023cxc, XENON:2023iku} (gray and gray dashed).}
   \label{fig:DMICA_sensitivity}
\end{figure}

The bottleneck for improving throughput in \cite{Snowden-Ifft:1995zgn} was the scanning speed of AFM used to explore the surfaces of mica. In our DMICA project, we leverage \cite{Snowden-Ifft:1995zgn}'s methodology but innovate by employing an optical profiler for mica surface scanning. This approach holds the potential to meet our objective of exploring mica with an exposure of 1 ton-year by surpassing previous scanning speed constraints. Our preliminary work has processed 524,765 $\mu$m$^2$ of mica, expanding the explored area by 6.5-fold compared to \cite{Snowden-Ifft:1995zgn}. The resulting pit-depth histogram, shown in Fig. \ref{fig:DMICA_histogram}, not only allows for a direct comparison with \cite{Snowden-Ifft:1995zgn}'s Fig.3(a) but also underscores the similarity in pit patterns, highlighting the capability of our experimental setup to replicate basic aspects of previous findings.

\subsection{Backgrounds and Sensitivity Projection for DMICA}\label{sec:backgrounds_and_sensitivity_projection_for_dmica}
The primary challenge in paleo detectors lies in their inability to veto background noise, as the events of interest have already occurred. For mica, the main background comprises alpha recoil tracks (ARTs) from the $\alpha$ decay chain of radiogenic elements like Uranium-238. In \cite{Snowden-Ifft:1995zgn}, the shallowest bins of the pit-depth histogram were experimentally identified as the region of interest, where pits from ARTs are unlikely, whereas pits from DM recoils were expected. However, expanding the scanning area beyond \cite{Snowden-Ifft:1995zgn} introduces the potential for additional background floors in the ARTs-free region: the radiogenic neutron floor \cite{Snowden-Ifft:1996dug} and the Thorium-234 floor, arising from the initial $\alpha$ decay in the Uranium-238 chain \cite{Collar:1995aw}. 

It is crucial for DMICA, which significantly extends the scanning area beyond \cite{Snowden-Ifft:1995zgn}, to evaluate these backgrounds. By incorporating them using \texttt{paleoSens}/\texttt{paleoSpec} \cite{Baum:2018tfw,Drukier:2018pdy,Baum:2019fqm,Baum:2021jak}, we projected DMICA's sensitivity in Fig.~\ref{fig:DMICA_sensitivity}. Despite a similar exposure, DMICA's upper limit is a few orders of magnitude higher than XENON1T's. This discrepancy is likely due to the two additional background floors, the limited number of ARTs-free bins in the projection, and the mica's low detection efficiency (e.g. Fig.~44 in \cite{Baum:2023cct}). To improve DMICA's performance, given that these backgrounds cannot be vetoed and boosting mica's detection efficiency presents challenges, we must adopt experimental strategies to secure and augment the ARTs-free bins (see Fig. \ref{fig:DMICA_histogram}), such as extending the etching time as suggested in \cite{Snowden-Ifft:1995zgn}.

\subsection*{Acknowledgements}
SH is grateful to Prof. Snowden-Ifft for his invaluable insights and discussions, which have significantly shaped the development of project DMICA. SH is also thankful to Dr. Baum for his guidance on the usage and details of \texttt{paleoSpec/paleoSens}.

\FloatBarrier
\newpage

\section{Towards a Prototype Paleo-Detector for Supernova Neutrino and Dark Matter Detection}

Authors: {\it Emilie LaVoie-Ingram, Greg Wurtz, Chris Kelso, and Zane Cable} 
\vspace{0.1cm} \\
University of North Florida
\vspace{0.3cm}\\
At the University of North Florida (UNF), we are interested in revealing nuclear recoil damage tracks in prototype paleo-detector minerals with various etching methods and measuring tracks with laser confocal and atomic force microscopy, ultimately searching for traces of core-collapse supernova neutrinos and dark matter. Preliminary experiments include plasma etching halite, Muscovite mica, and Phlogopite mica to analyze track etch behavior and density. The most extensively studied of these minerals is halite. We have revealed tracks in halite with argon plasma etching, imaged tracks with laser confocal microscopy, and developed an initial mathematical model comparing experimentally collected data to theoretical background models developed by Refs. \cite{Baum:2018tfw,Drukier:2018pdy,Baum:2019fqm}.

\begin{figure}
  \centering
  \includegraphics[width=.45\textwidth]{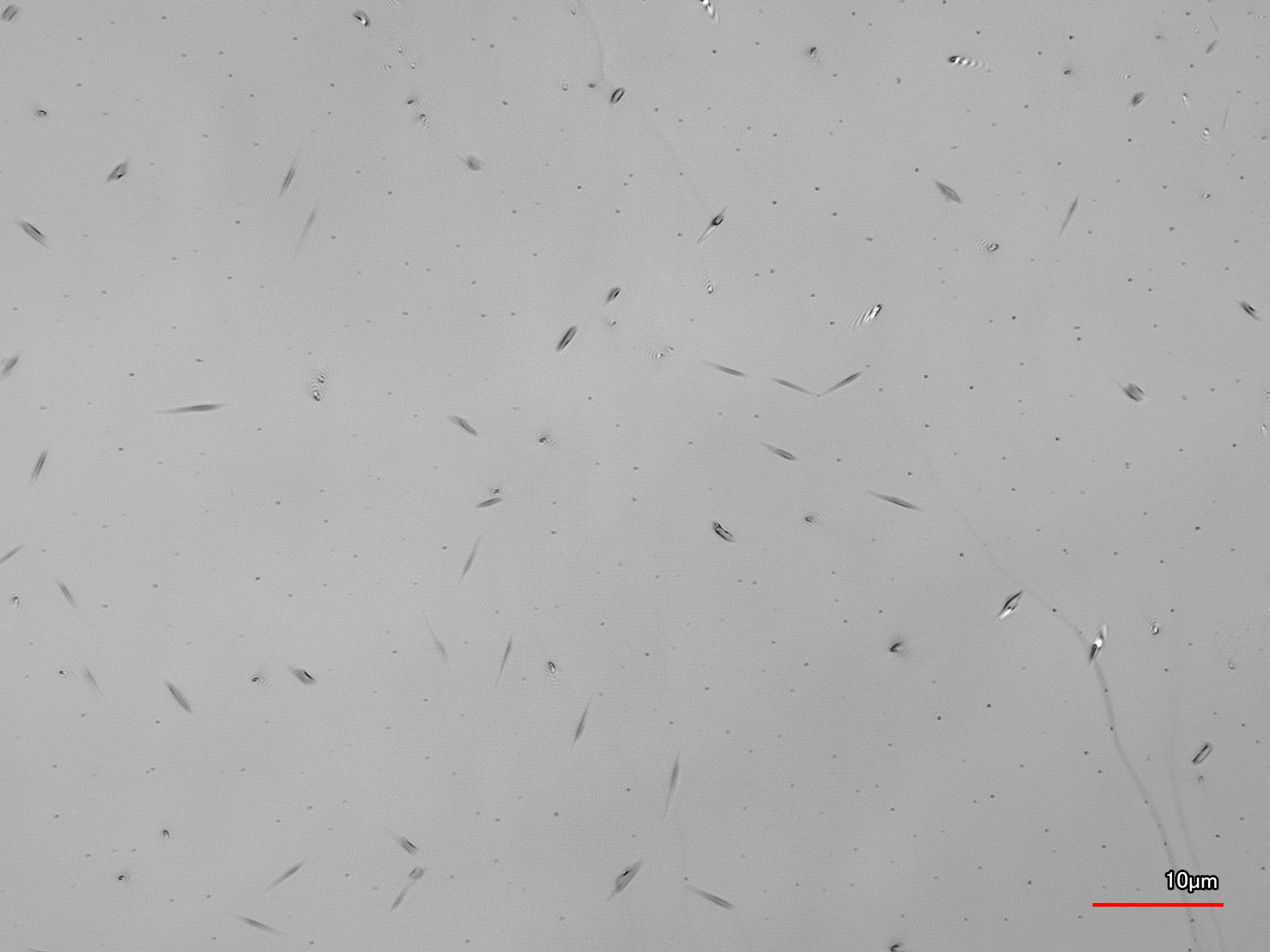}
  \hspace{0.5cm}
  \includegraphics[width=.45\textwidth]{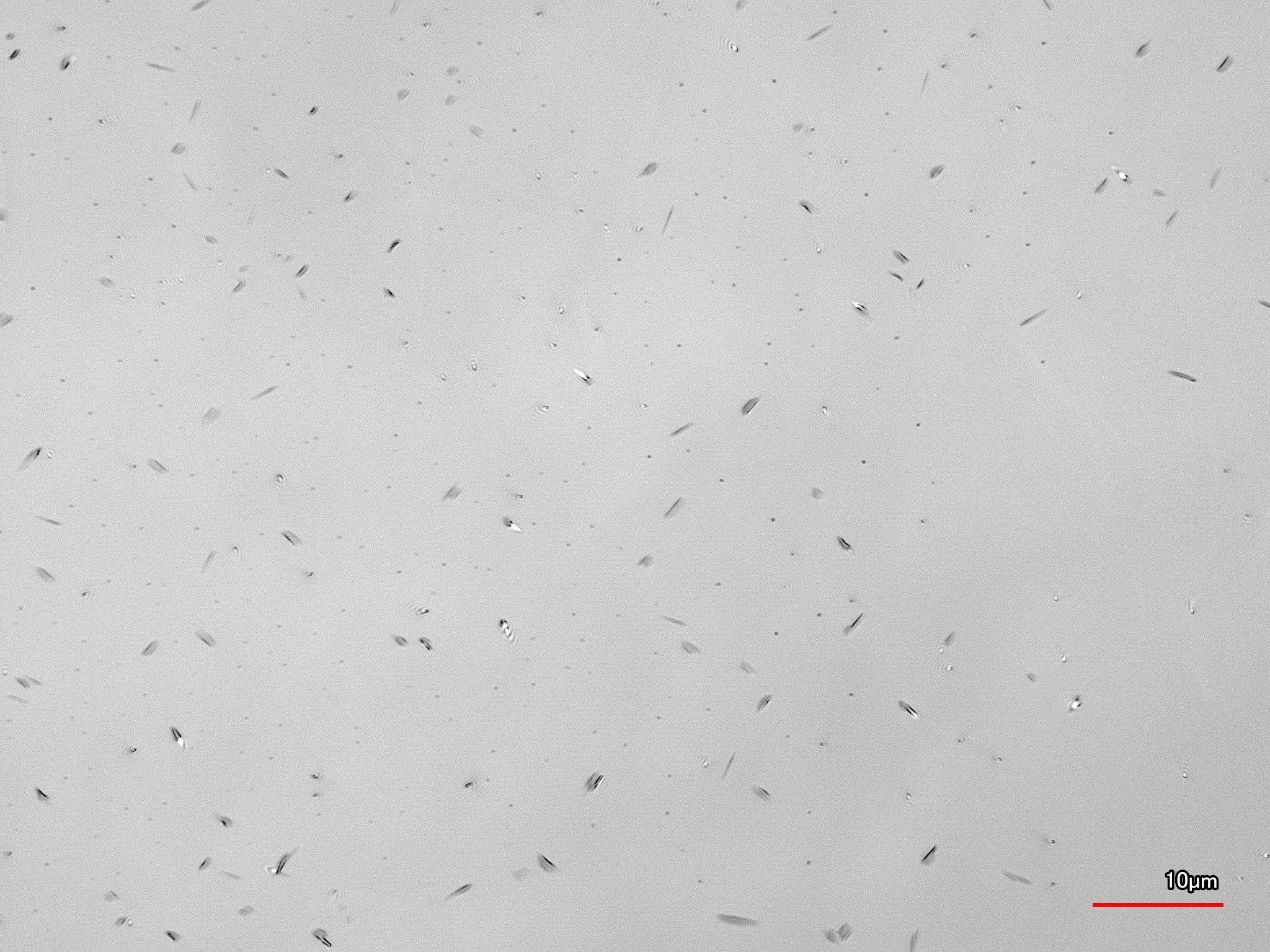}

  \caption{Nuclear recoil tracks revealed in Argon plasma etched halite, imaged with the Keyence VK-X1000 Laser Confocal Microscope at UNF's Materials Science and Engineering Research Facility (MSERF).}
  \label{fig:halitetracks}
\end{figure}

Our halite samples were etched with argon plasma at 100 W power, 0.2 mbar pressure, for 10 minutes each, using the Department of Physics' Diener Pico plasma cleaner. Tracks were revealed across many samples, examples as shown in Figure \ref{fig:halitetracks}. Across all etched samples, we determined an average track density in halite of approximately $\sim$ 6.8 $\times 10^6$ cm$^{-2}$. Tracks were automatically counted and measured using a Python-based automatic track detection algorithm, taking into consideration resolution limitations and track geometry with respect to the cleave surface, and then converted back to an approximate original length using an experimentally determined etch rate of halite under argon plasma. Once experimental data was collected, we compared our data against theoretical track background models developed by Refs. \cite{Baum:2018tfw,Drukier:2018pdy,Baum:2019fqm}. The model included radioactive $^{234}$Th+$\alpha$ decays and neutron backgrounds as well as an anticipated efficiency function. The final model used to fit to our experimental data as a function of track length ($x$) was written as:

\begin{equation}
    \frac{1}{1 + e^{-c(x - d)}} \times A \left[B e^{-\frac{1}{2} \left(\frac{x - x_0}{\sigma} \right)^2} + e^{-\beta x} \right]\,{\rm nm}^{-1}\,{\rm kg}^{-1}\,{\rm Myr}^{-1},
    \label{eq:math_model}
\end{equation}
where \textit{c} is the slope of the efficiency function, \textit{d} is the track length where our efficiency function is at 50\% , \textit{A} is the overall normalization of the thorium and neutron backgrounds, \textit{B} is the relative strength of of the Th-$\alpha$ peak background compared to the neutron background, $\sigma$ is the width of the Th-$\alpha$ peak, $x_0$ is the central value of the Th-$\alpha$ peak, and $\beta$ is the rate of decay of the exponential neutron background. The addition of a Sigmoid efficiency function accounts for limitations from resolution of our readout technology, limitations from the optical properties of halite, errors in etching rate measurement and conversion, limitations of the automatic track detection code, etc. A chi-squared goodness of fit test was used to determine the best fit between this mathematical model and our experimental data. 

\begin{figure}
    \centering
    \includegraphics[width=0.98\textwidth]{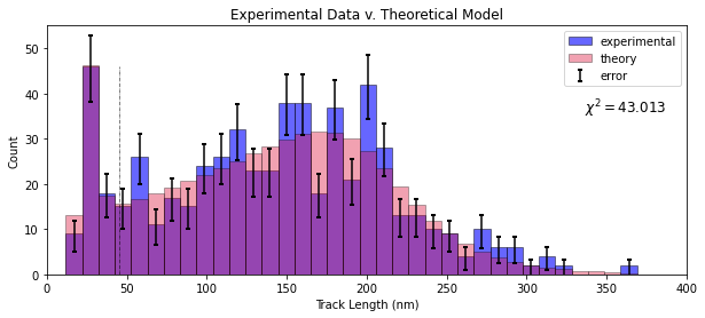}
    \caption{Experimentally collected track lengths in halite (blue) overlaid with our theoretical mathematical model (pink) based on Refs. \cite{Baum:2018tfw,Drukier:2018pdy,Baum:2019fqm} with variables determined by the minimum chi-square test statistic derived, 43.013. Error bars are computed based on Poisson fluctuations of the count number.}
    \label{fig:track_model}
\end{figure}

The best fit model produced $c=0.055$\,nm$^{-1}$, $d=200$\,nm, $A=6.2\cdot 10^5$\,nm$^{-1}$\,kg$^{-1}$\,Myr$^{-1}$ with the minimum $\chi^2=43.0$. Figure~\ref{fig:track_model} shows the experimental halite data along with our best fit model. This is a promising initial result comparing experimental data to theory. 

Future work will involve collecting more data (track measurements) and trying to obtain samples with better known properties such as background concentration, thermal history, and depth of extraction.  Optimization of both the etching conditions of halite and the automatic track detection algorithm is continuing. We intend to replicate similar procedures for Muscovite and Phlogopite mica to contribute to our understanding of backgrounds and measurement methodology for these potential paleo-detector minerals.

\subsection*{Acknowledgements}
This work is supported by the NASA Florida Space Grant Consortium Master's Fellowship.

\FloatBarrier
\newpage

\section{Progress toward a solid-state directional dark matter detector}

Authors: {\it Daniel~Ang, Xingxin~Liu, Jiashen~Tang, Maximilian~Shen, Reza~Ebadi, \\ Ronald~Walsworth}
\vspace{0.1cm} \\
Quantum Technology Center, University of Maryland
\vspace{0.3cm}

\subsection{Introduction}
The next generation of WIMP dark matter detectors are expected to approach the ``neutrino fog," where coherent scattering of solar neutrinos will be observable, making it challenging to extract the WIMP signal~\cite{OHare:2021utq}. However, due to the differing expected angular distributions of the WIMP and solar neutrino fluxes, the capability to determine the direction of an incoming particle would allow the rejection of solar neutrinos and provide insights into the cosmological origin of a WIMP signal~\cite{Mayet:2016zxu,Vahsen:2021gnb}. This could be performed by a hybrid ``conventional-directional" solid-state dark matter detector where nuclear recoils are registered in real-time by conventional readout techniques and directional information is extracted by high-resolution imaging of the resulting crystal damage track inside the material~\cite{Rajendran:2017ynw,Marshall:2020azl,Ebadi:2022axg}. Such a detector is particularly promising due to its higher target density compared to gaseous or emulsion detectors~\cite{Vahsen:2020pzb,Shimada:2023vky,Umemoto:2023hmt}, the availability of materials with favorable semiconductor properties and low nuclear mass~\cite{Kurinsky:2019pgb,Griffin:2020lgd}, and the ability to utilize color centers in the material for fast and precise identification and characterization of the damage track~\cite{Marshall:2020azl,Ebadi:2022axg}. 

\subsection{Detector scheme}
\begin{figure}
   \centering
   \includegraphics[width=0.6\textwidth]{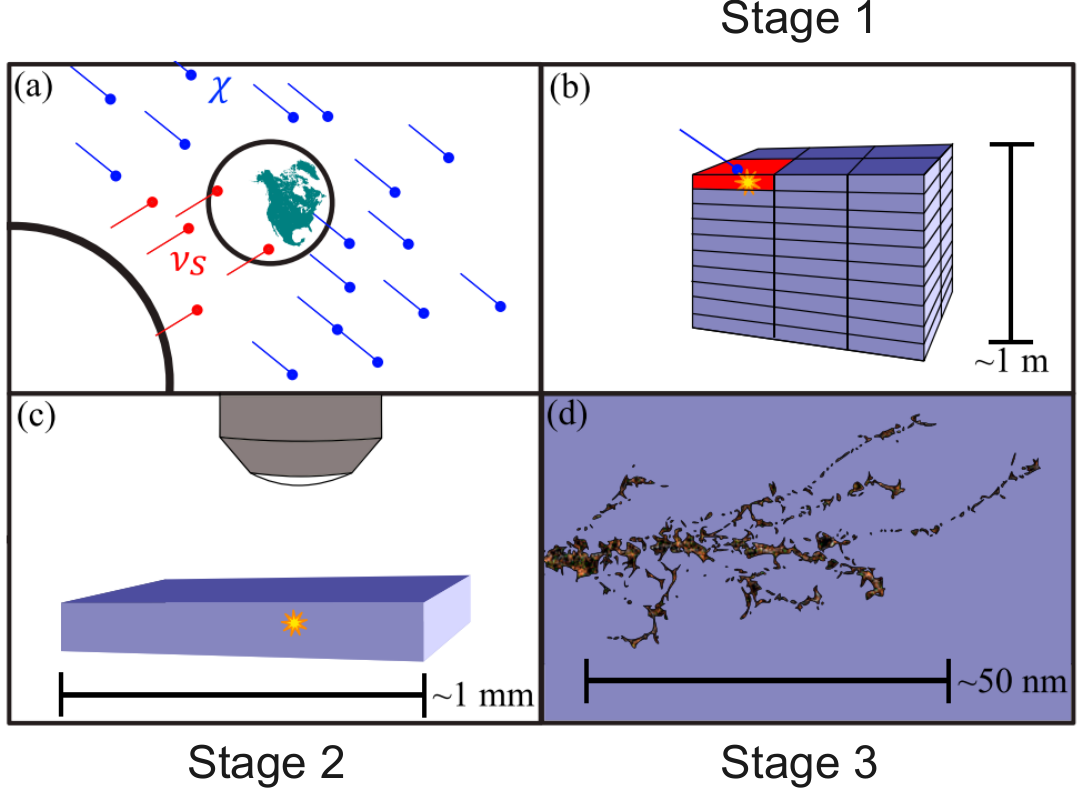}
   \caption{Multi-stage detection in a solid-state directional matter detector. a)~Solar neutrinos and WIMPs stream through the earth with distinct preferred directions. b)~Stage~1: WIMP or neutrino particles interact with a macroscopic detector, causing a nuclear recoil recorded in real-time with conventional readout methods. c)~Stage~2: micron-scale localization with a QDM. d)~Stage~3: nanoscale imaging of the damage track. Figure adapted from~\cite{Marshall:2020azl}.}
   \label{fig:dmdetectionscheme}
\end{figure}

The detector would consist of a $\sim$m$^3$ volume of diamond or silicon carbide (SiC) divided into numerous removable millimeter-scale segments. Occasionally, a WIMP or neutrino interacts with the detector mass. Simulations with diamond indicate that a $\sim$1-100\,GeV WIMP particle would lead to a $\sim$10-100\,keV nuclear recoil, producing a $\sim$10-100\,nm long-lasting damage track~\cite{Rajendran:2017ynw}. Localization and measurement of this signal would proceed in three stages (Fig.~\ref{fig:dmdetectionscheme}). In stage~1, the nuclear recoil is recorded in real-time using charge, phonon, or photon collection similar to those in existing semiconductor-based detectors~\cite{LopezAsamar:2019smu,DAMIC:2021crr}. The location of the interaction is also identified and the corresponding detector segment is removed for further analysis. 

In stage~2, the location of the damage track within the mm-scale detector segment is identified with micron-scale precision. Two readout strategies are possible~\cite{Ebadi:2022axg}. The first is optical strain spectroscopy of pre-existing color centers in the material, looking for a strain spike in color centers located near the damage track. The second is to use material with few existing color centers and detect the fluorescence from new centers created by the damage track. Both of these methods can be applied to color centers such as nitrogen vacancies (NVs) in diamond~\cite{Doherty:2013uij,Kehayias:2019,Marshall:2021xiu} or divacancies in SiC~\cite{Castelletto:2020ggk,Koehl:2011aln,Falk:2014,Udvarhelyi:2018}.

Having identified the $\sim$$\mu\mathrm{m}^3$ voxel containing the damage track, we proceed to stage~3, where nanometer-scale imaging techniques such as X-ray diffraction spectroscopy~\cite{Winarski:2012,Marshall:2021kjk} or superresolution spectroscopy of color centers~\cite{Maurer:2010,Jaskula:2017,Marshall:2020azl} are applied to measure the shape, orientation, and length of the damage track. This information can then be correlated with stage~1 real-time data to determine the direction and origin of the signal. In order to keep up with the anticipated event rate of a full m$^3$-scale detector, stages 1-3 need to be completed within a conservative benchmark time of 3\,days~\cite{Ebadi:2022axg}.

\subsection{Recent experimental progress and future outlook}
Here, we report recent progress and future outlook toward developing such a detector at the Quantum Technology Center (QTC) at the University of Maryland. Due to the maturity of conventional (stage 1) detection techniques with semiconductor materials~\cite{LopezAsamar:2019smu,DAMIC:2021crr} and ongoing efforts on implementing similar techniques in diamond detectors \cite{Tarun:2015,Kurinsky:2019pgb,Canonica:2020omq,Abdelhameed:2022skh,CRESST:2023uel}, our research has focused on stage 2 and 3 detection using NV centers in diamond, a leading platform for quantum sensing~\cite{Doherty:2013uij,Barry:2019sdg}.

\subsubsection{Stage 2: micron-scale localization}
For stage~2, precise, high-throughput optical strain spectroscopy can be performed with a quantum diamond microscope (QDM)~\cite{Levine:2019ocn}. To localize a damage track from a 10\,keV nuclear recoil, a strain sensitivity of at least $1\times10^{-7}/\sqrt{\mathrm{Hz}\,\mu\mathrm{m}^{-3}}$ with 1\,$\mu$m$^3$ voxel size is required~\cite{Marshall:2020azl}. In 2022, we demonstrated strain sensing on a confocal QDM with a sensitivity of 5(2) $\times~10^{-8}/\sqrt{\mathrm{Hz}\,\mu\mathrm{m}^{-3}}$, exceeding this requirement~\cite{Marshall:2021xiu}. For higher imaging throughput, 2D strain sensing was also performed on a wide-field QDM with 150 $\times$ 150\,$\mu$m$^2$ field of view and similar levels of strain sensitivity. With chemical vapor deposition, NV-rich diamonds can be synthesized with low background strain~\cite{Edmonds:2021cxl}, favorable for localization of strain spikes caused by damage tracks.

However, 2D strain imaging lacks depth resolution, and WIMP damage tracks can randomly occur within the entire volume of the diamond detector segment.  Our current research is focused on developing full 3D strain imaging capability using light sheet microscopy ~\cite{Olarte:2018, Vladimirov_MesoSPIM2023}. By shaping the excitation laser light, a light sheet QDM (LS-QDM) would permit interrogation of a $\mu$m-scale-thick layer of NV centers~\cite{Horsley_lightsheet_2018}. As an initial demonstration, we aim to construct an LS-QDM with $\sim$5 $\mu$m resolution, $\sim 100\times100\,\mu{\rm m}^2$ field of view, and strain sensitivity comparable to Ref.~\cite{Marshall:2021xiu}. In the longer term, achieving 1\,$\mu$m resolution is feasible~\cite{glaser_exaSPIM_2023}. For the alternative stage~2 readout method, which looks for fluorescence from NV centers created in the vicinity of the damage track, conventional LSMs akin to those used in other mineral detectors~\cite{Cogswell:2021qlq,Baum:2023cct} will be useful.

\subsubsection{Stage 3: nanoscale track imaging}
For stage~3, nanoscale imaging of diamonds using scanning X-ray diffraction microscopy (SXDM) has been explored at the Argonne National Laboratory. In this technique, a beam of X-rays is focused to a 10-25\,nm spot inside a sample held at a Bragg angle~\cite{holtNanoscaleHardXRay2013}. As the beam is scanned, local differences in the resulting diffraction pattern encodes the crystal structure of the diamond. In the 2021 experiment~\cite{Marshall:2021kjk}, strain sensitivity of 1.6$\times10^{-4}$ was achieved, which would be sufficient to detect 10\,keV recoils. A limited survey of low-strain regions of the sample found no natural strain features which would impede detection of nm-scale damage tracks. Future efforts will be geared towards performing SXDM of artificial damage tracks with similar spatial and strain resolution. 

An alternative stage~3 readout method is to perform super-resolution imaging of color centers using techniques such as spin-RESOLFT~\cite{Maurer:2010}. This technique has been used to perform 2D magnetometry with $\sim$20\,nm resolution~\cite{Jaskula:2017}, and is relatively straightforward to extend for strain sensing. Z-resolution can be obtained by combining this technique with super-resolved Fourier magnetic gradient imaging~\cite{Arai:2015,Marshall:2020azl}. By mapping the position of all the color centers inside the region of interest, pattern recognition algorithms can be used to reconstruct the damage track~\cite{Rajendran:2017ynw}. Compared to X-ray microscopy, a significant advantage of this readout method is that it can be performed locally at the detector site with a tabletop setup. Super-resolution NV imaging techniques have been previously developed in our group~\cite{Maurer:2010,Arai:2015,Jaskula:2017,Zhang:2017usf}, and construction of a super-resolution-enabled confocal QDM at the QTC is currently in the preliminary stages.

\subsubsection{Artificial damage tracks for sensitivity characterization}

To characterize the sensitivity of the aforementioned imaging techniques, it is crucial to have methods to generate injected signals in a diamond sample. We have an ongoing collaboration with the Ion Beam Lab at Sandia National Laboratory, where their microbeam facility can implant low numbers of carbon ions on the surface of a diamond sample with a spot size of $\sim$1\,um and minimum energy of 0.8\,MeV~\cite{Titze:2022}. In situ ion counting can be used to significantly reduce the ion intensity uncertainty in an ion pulse, allowing for deterministic implantion of single ions~\cite{Pacheo:2017,titzeSituIonCounting2022}. Another method to produce artificial tracks is to irradiate the diamond sample with energetic neutrons~\cite{IAEA:2012}. Preliminary simulations with MCNP~\cite{shultisMCNPPrimer2011} indicate that irradiation with a monoenergetic 14\,MeV neutron source would produce a small but sufficient number of $>0.1$\,MeV nuclear recoils in diamond. Unlike ion implantation, which is restricted to the surface, neutron irradiation can create tracks inside the sample and would be useful to study the effects of neutron backgrounds which are common in mineral detectors~\cite{Drukier:2018pdy,Baum:2023cct}. 

\subsection{Conclusion}
In conclusion, a three-stage conventional-directional solid-state dark matter detector is a promising prospect for detecting WIMP dark matter beyond the neutrino fog, by combining techniques in conventional real-time nuclear recoil detection and quantum sensing with color centers. Current research at the QTC aims to extend and fine-tune state-of-the-art quantum sensing techniques to demonstrate the technical feasibility of such a detector within the next few years, including determining the best readout method for each detection stage. Although current research has predominantly concentrated on NVs in diamond, development of similar techniques with color centers in silicon carbide should also be explored in the future, given its potential for easier scalability.

\subsection*{Acknowledgements}
This work was supported by the Argonne National Laboratory under Award No. 2F60042; the DOE QuANTISED program under Award No. DE-SC0019396; the Army Research Laboratory MAQP program under Contract No. W911NF-19–2-0181; the DARPA DRINQS program under Grant No. D18AC00033; the DOE fusion program under Award No. DE-SC0021654; and the University of Maryland Quantum Technology Center.

\FloatBarrier
\newpage

\section{Search for Dark Matter \& Neutrino Induced Tracks in Minerals}

Authors: {\it Alexey Elykov} 
\vspace{0.1cm} \\
Karlsruhe Institute of Technology
\vspace{0.3cm}\\
With dark matter (DM) still eluding detection by large-scale experiments, and in light of the technical difficulties and expenses that are associated with up-scaling such detectors, a window has opened for new and daring ideas in the field. 
Additionally, the properties of neutrinos, which represent perfect messengers for multiple astrophysical phenomena, still puzzle the minds of contemporary physicists.
A novel idea in the field of particle detection is to take advantage of the advent of modern microscopy and computational techniques to read out and reconstruct micrometer ($\mu$m) and nanometer (nm) sized damage tracks produced by interactions of DM and neutrinos with nuclei of ancient minerals.
Residing in the depths of the Earth for millions of years, certain minerals should have accumulated these minute tracks, allowing us to use such minerals as \textit{paleo-detectors}.

Karlsruhe Institute of Technology (KIT) is one of the major scientific research institutes in Europe.
KIT hosts a large number of institutes with expertise in a wide range of research topics, including, among others, geology, material sciences, particle and astroparticle physics. 
Moreover, the Karlsruhe Nano Micro Facility (KNMFi) at KIT offers access to a wide range of state-of-the-art nm- and $\mu$m-scale manipulation and imaging techniques.
KIT is truly a unique nexus of expertise in numerous fields and cutting-edge microscopy instrumentation, making it the perfect place for a paleo-detector project.

\begin{figure}[h]
   \centering
   \includegraphics[width=1\textwidth]{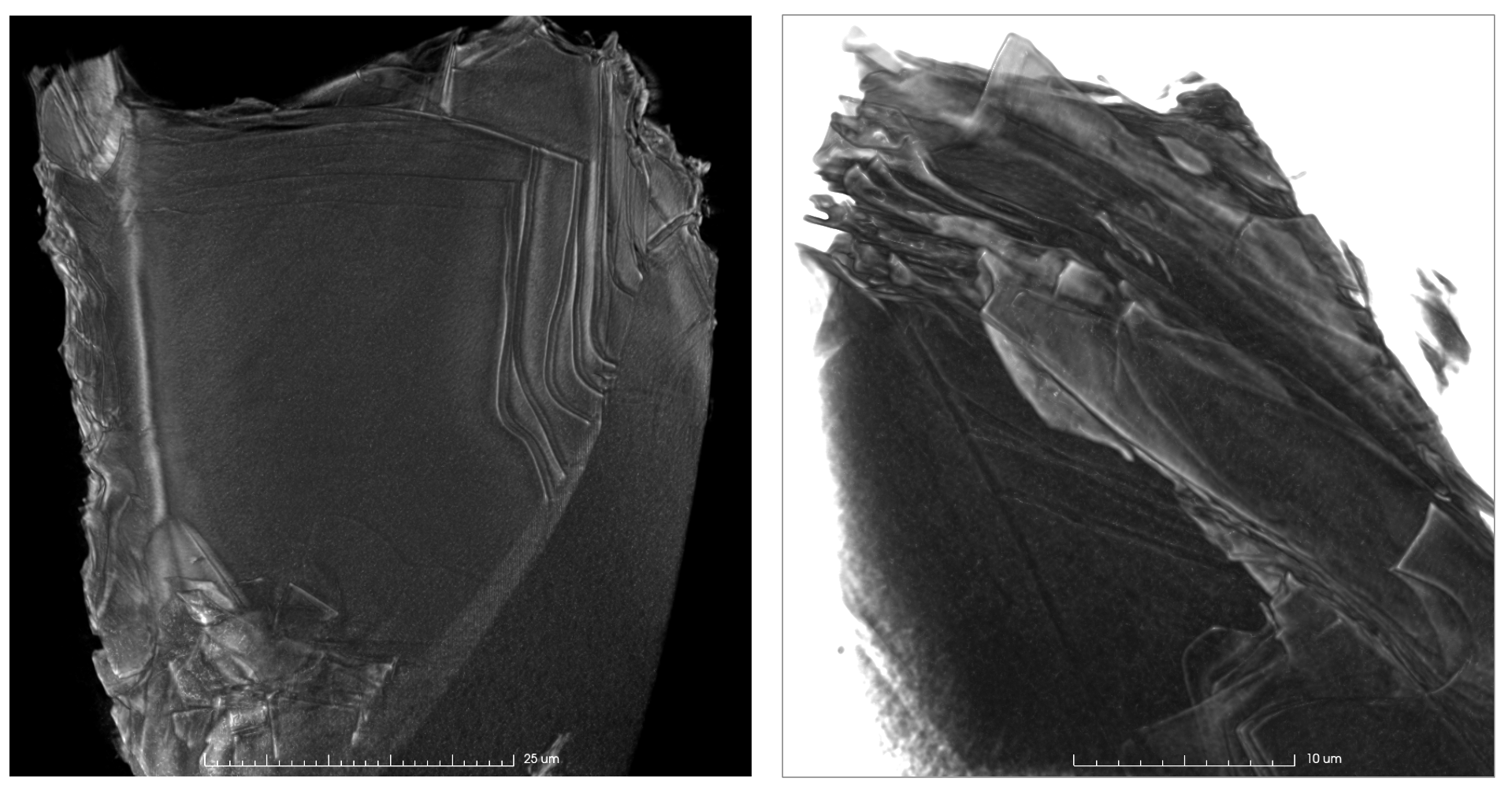}
   \caption{Muscovite mica samples imaged with nanoCT by Dr. Rafaela Debastiani at the Institute of Nanotechnology at KIT. One can easily discern sub-micrometer features in both samples, indicating the potential of this  technique for imaging a wide range of features in minerals spanning from large particle-induced tracks to natural defects.}
   \label{fig:sample}
\end{figure}

A pilot project at the Institute for Astroparticle Physics (IAP) at KIT, in collaboration with geologists from Heidelberg University and microscopy experts from several institutes at KIT aims to tackle key challenges of the concept of paleo-detectors. 
To that end, with the help of Prof. Ulrich A. Glasmacher we've obtained ``blank" samples of Muscovite Mica and Biotite samples irradiated with 11.4 MeV xenon ions.
As illustrated in Fig.~\ref{fig:sample}, these samples will be used to explore techniques for sample preparation and their subsequent imaging with nm- and $\mu$m-scale resolution using a variety of techniques and instruments.
Using the obtained data we will also aim to develop image processing and analysis techniques for track identification and correlation with expectations from GEANT4 and SRIM simulations.
Additionally, we are planning to study the effect of temperature on track survival in minerals using our small-scale lab.

Going beyond Mica, we will also aim to establish a list of realistic paleo-detector candidate minerals, not only from the perspective of their theoretical sensitivity to DM and neutrino induced interactions, but also by taking into account input from geology and microscopy.
Subsequently, if successful, these mineral samples could be imaged by employing the techniques developed using the aforementioned Muscovite and Biotite samples. 
The project at IAP KIT aims to tackle key challenges facing paleo-detectors, bringing us one step closer to the realization of the concept of paleo-detectors for neutrino and DM searches.       

\subsection*{Acknowledgements}
We thank Prof. Ulrich A. Glasmacher of the Institute of Earth Sciences,  University of Heidelberg for providing the mineral test samples and for the informative discussions. 
We thank Dr. Martin Peterlechner of the Laboratory for Electron Microscopy at KIT, for organizing and performing preliminary electron microscopy studies of Muscovite mica and Biotite.
We also thank Dr. Rafaela Debastiani of the Institute of Nanotechnology at KIT for preparation and imaging of Muscovite mica samples with nanoCT.
This work was partly carried out with the support of the Karlsruhe Nano Micro Facility (KNMF, www.knmf.kit.edu), a Helmholtz Research Infrastructure at Karlsruhe Institute of Technology (KIT, www.kit.edu). 
The Xradia 810 Ultra (nanoCT) core facility was supported (in part) by the 3DMM2O - Cluster of Excellence (EXC-2082/1390761711).
The presented project is supported in part through the Helmholtz Initiative and Networking Fund (grant agreement no.~W2/W3-118). 
We also gratefully acknowledge the support by the KIT Center Elementary Particle and Astroparticle Physics (KCETA) for this project.

\FloatBarrier
\newpage

\section{Past and present paleoneutrino research at University College London}

Authors: {\it Pieter Vermeesch and David Waters} 
\vspace{0.1cm} \\
University College London
\vspace{0.3cm}\\
Paleoneutrino research at University College London has pursued two approaches.

\subsection{Neutrino capture by thallium}

\textsuperscript{205}Pb is a radioactive lead isotope ($t_{1/2} = 17.3$~Myr) that is formed by (1) \textsuperscript{205}Tl($\nu$,e)\textsuperscript{205}Pb and (2) muon-induced \textsuperscript{205}Tl(p,n)\textsuperscript{205}Pb reactions in thallium-bearing mineral phases. At depths of more than 300\,m, the neutrinogenic \textsuperscript{205}Pb-production rate exceeds the muonogenic component. Ref. \cite{Freedman:1976exx} pointed out that the energy threshold of the \textsuperscript{205}Tl($\nu$,e)\textsuperscript{205}Pb reaction is exceptionally low. By measuring the \textsuperscript{205}Pb content of a deeply buried thallium deposit, it should therefore be possible to estimate the total (solar) neutrino dose accumulated over the age of the deposit. If the age of the deposit is known, then the neutrino flux can be calculated.

Thallium deposits are extremely rare. So rare in fact that only one suitable mineral locality has been identified: a 4.3\,Ma old lorandite (TlAsS\textsubscript{2}) deposit in Allchar, North Macedonia \cite{neubauer2009}. Unfortunately, the Allchar mine is only 120\,m deep, which means that the muonogenic \textsuperscript{205}Pb production rate exceeds the expected neutrinogenic component. However, if it can be demonstrated that erosion rates in the mountainous Allchar area are greater than 100\,m/Myr, then this would mean that the lorandite deposit spent most of its existence at depths of greater than 300\,m, so that the neutrino-signal may still exceed the muonogenic signal.

Average erosion rates over millennial year time scales can be estimated from the concentration of in-situ produced \textsuperscript{10}Be and \textsuperscript{36}Cl generated by spallation reactions of hadronic secondary cosmic radiation with target atoms in surface rocks. The higher the erosion rates, the lower the concentrations of these `cosmogenic nuclides'. An early attempt to use this technique found the \textsuperscript{36}Cl-concentrations of a single sample of carbonate to be high \citep{Dockhorn:1991it}, implying low erosion rates. However, this pilot study failed to account for the production of additional \textsuperscript{36}Cl by thermal neutron capture of \textsuperscript{35}Cl.

The first contribution of the UCL team to the field of paleoneutrino detection was to re-determine the erosion rates in the Allchar area using the latest insights in cosmogenic nuclide geochronology. We analysed 18~samples across a ${2}\times{3}$~km area around the Allchar mine \cite{Vermeesch:2018sgm}. The sample set included 11~carbonates and 4~volcanic rocks for \textsuperscript{36}Cl analysis, as well as 4~samples of modern river sand and gravel for \textsuperscript{10}Be analysis. The latter measurements allowed us to estimate the average erosion rate for the entire Majdanska River catchment. As expected, erosion rates vary across the area, with ridges being charactered by low erosion rates ($\sim$50\,m/Myr) and canyons by high erosion rates ($\sim$600\,m/Myr). Intermediate values of $\sim$120\,m/Myr are found on slopes and are inferred from the modern sediments. We are confident that these catchment-wide estimates accurately and robustly constrain the effective erosion rate in the Allchar area.

An erosion rate of 120\,m/Myr brings the lorandite experiment (LOREX, \cite{pavicevic1988}) back into the realm of feasibility. To make Ref. \cite{Freedman:1976exx}'s vision a reality will require further work to extract the Pb from the lorandite and measure the minute amounts of \textsuperscript{205}Pb by mass spectrometry \cite{Pavicevic:2010wpz}.

\subsection{Mineral Track Detection}

The discovery of coherent elastic neutrino-nucleus scattering (CE$\nu$NS) by Ref. \cite{COHERENT:2017ipa} started the current resurgence of paleoneutrino research. University College London is well placed to take part in this international effort thanks to its combined historic strengths in particle physics and fission track geochronology.

Two major hurdles must be cleared before the vision of CE$\nu$NS-based palaeo-detection can be realised. First, the  geological background rates (from cosmic radiation and ambient radioactivity) must be low and precisely known. Second, appropriate readout techniques must be developed to count the recoil tracks. 

We have performed initial experiments with hydroflouric acid etching of mica followed by optical and electron-microscopy, achieving promising results for the reconstruction of alpha-recoil tracks. We plan an ambitious programme of mineral scanning and track reconstruction, exploring techniques such as small-angle X-ray scattering, testing these techniques on samples exposed to natural cosmic-radiation and man-made sources.

The UCL work on mineral track detection is truly interdisciplinary; in addition to fundamental physics we will apply our techniques to diverse fields including geochronology and archaeometry.

\FloatBarrier
\newpage

\section{Activities at Queen's University}

Authors: {\it Levente~Balogh, Joseph~Bramante, Yilda~Boukhtouchen, Audrey~Fung, Matthew~Leybourne, Thalles~Lucas, Sharlotte~Mkhonto and Aaron~C.~Vincent}
\vspace{0.1cm} \\
Queen's University 
\vspace{0.3cm}\\
Work at Queen's University is split between experimental and theoretical investigation. The choice of target mineral in searches for WIMP and neutrino paleosignals is important: minerals must be resilient against annealing at geological temperatures and pressures, and concentration of radioactive elements must be as low as possible to minimise background contributions.  The group has identified two candidate minerals: olivine ((Mg,Fe)$_{2}$SiO$_{4}$) and galena (PbS). Work is underway in our geochemical labs to characterize background U and Th contributions, as well as at a tandem accelerator located at the Reactor Materials Testing Laboratory (RMTL) at Queen's University. 

Although the effect of dark matter (a non-relativistic, neutral particle) is in principle very different from a (relativistic, charged) proton beam, once the primary interaction has taken place, the subsequent displacement of lattice atoms should not be too different. Our aim is to produce a set of tracks from proton irradiation of a sample, and compare that set of tracks to a predicted track spectrum computed using Monte Carlo tools. This baseline will then allow a more reliable mapping of any signal (or lack thereof) in geological samples to the dark matter parameter space. 

\subsection{Experimental work}
\begin{figure}
    \centering
    \includegraphics[width=0.49\textwidth, trim={3.5cm 0cm 3.5cm 0cm}, clip]{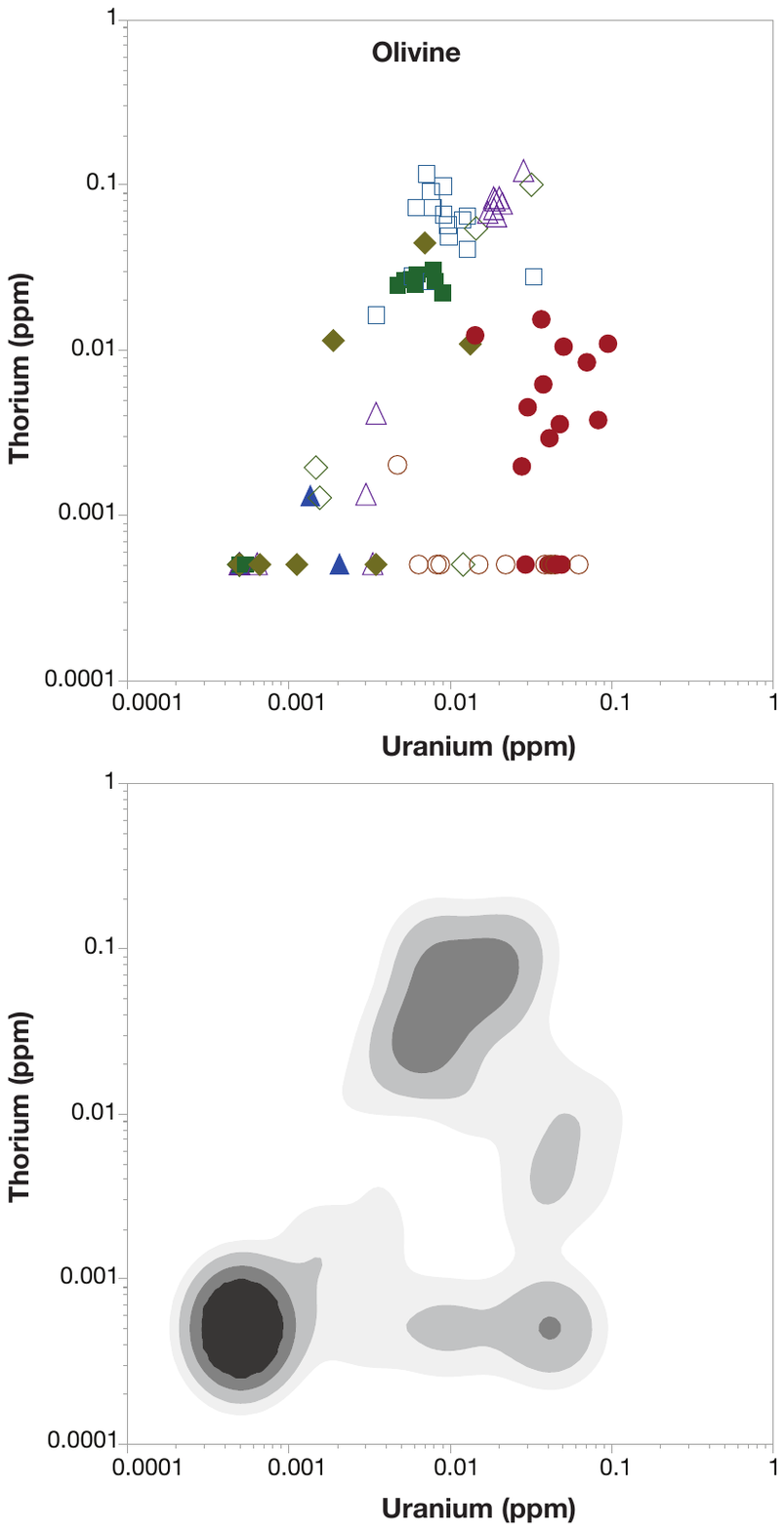}
    \includegraphics[width=0.49\textwidth, trim={3.5cm 0cm 3.5cm 0cm}, clip]{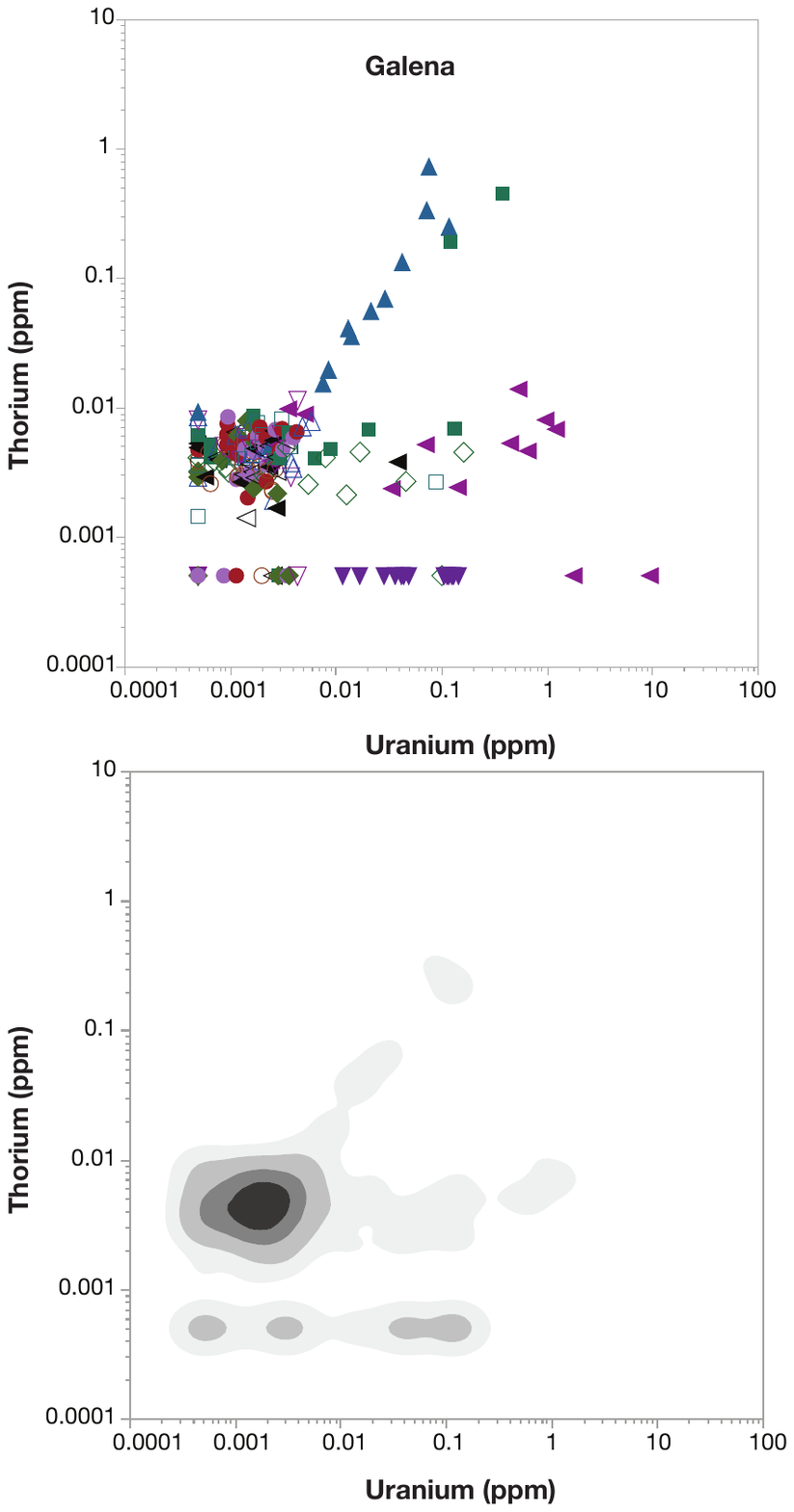}
    \caption{Concentration of uranium and thorium measured in samples of olivine (left) and galena (right). Note that the detection threshold is 0.5 ppb, suggesting much lower concentration in many samples examined.}
    \label{fig:UThgaloli}
\end{figure}

Samples of galena and olivine have been processed using an inductively coupled plasma mass spectrometry (ICP-MS). Preliminary results show very modest $(\sim$ ppb) Th and U concentrations in the galena and olivine samples. The instrument used was limited to 0.5 ppb sensitivity, and many samples returned values at this threshold, suggesting even lower concentrations.  These results are shown in Fig.~\ref{fig:UThgaloli}.

An olivine sample was prepared and irradiated in a 5 MeV proton beam at 0.25 dpa (displacement per atom). This relatively low radiation damage level, compared to dose levels typically imparted to metallic materials, was still too large for the olivine sample, leading to a loss of the natural crystalline structure in the mineral. The unirradiated and irradiated olivine samples were investigated by Transmission Electron Microscopy (TEM) at RMTL. TEM revealed that the unirradiated olivine is polycrystalline, consisting of grains of a few micrometers in size with multiple twin boundaries running through each grain. The twin boundaries in the grains/crystallites define rectangular-shaped single-crystal-like domains having a typical size of a few micrometers x a few tenths of micrometers. After 0.25 dpa irradiation dose these crystalline features has been transformed to a highly distorted structure consisting of amorphous, i.e., non-crystalline, domains and regions consisting of nanocrystals of ~100 nm with a very high density of planar/twin defects. According to the TEM images, the average distance of planar faults in the 0.25 dpa irradiated sample is less than ~10 nm. Within this highly distorted crystal structure a few track-like objects were identified, but results are too preliminary to tell if these can be identified with nuclear recoil-induced defects. The next steps will be to irradiate the olivine samples to lower doses and investigate the irradiation damage with TEM.

\subsection{Modelling track length distribution with \texttt{TRIM}}

\begin{figure}
   \centering
   \includegraphics[width=0.32\textwidth]{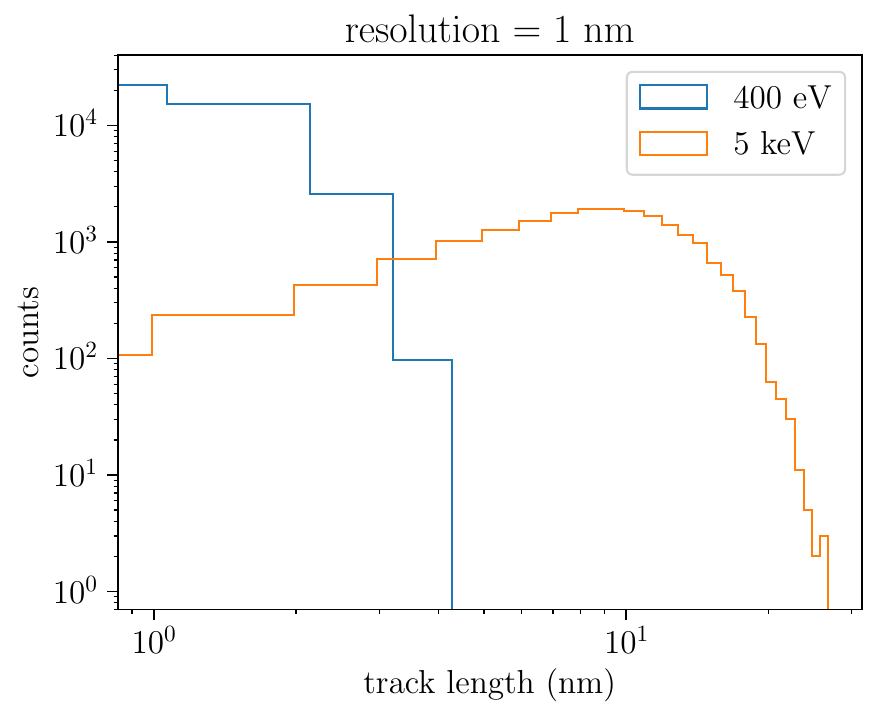}
   \includegraphics[width=0.32\textwidth]{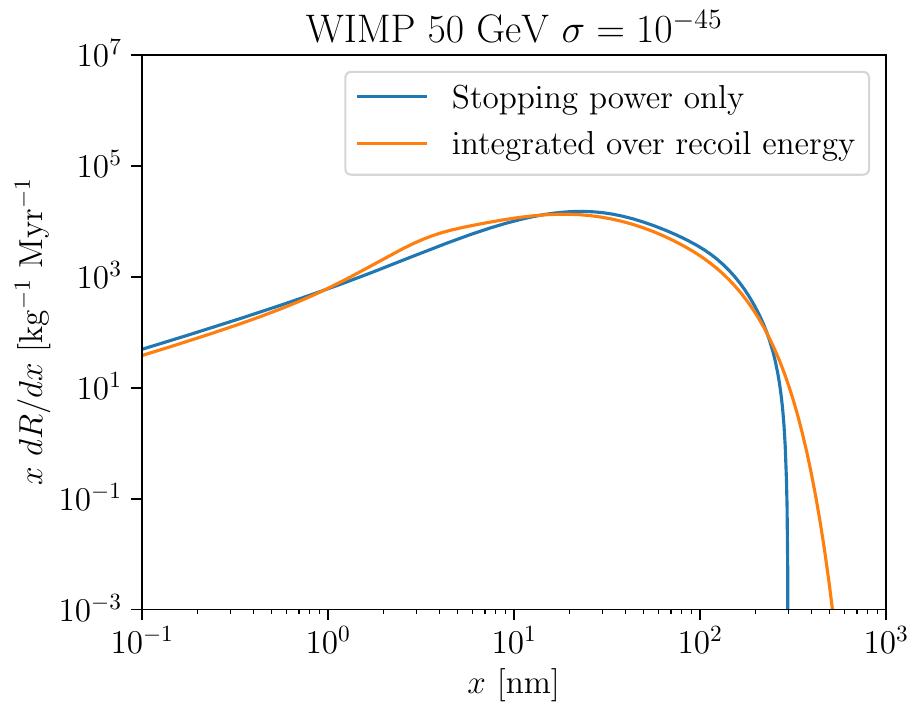}
   \includegraphics[width=0.32\textwidth]{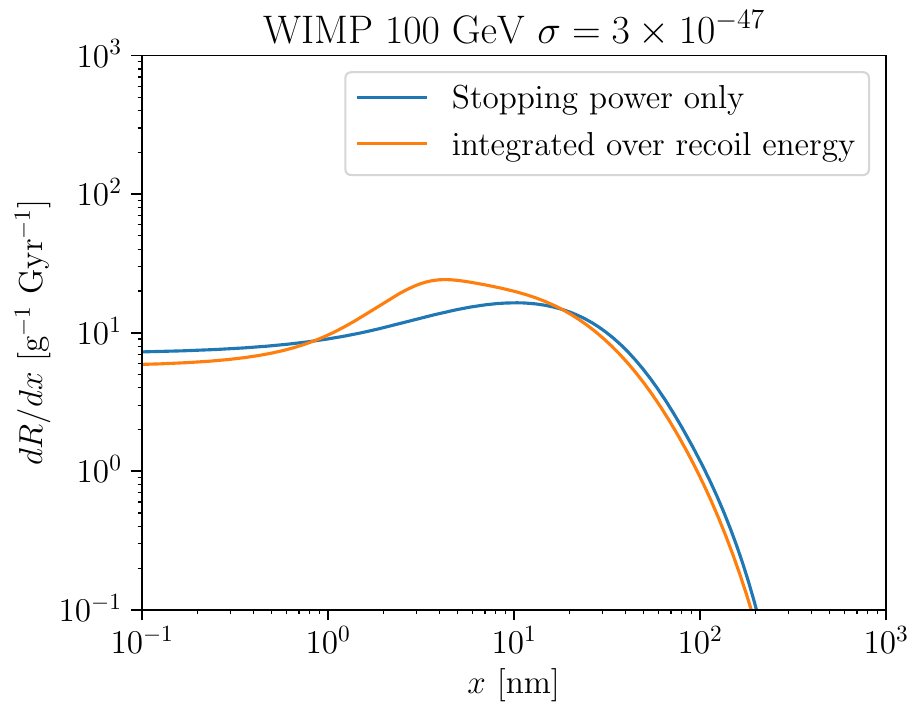}
   \caption{Left: Track length distributions for $400$eV and $5$keV recoiled ions obtained from \texttt{TRIM}. Middle: Differential recoil rate for $50$GeV WIMP showing additional measurable track lengths beyond ``cut-off". Right: Differential recoil rate for $100$GeV WIMP.}
   \label{fig:track_dist}
\end{figure}

One of the key advantages of paleo-detection is its large exposure, but that also means that paleo-detectors are exposed to large irreducible background that can only be mitigated through accurate modelling of tracks induced by both backgrounds and desired signals. Ref.~\cite{Drukier:2018pdy,Baum:2021jak} have built a general framework to model the track length distributions that assumed a one-to-one relation between stopping range of the recoiled ion and its recoil energy (we call this `stopping power only'). We  improve upon this with the aid of \texttt{TRIM}, a Monte Carlo program that models the transport of ions in matter. We found that a single recoil energy could induce a wide range of track lengths as shown in the leftmost panel of figure \ref{fig:track_dist}: the distributions are far from gaussian. To account for this, we computed the track length distribution by integrating the rate over all possible recoil energies:
\begin{equation}
    \label{eq:mydrdx}
    \frac{dR}{dx}(x) = \sum_N \textcolor{red}{\int dE_R P_N(E_R\vert x)} \left( \frac{dR}{dE_R} (E_R) \right)_N.
\end{equation}
The red part in Equation~\eqref{eq:mydrdx} is the probability $P_N(E_R\vert x)$ of a track $x$ that is induced by recoil energy $E_R$, can be inferred from simulation results from \texttt{TRIM}. Depending on the parameters of the signal of interest, some track length distributions show track lengths beyond the ``cut-off" length that is seen when modelled with the one-to-one recoil energy-track length assumption. The overall shape of the track length distribution is also modified. Using this new framework, we improve the modelling of WIMP-induced tracks and will extend our analysis to studying neutrino-induced tracks due to interactions via light mediators.

\subsection{Composite dark matter searches}

Models of heavy dark matter remain relatively unconstrained by direct detection experiments, as their low particle flux suppresses the event rate possible at detectors. Mineral searches for dark matter, taking advantage of the minerals' long exposure times, are well-suited to target heavy dark matter, and thus complement standard direct detection searches.

Dark matter composite states are a subset of heavy dark matter models that arise in theories of dark matter where some attractive dark force between constituent dark matter particles causes them to bind together in the early universe, analogously to Standard Model nucleosynthesis. Either during a period of assembly akin to Standard Model nucleosynthesis, or via formation in pockets of false vacuum after a first order phase transition, composite states can grow to radii much larger than a Standard Model nucleus. This results in a dark matter state that may scatter elastically with all nuclei it intercepts, or a more weakly-coupled composite states that scatters with a fraction of the SM nuclei it passes.

\begin{figure}
    \centering
    \includegraphics[width=0.65\textwidth]{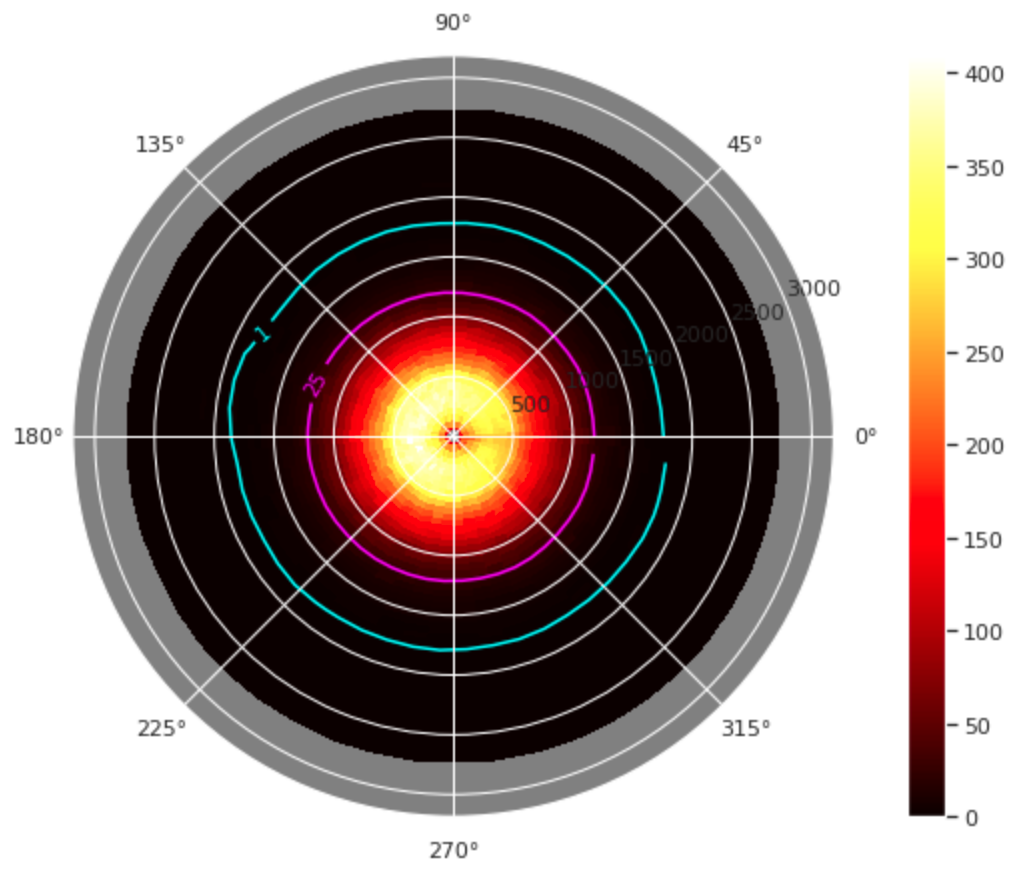}
    \caption{Average energy transferred to each galena atom (in eV) due to recoil cascades caused by a heavy composite dark matter particle of $5$ nm radius, as a function of distance from the composite centre. We take a cross-sectional view of the composite's path, with the composite at the origin shaded in grey.  We assume that all nuclei scatter elastically with the composite. The pink contour line at $25$ eV shows where the energy is sufficient to displace an atom from its crystal lattice. $1$ eV is an estimate of the energy required to reach the melting point of galena.}
    \label{fig:composite}
\end{figure}

Composite dark matter states are therefore expected to have a distinct signature in a mineral, leaving behind long tracks of distortions and defects in the crystal lattice, and potentially melting regions of the mineral along their path. At sufficiently-high masses, the energy deposited to the primary knock-on atoms is independent of the dark matter mass, and the expected size of the damage can be characterized by the radius of the composite (which is itself determined from the number of constituents expected in the theory). Using \texttt{TRIM}, we have modeled the primary and secondary recoil cascades caused by composites of various radii. An example plot of the energy imparted to atoms in galena from these cascades is shown in Figure \ref{fig:composite}. From there, we are working towards determining the expected radius of damage as a function of composite size.

\subsection{Conclusion}
A team of physicists, geochemists and materials engineers at Queen's are working on a variety of fronts to refine and validate this approach to detecting WIMP-like dark matter, neutrinos, as well as heavier, more exotic dark matter candidates.

\subsection*{Acknowledgements}
Work at Queen's is supported by the Arthur B. McDonald Canadian Astroparticle Physics Institute, NSERC, and the Ontario Government.

\FloatBarrier
\newpage

\section{Astrophysical neutrinos with paleo detectors}

Authors: {\it Shunsaku Horiuchi}
\vspace{0.1cm} \\
Virginia Tech
\vspace{0.3cm}\\
Neutrinos are unique cosmic messengers with the promise of views of the interiors of celestial phenomena and probes of fundamental physics in extreme environments. Yet their small cross sections also make neutrinos a challenge to detect. To date, neutrinos have been detected from a handful of astrophysical sources, including our Sun, supernova SN1987A, and active galaxies TXS 0506+056 and NGC 1068, with large-volume detectors in the kton to Gton scale target masses. Yet paleo detectors, with masses of only a few grams, can provide complementary probes with competitive sensitivities. Here, we review two case studies involving supernova neutrinos and solar neutrinos. 

Paleo detectors open a new window for supernova neutrino searches. Core-collapse supernovae emit copious neutrinos of all flavors in the tens of MeV energy range~\cite{Mirizzi:2015eza}. If a core-collapse supernova were to occur in the Milky Way or nearby galaxy, current and future neutrino detectors will yield significant neutrino detections~\cite{Scholberg:2012id}. However, given the transient nature of core-collapse supernova, one must wait for Nature's co-operation; the Milky Way supernova rate is a few per century~\cite{Rozwadowska:2020nab}. In parallel, a significant effort is underway to detect the diffuse supernova neutrino background arising from all core-collapse supernovae in the past~\cite{Beacom:2010kk,Lunardini:2010ab}. This signal is constant in time and isotropic in the sky, and with a predicted peak flux of a few ${\rm cm^{-2}\,s^{-1}\,MeV^{-1}}$, is detectable by Super-Kamiokande~\cite{Li:2022myd,Ando:2023fcc}. The key feature of paleo detectors is that they record neutrinos on time scales longer than the inverse of the Milky Way supernova rate. Recorded over such long time scales, paleo detectors are sensitive to the mean supernova neutrino emission from many past core collapses, similar to the diffuse supernova neutrino background modulo different distances. In fact, the so-called Galactic supernova neutrino background dominates over the diffuse supernova neutrino background by more than an order of magnitude~\cite{Baum:2019fqm}. This signal is accessible only by the extreme long time window offered by paleo detectors. 

The motivation for detecting supernova neutrinos with paleo detectors is multi-fold. On the one hand, the core-collapse supernova rate of the Galaxy over giga-year time scales, on which the Galactic supernova neutrino background depends, remains uncertain. Thus, the identification of supernova neutrinos in paleo detectors will inform whether the Milky Way underwent, e.g., a strong burst of star formation in the past~\cite{Baum:2019fqm}. On the other hand, paleo detectors can open a unique window to study the heavy lepton flavor of supernova neutrinos. Obtaining full-flavor information is crucial for supernovae, not only from an energetics perspective but also in light of the complicated oscillation physics involved~\cite{Horiuchi:2018ofe}. Paleo detectors detect supernova neutrinos via Coherent Elastic neutrino-Nucleus Scattering (CE$\nu$NS), a process sensitive to the total neutrino flux. By contrast, Super-Kamiokande is currently searching for the $\bar{\nu}_e$ component of the diffuse supernova neutrino background using inverse-beta interaction on free protons ($\bar{\nu}_e + p \to e^+ + n$), while the Deep Underground Neutrino Experiment (DUNE), currently under construction, will allow a search for the $\nu_e$ component through charge-current interaction on argon~\cite{DUNE:2020ypp}. For the heavy lepton flavor neutrinos, direct dark matter detectors can be utilized for their flavor-blind coherent scattering, however the sensitivities are orders of magnitudes above the predicted fluxes~\cite{Suliga:2021hek}. By contrast, paleo detectors can even allow a \emph{detection} of the heavy lepton flavor component of thousands of core collapses under favorable conditions~\cite{Baum:2022wfc}. 

Paleo detectors also open a new window into solar neutrinos. Our sun was the first extraterrestrial neutrino source detected, and in the decades since discovery, much of the solar neutrinos have been successfully measured~\cite{Haxton:2012wfz}; most recently, the CNO neutrino flux by Borexino~\cite{BOREXINO:2020aww}. However, all the measurements to date have been of the present-day solar neutrino flux. This contrasts with paleo detectors that are sensitive to the solar neutrino flux in the past mega- to giga-years. Among the various solar neutrino components, the $^8$B neutrino flux is the most detectable with paleo detectors, due to its high energy which helps separate it from backgrounds~\cite{Tapia-Arellano:2021cml}. Thus, by combining a collection of paleo detectors, each from different epochs of Earth's history, a time-profile of solar neutrinos can be obtained. 

The evolution of the Sun is interesting for various reasons. The Sun is active on multiple time scales. For example, by comparing the present-day solar neutrinos with the present-day solar radiation, a comparison of the energy outputs of the sun at the present day with that roughly $10^4$--$10^5$ years ago, corresponding to the time scale for energy transport within the solar interior, can be performed. On longer time scales, the evolution can be modeled with the solar standard model (SSM), a theoretical tool to model the solar interior~\cite{Bahcall:1998jt}. Recently, extended solar standard models have been developed in order to address discrepancies between the SSM and observations. One of the most widely explored is the solar metallicity problem, where the SSM struggles to match new measurements of solar elemental abundances and helioseismology~\cite{Haxton:2012wfz}. Possibilities include relaxing some of the assumptions in the SSM, e.g., that the Sun is chemically homogeneous with no mass loss or gain during evolution. It would be interesting to explore how the evolution of the solar neutrinos on geological time periods could inform extended SSMs. 

\subsection*{Acknowledgements}

The work of SH is supported by the U.S.~Department of Energy Office of Science under award number DE-SC0020262, NSF Grant No.~AST1908960 and No.~PHY-2209420, and JSPS KAKENHI Grant Number JP22K03630 and JP23H04899. This work was supported by World Premier International Research Center Initiative (WPI Initiative), MEXT, Japan.

\FloatBarrier
\newpage

\section{Solar Neutrino Flux evolution over gigayear timescales}

Authors: {\it Natalia~Tapia-Arellano}
\vspace{0.1cm} \\
University of Utah
\vspace{0.3cm}

\subsection{Introduction}
Solar Standard Models (SSM) are essential for predicting and studying solar neutrino fluxes. However, the most recent measurement of element abundances or metallicities in the photosphere has led to a new contradiction for these models. Lower metallicities in Solar Standard Models result in lower predicted solar core temperatures, which is in contrast to what helioseismology measurements suggest about the internal structure of the Sun, from the location of the radiative and convective zones~\cite{Haxton:2012wfz, Serenelli:2009yc}.

Different metallicities and temperatures affect the solar neutrino fluxes. By studying the variation of the different neutrino fluxes throughout the history of the solar system, we can test and set constraints on Solar Standard Models. The 8B neutrinos are one of the fluxes that are highly dependent on the core temperature as they are produced near the core. 
We look at the possibility that the same mechanism forming tracks for incoming dark matter particles, can also form tracks from incoming Solar neutrinos.
The information gathered from neutrino tracks formed in paleo detectors can help us understand the last couple of gigayears of the Sun's history and provide valuable information for the study of Solar Standard Models, which conventional direct detection experiments do not offer.

\subsection{Number of event estimate}
In order to study the time variation of solar neutrino fluxes, we have modeled the variation of Solar neutrinos flux using the {\tt MESA} code version r12115~\cite{Paxton:2010ji} following the procedure outlined in Farag et al. 2020~\cite{Farag:2020nll}. We have used their solar models which are calibrated to reproduce the present-day neutrino fluxes. Our model considers the final age of the Sun to be $t_\odot = 4.568$ Gyr, and its radius to be $R_\odot = 6.9566 \times 10^{10}$ cm. The photon luminosity $L_\gamma = 3.828 \times 10^{33}$ erg/s, and its surface metallicity is Z/X. We have computed two SSMs to match two different abundances of heavy metals at the surface of the Sun: Z/X = 0.0229 for Grevesse \& Sauval 1998 (hereafter GS)~\cite{Grevesse:1998bj} and Z/X = 0.0181 for Asplund et al.~2009 (hereafter AGSS)~\cite{Asplund:2009fu}.

To model the reactions emitting neutrinos, including pp, pep, 7Be, and 8B, we have used the built-in MESA reaction network {\tt add\_pp\_extras}. For the CNO chain, we have used the {\tt  add\_cno\_extras} and {\tt add\_hot\_cno} to compute the neutrinos from 13N, 15O, and 17F.

Using {\tt paleopy}~\cite{Baum:2019fqm}, we have extracted the track length spectra for 8B Solar Neutrinos and by using the time variation described above we obtain a re-scaled neutrino flux for different time slots as shown in figure~\ref{fig:fig1} and figure~\ref{fig:fig2}.

In the scenario in which we can obtain our rock samples from deep underground ($\sim 6$ km) we can neglect the cosmic ray track signal and the main background will be attributed to neutron tracks from Uranium concentration which can form tracks due to $\alpha$ decay and spontaneous fission, low radioactive contamination is a crucial part of
material selection.

\subsection{Results}

\begin{figure}[h]
\begin{subfigure}{0.5\textwidth}
\includegraphics[width=\linewidth]{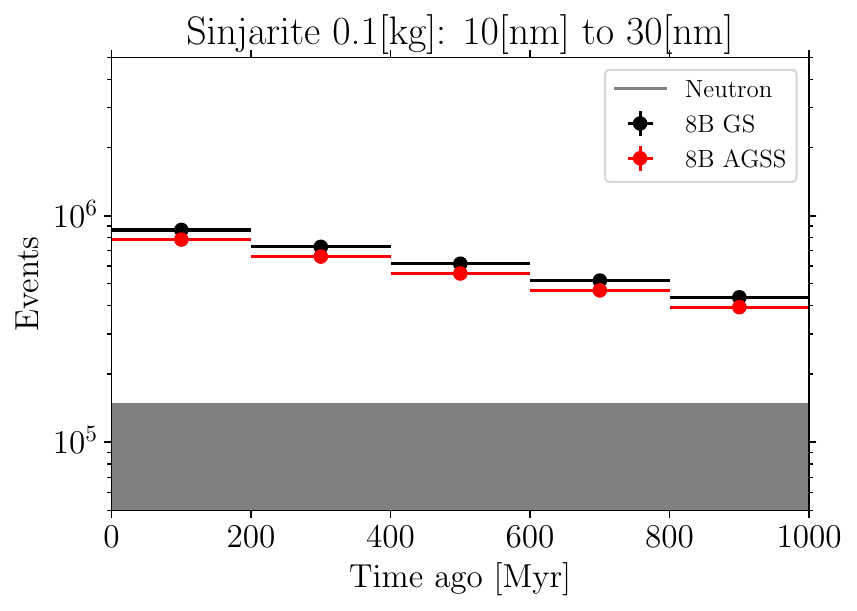} 
\caption{}
\label{fig:fig1a}
\end{subfigure}
\begin{subfigure}{0.5\textwidth}
\includegraphics[width=\linewidth]{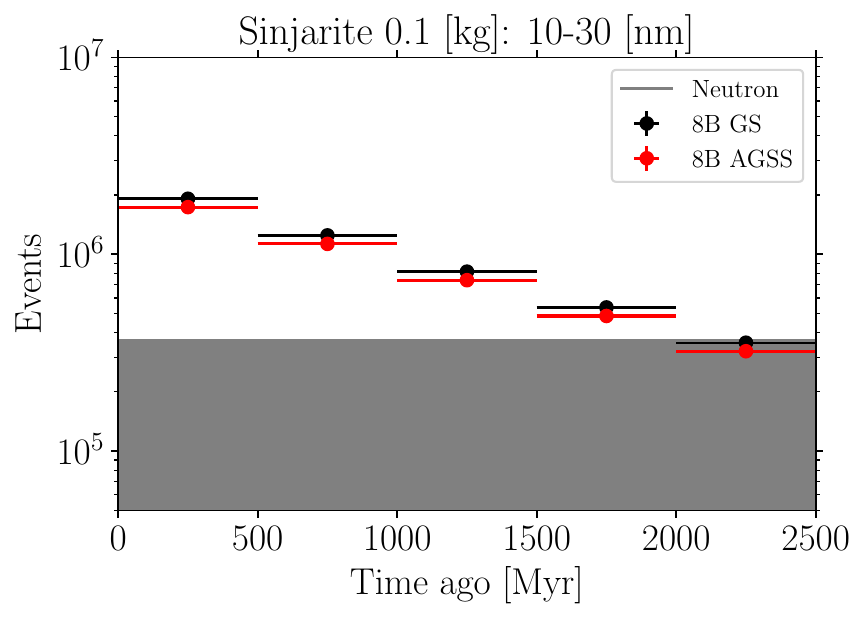}
\caption{}
\label{fig:fig1b}
\end{subfigure}

\caption{Number of events summed over track lengths of 10 to 30\,nm, for 0.1\,kg of Sinjarite. The black dots represent the GS SSM and red represents AGSS SSM. The shaded region is where the neutron backgrounds will dominate the events and are summed over the same time windows. (a) Number of events per 200\,Myr time bins. (b) Number of events per 500\,Myr time bins.}
\label{fig:fig1}
\end{figure}

\begin{figure}[h]

\begin{subfigure}{0.5\textwidth}
\includegraphics[width=\linewidth]{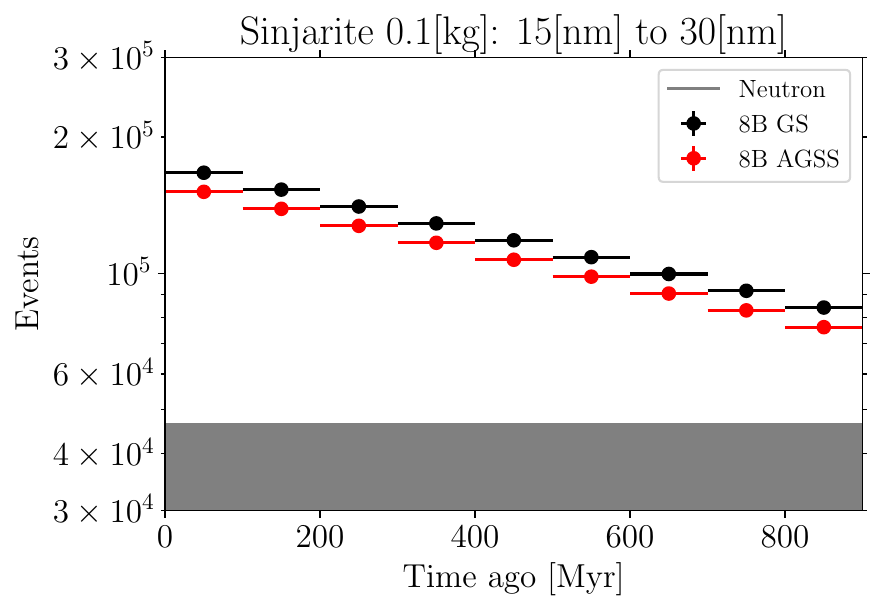} 
\caption{}
\label{fig:fig2a}
\end{subfigure}
\begin{subfigure}{0.5\textwidth}
\includegraphics[width=\linewidth]{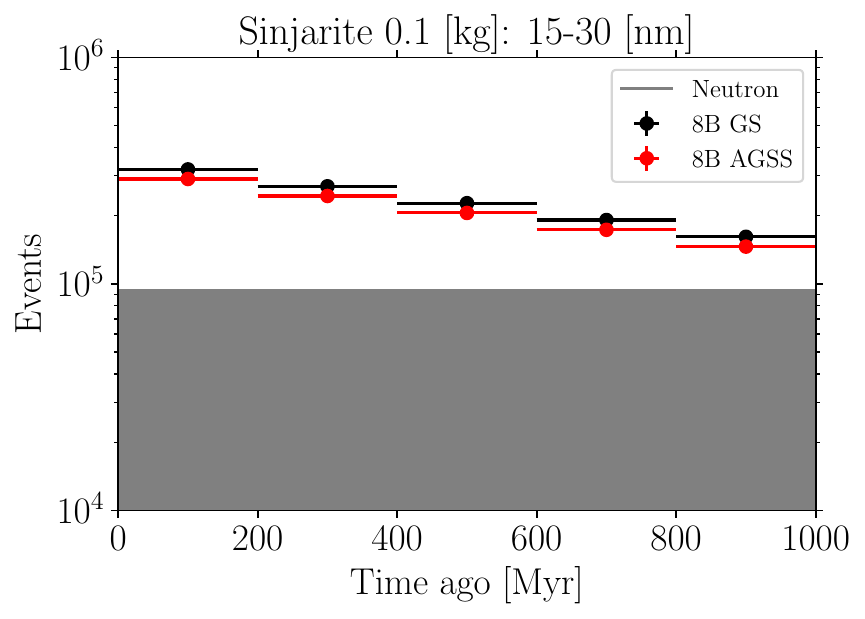}
\caption{}
\label{fig:fig2b}
\end{subfigure}

\caption{Same as figure~\ref{fig:fig1}, but in this case the number of events are summed over track lengths of 15 to 30\,nm, for 0.1\,kg of Sinjarite. (a) Number of events per 100\,Myr time bins. (b) Number of events per 200\,Myr time bins.}
\label{fig:fig2}
\end{figure}

We show our results for different time window selection in figure~\ref{fig:fig1} and figure~\ref{fig:fig2}, for a sample of 0.1 kg\,of Sinjarite. First, in the case in which read out methods can reach a sensitivity of 10\,nm, we integrate up to 30\,nm track lengths in figure~\ref{fig:fig1}, and we show two time windows, of 200\,Myr (figure~\ref{fig:fig1a}) and 500\,Myr (figure~\ref{fig:fig1b}). The constant flux of neutrons, representing the background, is summed up in the same fashion. The two SSM of choice are shown in black dots which represent the GS SSM and red dots represent AGSS SSM.

In a more realistic scenario, readout methods predict a sensitivity of up to 15\,nm, and we show our results for this case in figure~\ref{fig:fig2}. In this case we take a smaller time window of 100\,Myr (figure~\ref{fig:fig2a}), and 200\,Myr (figure~\ref{fig:fig2b}) as in figure~\ref{fig:fig1}.

As can be seen in our results, most of the cases presented here offer a favourable signal to background ratio, and the difference between GS and AGSS is relatively time independent. For our 200\,Myr time bin width result (figure~\ref{fig:fig2b}), the difference between the GS and AGSS metallicity models is $\sim 10^{4}$ tracks per time bin, i.e., approximately $\sim 10$\% of the total number of tracks. The total number of tracks is $\sim 1.17 \times 10^{6}$ and $\sim 1.06 \times 10^{6}$, for GS and AGSS, respectively, if we sum over the last Gyr. If we assume only Poisson errors, then $\sigma \sim 10^3$, and a difference of 10\% due to the metallicity models would be easy to measure.

\subsection{Conclusions}

We have considered SSMs with two metallicities (GS and AGSS) to study the detectability of the boron-8 ($^8$B) solar neutrino flux over gigayear timescales using paleo detectors. 
Our default setup is a paleo detector composed of a collection of 0.1\,kg sinjarite crystals of different ages, ranging up to 1\,Gyr old and aged with 100\.Myr and 200\,Myr resolution. We also consider a sample reaching 2.5\,Gyr with 200\,Myr and 500\,Myr age-dating resolution. We explored track lengths of 10--30\,nm and 15--30\,nm for detecting the $^8$B neutrino flux. We found that up to 1--1.5\,Gyr, the $^8$B signal to background ratio is favourable for measuring the time-evolution of the $^8$B neutrino flux for a realistic case of 15\,nm readout method resolution and for {gaining insight about solar models motivated by GS and AGSS metallicities.}

The main background to solar neutrinos is caused by fast neutrons, originating from radioactive sources within the minerals. The normalization of this background solely depends on the original concentration of radioactive materials. We follow Ref.~\cite{Drukier:2018pdy} and adopt a Uranium concentration of 0.01 parts per billion for sinjarite. Since the neutron background rate scales with this concentration, the detectability of the $^8$B neutrino flux with sinjarite requires the concentration to be not more than a few times 0.01 parts per billion.

To sum up, the use of paleo detectors can provide novel avenues for exploring and verifying SSMs.

\subsection*{Acknowledgements}
The work of Natalia Tapia-Arellano is supported by the University of Utah.

\FloatBarrier
\newpage

\section{Supernova Burst Detection with Mineral Detectors}

Authors: {\it Kate~Scholberg} 
\vspace{0.1cm} \\
Duke University 
\vspace{0.3cm} \\
Stellar core collapses, which occur at the rate of a few per century
per galaxy, produce copious neutrinos of all flavors in the few- to
few-tens-of-MeV range.  The neutrino emission for a given core
collapse is in the form of a prompt burst of a few tens of seconds in
duration that precedes the subsequent astronomical supernova fireworks
by hours or days.  The bright neutrino bursts from nearby (i.e., the
Milky Way and Local Group) core collapses dominate the diffuse
supernova neutrino background (DSNB) during this tens-of-second
timescale by nine or ten orders of magnitude; in fact the galactic
supernova neutrino background (GSNB) dominates over the DSNB when
integrated over time in a paleodetection context~\cite{Baum:2023cct}.
A measured supernova burst signal will bring us vast and rich
knowledge of both the astrophysical event (because weakly interacting
neutrinos bring information from otherwise obscured regions) and of
neutrino properties~\cite{Scholberg:2012id,Mirizzi:2015eza}.
Furthermore, the promptness of the burst enables an early warning of an
impending supernova; the Supernova Early Warning System (SNEWS)
network has implemented this alert in a global detector
network~\cite{SNEWS:2020tbu}.

Both charged-current (CC) and neutral-current (NC) neutrino-matter
interactions can be exploited by terrestrial detectors to observe
supernova neutrino bursts.  For the $\nu_{\mu,\tau}$ and
$\bar{\nu}_{\mu,\tau}$ components of the flux, only NC interactions
can be employed for detection, because the supernova neutrino flux is almost
entirely below CC threshold for muon and tau lepton production.  In a
given target material, coherent elastic neutrino-nucleus scattering
(CEvNS) interactions typically dominate over CC interactions by a
couple of orders of magnitude, thanks to the coherent enhancement of
the cross section per target nucleus that scales as $N^2$, where $N$
is the number of neutrons in the nucleus, as well as the fact the
CEvNS is NC and hence sensitive to the full supernova flux.  However,
CEvNS events, while copious, can be observed only via the energy loss
of the recoil nuclei, which typically have only tens of keV.  In
contrast, while CC events ($\nu_e+(N,Z)\rightarrow (N-1,Z+1) +e^-$,
$\bar{\nu}_e+(N,Z)\rightarrow (N+1,Z-1) +e^+$), are much lower rate
than CEvNS, the observable energy deposition is 2-3 orders of
magnitude higher than that of the nuclear recoils, as the final-state
lepton takes most of the neutrino energy.  Elastic scattering on
electrons (eES) has the lowest cross section.

CEvNS interaction rates per detector mass scale roughly as $N$, whereas
typically CC and eES rates per detector mass are roughly independent
of target type (with variation of a factor of a few in CC rates due to
variations in nuclear matrix elements).
There are order-of-magnitude uncertainties on the flux at a given
distance, but typically one expects $10^{-2}-1$ CC events per ton of
target material for a core collapse at 10~kpc (just beyond the center
of the galaxy), with inverse square scaling with supernova distance.
For CEvNS, one expects of the order of 10 interactions per ton of
target material at 10 kpc.

It has been well recognized in the past decades that dark matter WIMP
detectors are sensitive to supernova bursts via CEvNS (e.g.,
~\cite{Horowitz:2003cz,Lang:2016zhv}), and are suitable for real-time
detection of supernova bursts.  What about mineral detectors?  Mineral
targets in the Earth will record a few tens of fresh nm to $\mu$m
tracks per ton within tens of seconds for a 10-kpc supernova.  Is it
conceivable to find these tracks promptly?  Clearly, this is a
challenging task, but one can make a back-of-the-envelope estimate of what
it would take to do this: likely ton scale or more of target material;
nm track resolution; and ability to scan, monitor or otherwise
interrogate the material on a very short timescale.  How short is
short enough?  In the most ideal case, minutes would be desirable, to
enable input to an early alert, but hours or longer might still
produce useful prompt information.  Of course, even longer (days,
weeks, months) might still yield physics information -- better late
than never.  For prompt response, however, one needs at least ton/hour
scanning capability.  Furthermore, fresh burst tracks will be
superimposed on long-timescale integrated tracks (the physics target
for other mineral detection searches).  Therefore freshly
annealed/blank material is best for supernova burst detector.  Also
needed are low ambient and cosmogenic background (so, an underground
location).  An external prompt trigger such as SNEWS could initiate
readout.

A prompt mineral supernova burst detection is clearly a tall order.  What
would we learn? A measured distribution of tracks would
provide useful contribution to the collected world supernova sample.
The rate of CEvNS for a known target material provides an all-flavor
flux.  Spectral information available in the distribution of track
lengths would be valuable.  What value would there be in a
\textit{prompt} mineral scan?  A potential ``killer app'' is
\textit{directionality.}  A supernova burst detection is most helpful
to astronomers if the neutrinos can be used to point back to the
supernova with low latency, so that astronomers know where to look for
the supernova\footnote{And in fact not every core collapse may yield
  a bright supernova; formation of a black hole may result in a
  ``winked-out'' star. Directional information is still valuable in
  that case, for the hunt for the progenitor or for studies of
  neutrino oscillation in the Earth.}.

Neutrino pointing via anisotropic interactions is possible in large
(multi-kton-scale) detectors, primarily making use of eES.  Both
Cherenkov detectors~\cite{Super-Kamiokande:2016kji} and liquid argon time
projection chambers~\cite{DUNE:2020ypp} can point potentially with
few-degree resolution, and scintillator detectors have some modest
pointing ability~\cite{Fischer:2015oma}.  CEvNS events have a
well-defined angular distribution; a directional track determination
in a mineral detector could exploit this for supernova pointing.
Mineral detectors would suffer from a head-tail ambiguity in the CEvNS
recoil direction determination, which would weaken the pointing
resolution.  Nevertheless, pointing resolution at the few-degree
level, rivaling that for multi-kton-scale detectors, could be obtained
with tens-of-ton-scale directional CEvNS detectors.

\begin{figure}
   \centering
   \includegraphics[width=0.88\textwidth]{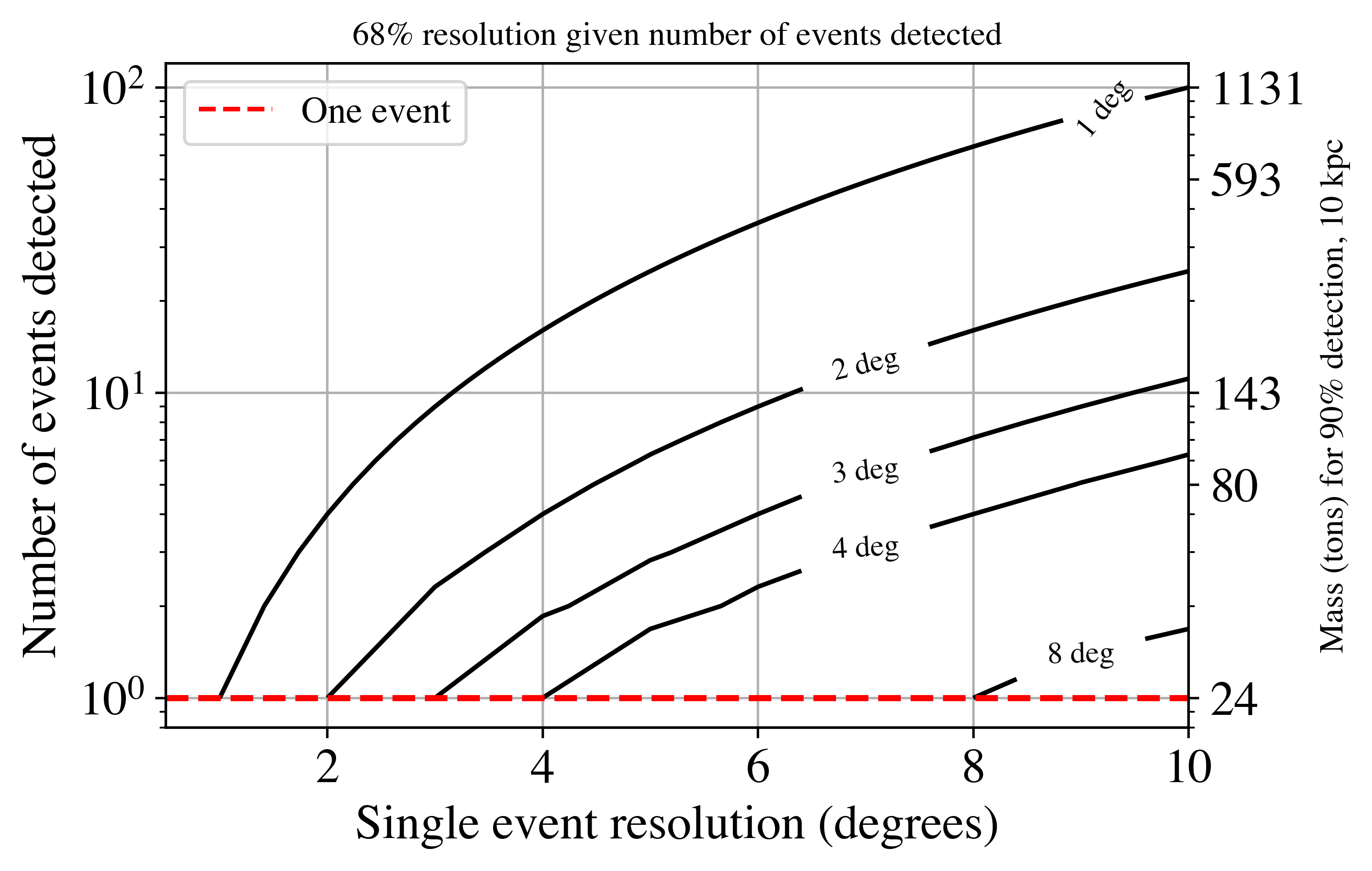}
   \caption{A simple statistical scaling showing detector mass necessary for a given
     supernova pointing resolution assuming fully reconstructed CC
     events, for which both nucleon and lepton in the final state are
     tracked.  Overall supernova-burst pointing resolution contours
     are shown as a function of number of events and intrinsic
     single-reconstructed-event resolution.  The right-hand axis shows
   the detector mass required for 90\% detection of at least the given
   number of events, assuming 0.1 supernova events/ton (typical for CC
   interactions at 10 kpc).}
   \label{fig:res-scaling}
\end{figure}

An even more ambitious concept is the idea of full kinematic
reconstruction of the incident neutrino four-momentum via measurement
of directionality of both lepton and recoil nucleon in the final state
of a supernova-neutrino-induced CC interaction.  An intrinsic
single-fully-reconstructed-event pointing resolution of several
degrees requires only a small number of events to obtain a good
overall burst pointing.  See Fig.~\ref{fig:res-scaling}.  Furthermore,
a single fully reconstructed event could break a head-tail ambiguity
via reconstruction of an event vertex. However, since typically only
the electron-flavor portion of the supernova flux is accessible via CC
interactions (primarily $\nu_e$, representing $\sim$1/6 of the flux;
$\bar{\nu}_e$ interactions are suppressed by Pauli blocking), the
event rate is such that this approach requires $\sim 6N$ times the detector mass
as for CEvNS (so, likely at least multi-ton scale at 10 kpc).  Ability
to reconstruct final-state leptons is needed, which might require some
additional hybridized real-time detection method (scintillation,
Cherenkov, etc).

These concepts are clearly ambitious and potentially unrealistic.
Nevertheless they are interesting to consider as mineral detection
technology and other fine-grained track reconstruction capabilities advance.

As a final note, an ``artificial supernova'' exists in Neutrino Alley
at the Oak Ridge National Laboratory Spallation Neutrino Source (SNS),
already used for the first CEvNS detection by the COHERENT
collaboration~\cite{Barbeau:2021exu}.  This source produces a high flux of
neutrinos from pion decay at rest with energies up to $\sim$50~MeV,
overlapping with the supernova-neutrino energy range, with fluence per
day at $\sim$20~m equal to two 10-kpc supernovas' worth of neutrinos.
While backgrounds are challenging for time-integrated signals,
real-time approaches could exploit the sharp SNS timing for background
reduction.

\subsection*{Acknowledgements}

The author is supported by the Department of Energy and the National
Science Foundation and thanks Janina Hakenm\"uller for input for the
talk at this workshop.

\FloatBarrier
\newpage

\section{Towards Imaging Atmospheric Neutrinos with Paleo-detectors at University of Michigan}

Authors: {\it J.~Spitz, K.~Sun, A.~Calabrese-Day, C.~Little, K.~Ream, and A.~Takla}
\vspace{0.1cm} \\
University of Michigan
\vspace{0.3cm} \\
Atmospheric neutrino imprints in minerals can provide unique sensitivity to changes in the cosmic ray rate over the past $\sim$1~billion years~\cite{Jordan:2020gxx}. For example, if Earth in its trajectory around the galaxy encountered a transient event (e.g. caused by a nearby supernova or neutron star merger), the resulting increased cosmic ray rate, and therefore atmospheric neutrino interaction rate, information has been recorded, and is potentially measurable, with paleo-detectors. Gradual changes (e.g. originating from slowly evolving star formation or supernova rates) in the cosmic ray rate would also be discernible using this technique. For comparison, the deposition of cosmogenic nuclides (e.g. $^{10}$Be) in the crust provides a history of the cosmic ray flux going back as far as 10~million years (see, e.g., Ref.~\cite{Lal1967}). Cosmic ray rate measurements dating back 1~billion years have been attempted, based on $^{40}$K ($\tau_{1/2}=1.4$~Gyr) abundances in meteorites, but these studies can only infer the cosmic ray exposure integrated over a given sample’s dynamic location history, including when the rock fell to Earth, if it originated from an asteroid break-up event, etc., which is difficult to infer and varies considerably from sample to sample~\cite{Wieler:2013}.

Among the identifiable use cases for paleo-detectors in the context of particle physics and astrophysics, atmospheric neutrino studies are the most easily accessible with current technology, and represent an excellent ``jumping off'' point for more challenging searches. Simply, both the event rate and typical track size are significantly larger and the background is smaller for atmospheric neutrino events, as compared to dark matter and solar/supernova neutrinos. To set the scale of mass needed for an atmospheric neutrino study, about 10$^4$ primary nuclear recoils are expected in a 1~Gyr old, 100~g sample, assuming today's atmospheric neutrino flux on Earth; if secondary nuclear recoils (also considered signal) are included, initiated by primary hadrons produced in the neutrino interactions,  about 6$\cdot$10$^4$ tracks in a ``background-free" region with lengths 2~$\mu$m~$\leq$~x~$\leq$~20~$\mu$m and 50~$\mu$m~$\leq$~x~$\leq$~1~mm, avoiding radiogenic neutrons ($<$2~$\mu$m) and $^{238}$U spontaneous-fission daughter (20-40~$\mu$m) backgrounds, are expected~\cite{Jordan:2020gxx}. 

At Michigan, we are working to develop a strategy for efficiently imaging atmospheric neutrino induced damage tracks in olivine samples involving a combination of soft X-ray micro computed tomography for coarse, all-sample imaging and Transmission Electron Microscope (TEM; $\sim$sub-Angstrom resolution is achievable) for fine, sub-sample imaging. The general idea is to identify track candidate regions-of-interest (ROI) with the soft X-ray micro-CT device, including phase-contrast imaging, and then to interrogate these with a combination of Focussed Ion Beam (FIB) and Scanning Electron Microscope (SEM) for sub-sample preparation, and TEM, to determine the ROI properties at ultra-high-resolution. We are also actively utilizing our Ion Beam Laboratory, featuring 1.7~MV and 3~MV tandem particle accelerators capable of accelerating, O, Mg, Fe, Si, and others for mimicking atmospheric neutrino induced nuclear recoils. 

\subsection*{Acknowledgements}
The work of the University of Michigan group is supported by the Gordon and Betty Moore Foundation.

\FloatBarrier
\newpage

\section{Q-ball Search with the Muscovite Mica}

Authors: {\it Tatsuhiro~Naka$^1$, Yuki~Ido$^2$, Takenori~Kato$^2$}
\vspace{0.1cm} \\
$^1$Toho University, $^2$Nagoya  University
\vspace{0.3cm}

\subsection{Composite Dark Matter and Q-ball}
 
Various current cosmological observations reveal the components of the universe, and the dark matter which constitutes approximately 80$\%$ of the matter component is very important target to understand for nature science because that is anticipated to be a new particle not accounted for in the standard model, as there are no candidates within it. 
Currently, there are many theoretical candidates for dark matter through very wide mass range. In this context, we focus on very heavy dark matter in the form of composite states. Q-balls, quark nuggets, monopoles, and others are mentioned as potential candidates. 
The composite state of dark matter exhibits a characteristic ultra-heavy mass, approximately $10^{15}\,\text{GeV}/c^{2}$ or heavier. For such ultra-heavy dark matter, the flux reaching Earth should be $10^{-13}/\text{cm}^{2}/\text{sec}$ or less. This corresponds to $\mathcal{O}(1)$ or less as the number of events passing through the detector with a scale of m$^{2}$. The Paleo detector overcomes this challenge through exposure over geological timescales.

In this context, we focus on the Q-ball, which can exist stably as either the main contributor or only a part of dark matter. The verification that Q-balls exist give the indication to important subjects in the particle physics such as the generation of baryon and lepton number through the Affleck-Dine mechanism~\cite{Affleck:1984fy} in early universe and beyond standard model such as the MSSM theory~\cite{Dine:1995uk}. The Q-balls  can adopt states both with and without electric charge. Particularly, charged Q-balls can also exist as part of dark matter by capturing electrons around the positively charged Q-ball core with a femtometer-scale size. Stable conditions resembling an "atom" are acceptable, and the core charge is expected to have the inverse of the fine structure constant ($\alpha^{-1}=137$) and carry +$\mathcal{O}(1)$ ions~\cite{Hong:2016ict,Hong:2017qvx}. 
 
Such quantum state should be passing through the objects even though very high ionization loss, however it should not stop in the material due to ultra-heavy mass. Then, such events remain the long track in the paleo detector.  Especially, the muscovite mica is promising device from the cleavage, transparency and track formation threshold. In terms of track formation threshold, several previous studies already exist such as~\cite{Fleischer:1967zz}, but it is important to cross check and directly observe the tracks by ourselves. Therefore, as first step, we checked the track formation ability of the muscovite mica using accelerated heavy ions. 

\subsection{Track Formation Ability with the Muscovite Mica}

\begin{figure}
   \centering
   \includegraphics[width=1\textwidth]{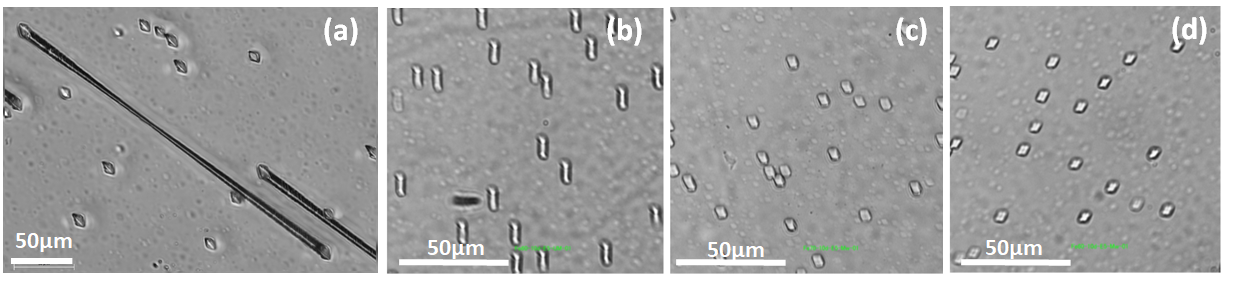}
   \caption{Phase contrast optical microscope image for the etched tracks induced by each ion beam on the muscovite mica. (a) is a image of penetrating track of Xe ion beam of 290 MeV/u, (b), (c), (d) are Fe ion tracks of 60, 6, 3 MeV respectively, and incident angle of 10 deg. to horizontal plane of the mica plate.}
   \label{fig:example_image}
\end{figure}

\begin{figure}
    \centering
    \includegraphics[width=0.7\linewidth]{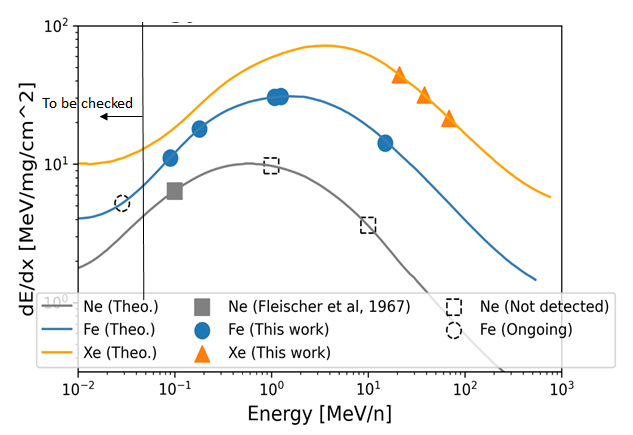}
    \caption{Summary of track formation ability of ionization loss to energy for each ions. Solid lines for each color are corresponds to theoretical value, and triangle (orange) plots are result of Xe ion, and circle (blue) plots are result of Fe ions in this work. square (gray) plot is previous work by~\cite{Fleischer:1967zz}, and lower energy calibration is now on going.}
    \label{fig:ability}
\end{figure}

For the exotic particle search such as the dark matter, detector calibration is required for final confirmation of identification of signal and the evaluation of kinematics. Especially, as the charged Q-ball should pass through the mica like long ion tracks, the calibration of detection ability and the image quality using the optical microscope should be required.  As first step, we evaluated the track formation ability for the mica by ion-beam of Iron and Xenon. Iron beam of 500\,MeV/u and Xe ion beam of 290\,MeV/u were exposed at the HIAMC of the National Institute for the Quantum and Technology (QST), Japan, and evaluated the track formation threshold by making the Bragg peak in the mica and  backing trace from the peak.  And, $\mathcal{O}$(10) MeV ions were utilized the Tandem accelerator of Japan Atomic Energy Agency (JAEA), Japan. Here, Fe$^{6+}$ of 60 and 70\,MeV were directly exposed on the mica in the vacuum chamber, and the ions attenuated to approximately 3 and 6\,MeV were also exposed by thin film attenuator. Example image by phase-contrast optical microscope was shown in Fig~\ref{fig:example_image}.  For example, passing through track expected as the Q-ball was observed like Fig~\ref{fig:example_image}(a), and lower energy ions produced by the Fe ions were observed like Fig.~\ref{fig:example_image}~(b)--(d). Track formation ability of the ionization loss respect to the ion energy was summarized in Fig.~\ref{fig:ability}. Here, etching treatment was unified the hydrogen fluoride of 80\,$\%$ for 80\,min soaking at 20${}^\circ$C. From Fig.~\ref{fig:ability}, we can understand the track formation threshold for the ionization loss was around 10\,$\text{MeV}/\text{mg}/\text{cm}^{2}$, and this value was consistent with previous research. 

We confirmed the track formation threshold of the mica for the optical microscope recognition. The ion-beam utilized in this test has ionization dominated energy loss, however the Q-ball is expected to have larger nuclear loss mechanism than ionization loss because of velocity of 10$^{-3}$ to light velocity. Actually, this region is not a extension of the track formation mechanism with ionization loss (\textit{i.e.}, ion-explosion model), therefore we should calibrate by lower velocity ions, and simulation study is need to understand the optical image of the tracks formed by such quite heavy particles. As important indication, the $\alpha$ recoil tracks are almost same energy loss mechanism with the charged Q-ball, therefore such particles should be observed in the mica.

\subsection{Study for Optical Microscope Readout System}

The Q-ball search are required the high speed scanning system to search wide area. Now, we constructed the new scanning system for the Paleo detector, mainly for the Q-ball tracks as first target, but it becomes applicable system for any other tracks in the mineral. In this system, high scanning speed system will be installed based on the system developed in the direction sensitive dark matter search with the super-resolution nuclear emulsion~\cite{Umemoto:2020rip,Shiraishi:2021jce}. Our first target is the speed of several hours per 100\,cm$^{2}$, and this speed enable to search the Q-ball search with the sensitivity for flux of 10$^{-20}$/cm$^{2}$/sec/sr within couple of weeks. This will be the highest sensitivity for the Q-ball search, and drastically improve to artificial detector.  This speed should be improved several orders more by upgraded microscope hardware in next step. And also, recent cutting-edge technologies of image processing using machine learning will contribute to the upgrade for such microscope system. 

\subsection*{Acknowledgements}
This work was supported by the Director's Leadership Grant, ISEE, Nagoya Univesity.  And, experiments of the ion beam exposure were supported by the Research Project with Heavy Ions at QST-HIMAC and the Tandem Accelerator, JAEA.

\FloatBarrier
\newpage

\section{Color center bulk spectroscopy in CaF$_2$}
\label{sec:bulk}

\begin{figure}[h]
    \centering
    \includegraphics[width=0.8\textwidth]{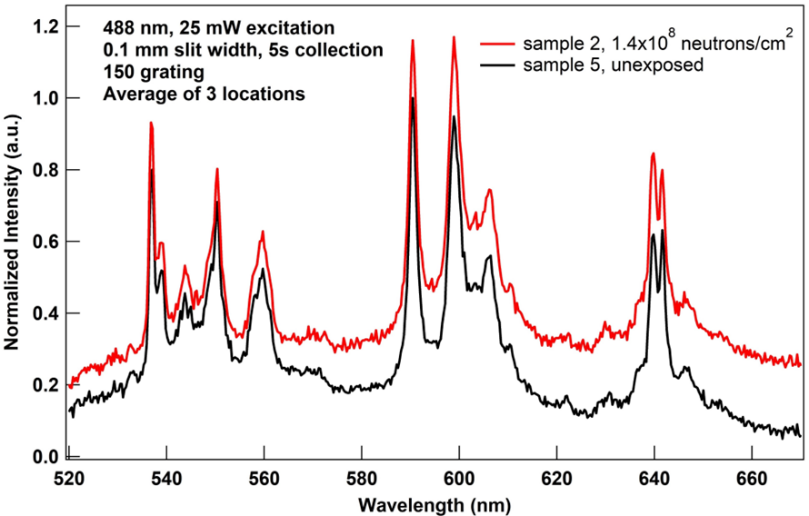}
    \caption{PL spectra comparing exposed and unexposed CaF2 crystals using a 488\,nm laser with a 25\,mW excitation power where the intensity is normalized to the unexposed sample.}
    \label{fig:spectrum}
\end{figure}

Authors: {\it Giti Khodaparast, Brenden Magill and Patrick Huber, \\ on behalf of PALEOCCENE}
\vspace{0.1cm} \\
Virginia Tech 
\vspace{0.3cm} \\
Color center (CC) creation in crystals offers a sensitive and durable way to detect dark matter (DM) where CCs can be created in transparent crystals, allowing direct and effective optical measurements such as photoluminescence, PL~\cite{Budnik:2017sbu}. Furthermore, transparent crystals such as CaF$_2$ offer great promises for both low-energy DM and neutrino detection. Here we present some preliminary observations obtained with spectroscopy.
Following the approach of Mosbacher {\it et. al.}~\cite{Mosbacher:2019igk} we selected CaF$_2$ as our first candidate where several color center emission peaks can be excited easily in the blue region rather than in more challenging optical regions such as UV. In Fig.~\ref{fig:spectrum} we present an example of the PL measurements at Virginia Tech from two CaF$_2$ crystals where one had been exposed to a total neutron flux of $1.4 \times 10^8$ neutrons/cm$^2$, using a 10\,mCi AmBe source where the gamma rays were blocked by a piece of lead foil. In the exposed crystal, the overall PL intensity has increased, and this change has persisted over 18 months in our oldest samples. 

\begin{figure}[h]
    \centering
    \includegraphics[width=0.9\textwidth]{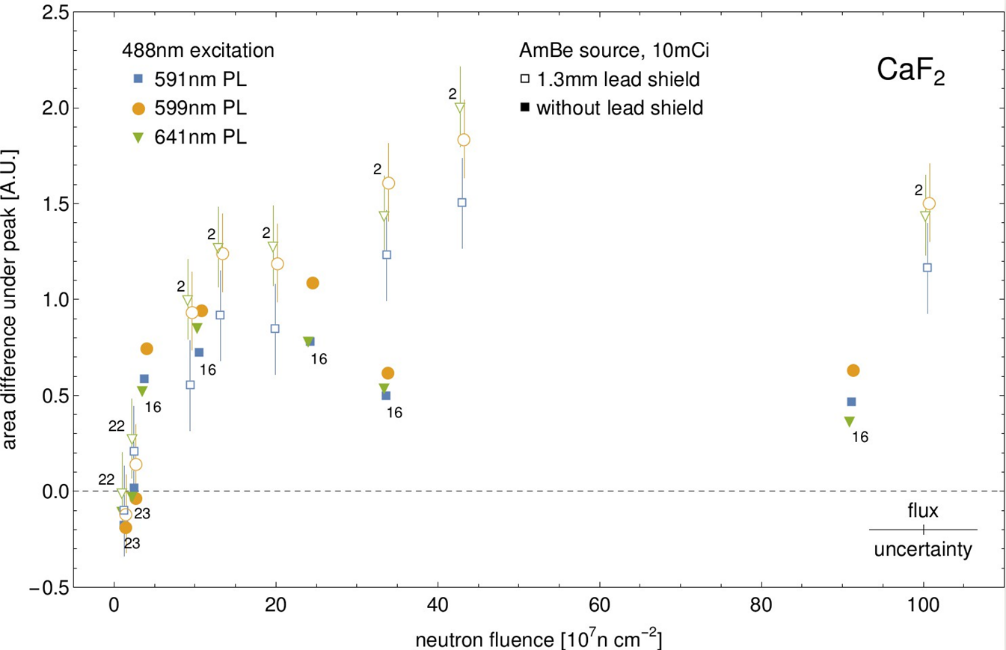}
    \caption{Variations related to the area under the curve around the 600 nm PL emissions as a function of neutron exposure for the samples exposed to neutrons (open squares) and the samples exposed to neutrons and gamma rays (filled squares).}
    \label{fig:sp2}
\end{figure}

To understand the PL enhancement (PLE) as a function of neutron dosage, we exposed our CaF$_2$ samples to neutrons and neutrons plus gamma rays over 200 days while stopping regularly to take spectra to determine the PLE. Each time spectra were taken under the exact experimental conditions and subsequently were normalized to the same unexposed sample. Furthermore, the areas under the three peaks from 590 to 610 nm were extracted from the observations, for the fixed values of neutron doses which were recorded using flux calibration from a Ni foil. Figure~\ref{fig:sp2} summarizes the PLE as a function of the variations in the neutron dosage. Here we observe that the PLE initially decreases slightly, then increases to a maximum followed by another decrease at very high doses. The initial dose region below a fluence of $20 \times 10^7$ neutrons/cm$^2$ displays a dramatic linear sensitivity to the strength of the exposure strength. 

\begin{figure}[h]
    \centering
    \includegraphics[width=\textwidth]{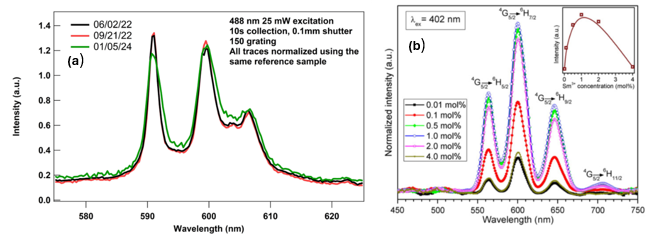}
    \caption{a) An example of the PL emissions from our CaF$_2$ taken with an excitation wavelength of 488 nm and power of 25 mW, presenting three measurements taken over a one-and-a-half-year period, displaying three peaks from 595 to 605 nm. B)  The PL measurements of a piece of Sm3+ doped glass displaying a similar tri-peak structure centered at the same wavelengths, adapted from Umamaheswari {\it et al.}~\cite{Umameheswari:2012}.
}
    \label{fig:sm}
\end{figure}

\begin{figure}[h]
    \centering
    \includegraphics[width=0.65\textwidth]{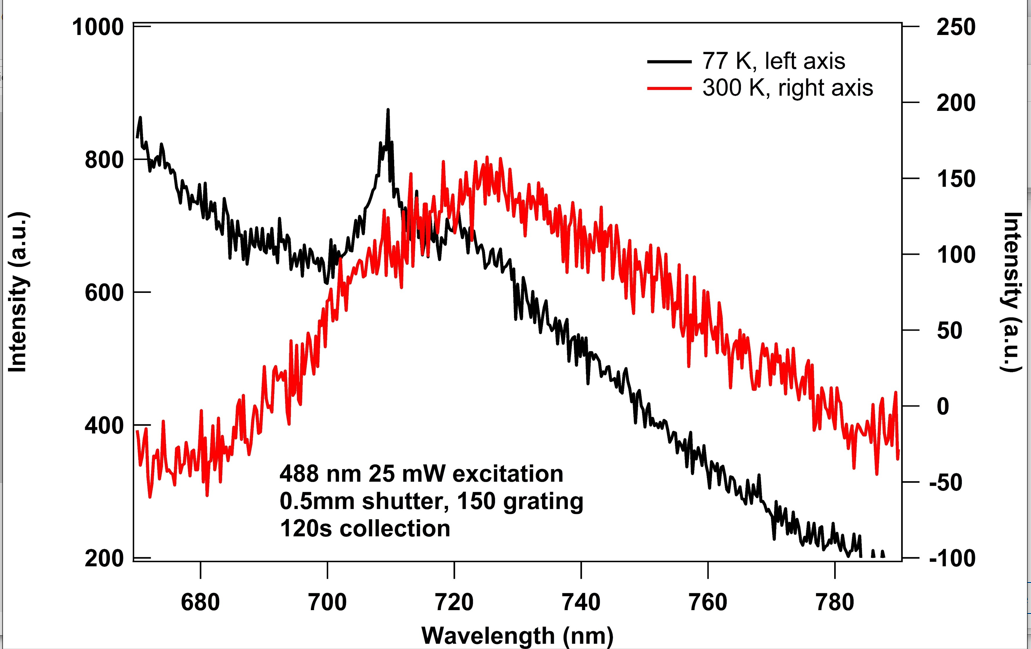}
    \caption{The PL spectra at room temperature (red trace) and 77 K (black trace) for one of our CaF$_2$ samples with an emission peak at 709 nm resolved at 77 K. This emission matches the expected low-temperature PL peak position for Sm3+ impurity.
}
    \label{fig:low}
\end{figure}

Our strong PL emissions and tunability can provide an important prospect to be employed for neutrinos or neutron detections. However, we have confirmed another potential source for the observed PL in our CaF$_2$ samples which is not related to vacancy-related CCs but rather to the rare earth contaminates in the crystal. 
In Fig.~\ref{fig:sm}a, we present an example of the PL from our samples from 590 to 630 nm in comparison to three of the principal PL peaks observed in Sm3+ doped glass, Fig.~\ref{fig:sm}b, by Umamaheswari, et. al. from Ref.~\cite{Umameheswari:2012}. As shown in Fig.~\ref{fig:sm}b, the Sm3+ emission lines centered at 600 nm display a three-peak structure like the PL emissions from our CaF$_2$ samples. While the peaks in Ref.~\cite{Umameheswari:2012} are spread out over a wider range than the peaks observed in our samples, the similarities were enough to raise the concern that the photoluminescent CCs from our crystals is due to impurities in the crystals and not vacancy creation. 
To address this question, we performed PL measurements on one of our neutron-exposed CaF$_2$ at 77 K, to explore the low-temperature emission around 709 nm PL peak, which is a characteristic of Sm3+ impurities.  In Fig.~\ref{fig:low}, we present the PL emission at 709 nm which could not have been resolved in the broad spectrum at room temperature and is the signature emission of Sm3+ impurities. Once we had this indication of Sm contaminants, we had an elemental analysis performed on a crystal from the original batch of CaF$_2$ crystals and found that they did contain trace amounts of Sm. 
These results now suggest that the observations presented in Fig.~\ref{fig:spectrum}, were the consequences of the activation of Sm3+ impurity sites by neutron exposure and not the production of lattice vacancies which can be used as a method employed for optical detections at room temperature with strong sensitivity to radiation doses.

\FloatBarrier

\subsection*{Acknowledgements}
The work of PH is supported by the U.S. Department of Energy Office of Science under award number DE-SC0020262 and by the National Nuclear Security Administration Office of Defense Nuclear Nonproliferation R\&D through the Consortium for Monitoring, Technology and Verification under award number DE-NA0003920.

\newpage

\section{Read-out of Color Center with Light sheet fluorescence microscopy}

Authors: {\it Gabriela~R.~Araujo$^1$, Patrick Huber$^2$, and N. Vladimirov$^1$ on behalf of PALEOCCENE}
\vspace{0.1cm} \\
$^1$University of Zurich, $^2$Virginia Tech 
\vspace{0.3cm} \\

As described in Section~\ref{sec:bulk}, color centers could serve as sensitive, low-threshold signals of dark matter and neutrino interactions within transparent crystals.  However, establishing color center imaging as a competitive method for dark matter detection requires fast imaging of individual color centers~\cite{Alfonso:2022meh, Cogswell:2021qlq}.

In the last ``Mineral Detection of Neutrinos and Dark Matter'' workshop (MD$\nu$DM'22), we reported the first tests of light-sheet microscopy imaging of radiation-induced color centers within transparent crystals~\cite{Baum:2023cct}. These measurements, conducted using the state-of-the-art mesoSPIM microscope, demonstrated clear fluorescence signals originating from color centers within gamma-ray irradiated CaF$_{2}$ crystals. In this work, we present advancements in understanding the causes of these color centers in CaF$_{2}$, upgrades to the mesoSPIM microscope, and progress in imaging neutron-induced color centers, along with the development of analysis methods for imaging structures within large microscopy datasets.

\subsection*{Color center imaging with the mesoSPIM light-sheet microscope}

Light sheet microscopy offers rapid 3D imaging of color centers within large (cm$^3$) samples without the need for slicing. The mesoSPIM initiative, an open-source project spanning over 20 microscopes worldwide, has been at the forefront of this field, enabling the imaging of samples up to 80 cm$^3$ with isotropic resolutions of a few micrometers and speed of under an hour per cm$^3$~\cite{mesoSPIM, Vladimirov:2023}. These high-throughput features, combined with the sharp imaging and reduced sample exposure characteristics of the light-sheet method, make the mesoSPIM a suitable readout for the PALEOCCENE concept \cite{Cogswell:2021qlq, Alfonso:2022meh}.

Using the mesoSPIM version described in Ref.~\cite{mesoSPIM}, we successfully obtained the first dedicated imaging of radiation-induced color centers within CaF$_{2}$. However, visualizing single and faint color centers induced by neutrons presented itself as a challenge. To address this challenge, the new version of the mesoSPIM, the Benchtop mesoSPIM, incorporated enhancements specifically designed for color center imaging. These enhancements include a 20x magnification lens and a custom-made crystal holder, alongside improvements addressing overall needs: improved image quality, reduced footprint, and lower cost~\cite{Vladimirov:2023}. 

With these upgrades, we successfully imaged track-like structures within CaF$_{2}$ crystals using 20x magnification. These structures, elongated concentrations of color centers spanning approximately \SI{20}{\um}, were consistently observed in repeated scans of the same region within the crystal's bulk volume. For their identification, an algorithm was developed to detect clusters of luminous pixels and match them in repeated scans. While the exact origin of these tracks is unclear, these initial findings demonstrate the Benchtop mesoSPIM's capability to confidently image color center structures within crystals.

\subsection*{Color centers in transparent crystals \& microscopy data analysis methods}

We initiated a campaign to investigate the type of color centers induced by gamma-ray irradiation of CaF$_{2}$ crystals and to probe the imaging of neutron-induced color centers in CaF$_{2}$, LiF and Sapphire. This campaign involved a large number of crystals obtained from several vendors and aimed to explore the variability in color center fluorescence levels measured from different crystal batches, materials, and the dependence of the signal on irradiation dose.

The crystals were imaged using the mesoSPIM and far-field spectroscopy both before and after irradiation with $^{60}$Co gamma-ray sources or AmBe neutron sources at varied doses. Measurements obtained using both techniques revealed a significant dependence of color center fluorescence levels on the batch of crystals. More detailed studies of crystal compositions and absorption and emission spectra indicated impurities as the primary causes of variability.

Initial studies of neutron-induced color centers showed a moderate level of signal above the background. However, confidently imaging small structures or single color centers requires estimation and exclusion of several background sources from the data. The small size of the structures being searched for, coupled with the substantial computing processing required to exclude small background sources, rendered the existing track-finding algorithm inefficient, as this program was developed to identify rather large ($\sim$\SI{20}{\um}) concentrations of color centers. Notably, each image captured by the benchtop mesoSPIM is 15 megapixels with 16-bit depth, generating approximately 30 GB of data for a scan of a cm$^3$ crystal at a few micrometer resolution. With repeated scans of each crystal and responses to multiple excitation wavelengths, this campaign generated terabytes of data.

Consequently, the automated code for identifying color center structures and tracks is undergoing upgrades, and it is planned for public release in the future. We foresee that this upgraded data analysis software, tailored for efficiently identifying tracks and structures in microscopy images, will be useful for those in the MD$\nu$DM community who are currently performing microscopy measurements of mineral detectors.  

\subsection*{Acknowledgements}
The work of PH is supported by the U.S. Department of Energy Office of Science under award number DE-SC0020262 and by the National Nuclear Security Administration Office of Defense Nuclear Nonproliferation R\&D through the Consortium for Monitoring, Technology and Verification under award number DE-NA0003920.

\FloatBarrier
\newpage

\section{Optical measurements of irradiated crystals for PALEOCCENE}

Authors: {\it Xianyi Zhang}
\vspace{0.1cm} \\
Lawrence Livermore National Laboratory
\vspace{0.3cm}

\subsection{Introduction}
Particle interactions, such as electron recoils and nuclear recoils, can cause damage by displacing ions from their original lattice, or disturbing the chemical bonds between the ions.  Nuclear recoils caused by coherent elastic neutrino-nucleus scattering (CEvNS) can leave unique crystal damages. These CEvNS caused damages can occur in the forms groups of color centers in the crystal. Color center formed in crystals is one of the result, after low energy nuclear recoil displaces nuclei its crystal lattice. The color centers will show different spectroscopic properties compared to the base material. Using spectroscopic and microscopic measurements, one can potentially observe clusters of color centers and reconstruct the particle information, such as detect CEvNS interaction from reactor neutrinos. Existing study~\cite{Cogswell:2021qlq} suggested 100-g scale crystal detector would take approximately three months to collect enough neutrino interactions to evaluate a fission reactor’s components. Such study also suggest long term passive detection to measure dark matter recoil is possible with crystal detector. In both cases, it is necessary to find crystal materials that are suitable to the long term detection of nuclear recoils from rare particles. PAssive Low Energy Optical Color CEnter Nuclear rEcoil collaboration, PALEOCCENE, is set to develop the passive crystal detection technology to search for reactor CEvNS and dark matter.
In this study, we searched for the crystal material that can be used for measuring nuclear recoil through the measurement optical damages. The impurity and damage in crystals can be detected through fluorescence and/or Raman spectra of materials. We worked to identify the signature wavelength in the spectroscopic measurements after the crystals are irradiated by neutron and gamma. To search for the suitable material for a long-term passive detector, we also exposed the crystals to aging, heating and UV from the sunlight. 

\subsection{Crystal irradiation and measurements}
Materials tested include CaF$_2$, LiF, MgF$_2$, BaF$_2$, and Al$_2$O$_3$. These are commonly used inorganic crystal scintillators. These materials have previously shown changes in fluorescence spectroscopy after neutron irradiation~\cite{Mosbacher:2019igk}. They are relatively low-cost materials with simple working environment requirements. Samples are grouped as control, neutron irradiated samples by $^{252}$Cf (neutron source) and gamma irradiated samples by $^{60}$Co (gamma source). The neutron and gamma fluence in the crystal samples are 10$^9$ n/cm$^2$ and 10$^{10}$ n/cm$^2$ respectively. 
The irradiation campaign was separated into three equal periods. Irradiated crystals were measured before and after each period to observe radiation dependence of optical properties. Using fluorescence spectrometer and Raman spectrometer, the excitation and emission spectra were measured on the entire crystal as well as the precisely spots inside of the crystals. Microscopic images with 455~nm exciting laser were taken in searching for actual crystal defects after neutron irradiation.

\subsection{Results}
\begin{figure}[htbp]
\centering
\includegraphics[width=.47\textwidth]{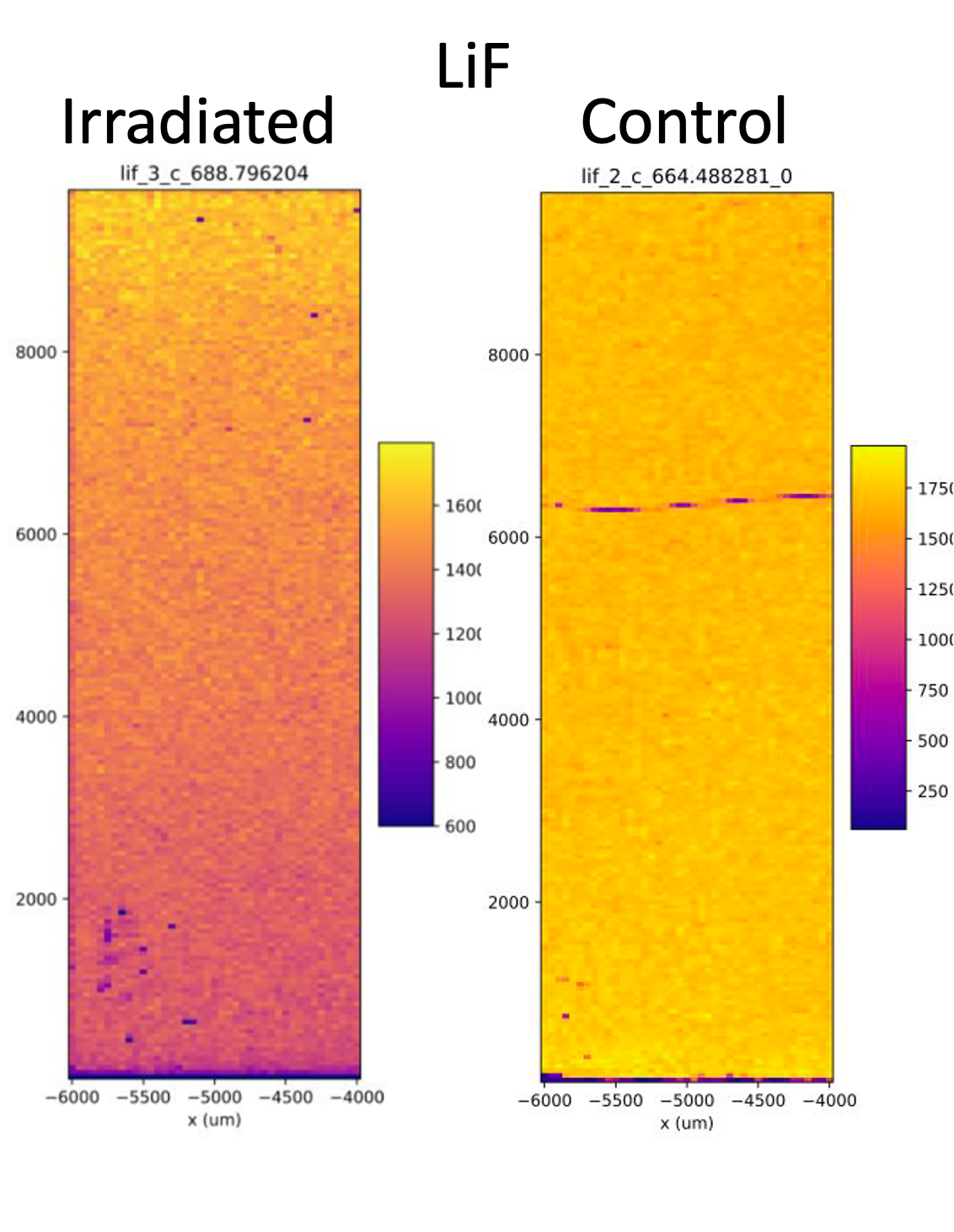}
\hspace{.02\textwidth}
\includegraphics[width=.48\textwidth]{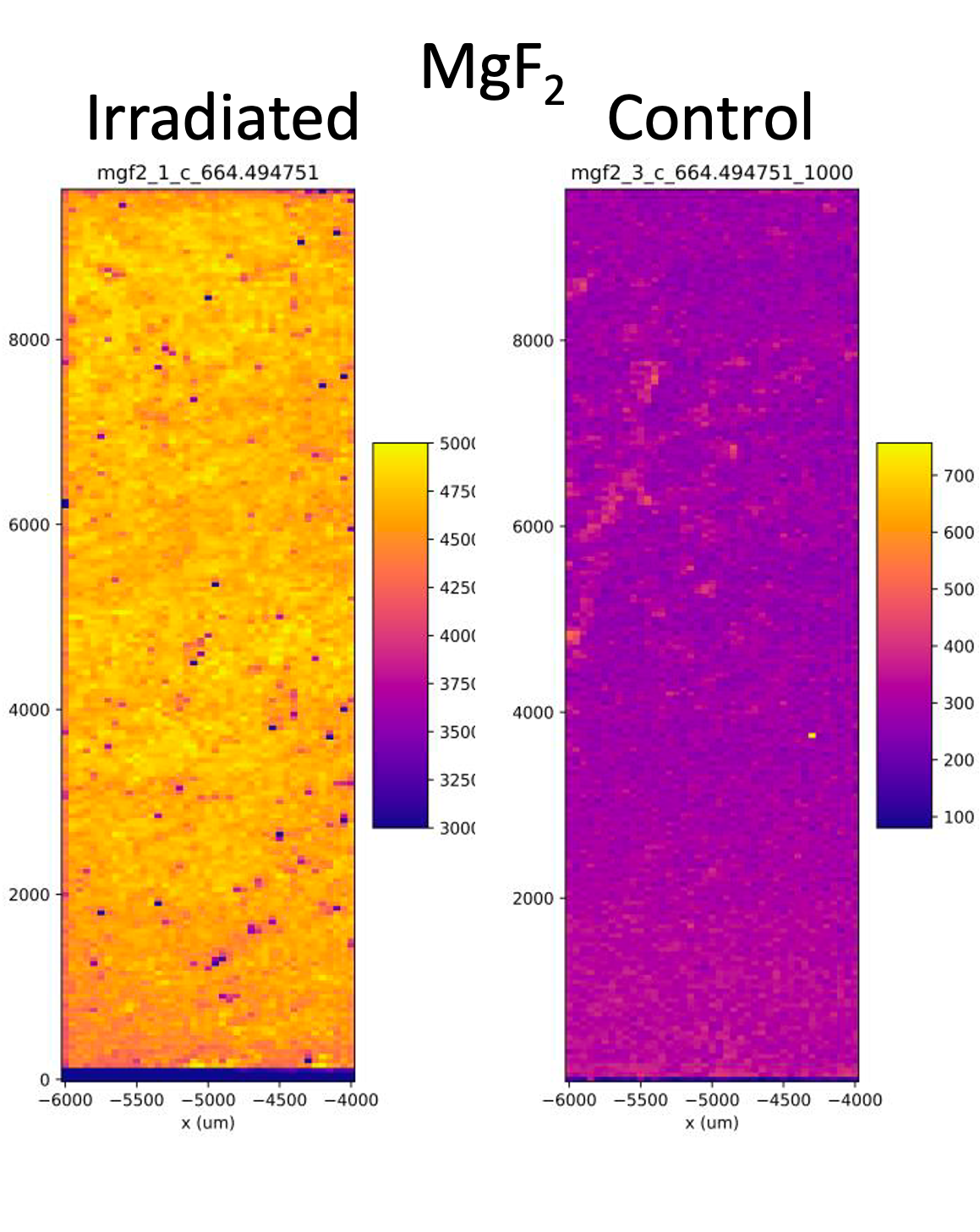}
\caption{Raman microscope image of baseline (no-radiation) subtracted compared to the control samples. Irradiated crystals indicates low amplitude spots.\label{fig:XZ_i}}
\end{figure}

In Fluorescence spectra, subtle rise of at different excitation-emission wavelengths were observed among all materials after irradiation. 
When excited by 455~nm laser, only LiF, MgF$_2$, and Al$_2$O$_3$ emitted Raman spectra with significant change between 450 nm and 700 nm, while the changeo in Al$_2$O$_3$ Raman intensity is not proportional to radiation dosage. Raman intensities between 650 nm and 750 nm of neutron irradiated samples were compared against gamma irradiated counterparts. LiF, MgF$_2$, and Al$_2$O$_3$ also showed largest neutron-gamma difference. The microscopic image of Al$_2$O$_3$ did not show observable crystal defects after irradiation. On the images of LiF and MgF$_2$ (Figure \ref{fig:XZ_i}), after subtracting pre-irradiation baseline, multiple location showed significant drop of Raman peak amplitude, suggesting change of materials at the dark spot. All current measurement results suggest LiF, MgF$_2$ are potential suitable material for the development of PALEOCCENE technology.

\subsection{Conclusion}
In current state of the study, the measurement and analysis have been mostly focused mostly on optical spectral changes in entire crystals. The material suitable for further investigation for the technology were identified, including LiF and MgF$_2$. While we could observe changes in spectral signatures both in Raman and fluorescent spectroscopy in the relatively larger dimension, the readout technologies are slow and cumbersome. The confocal Raman microscopic measurement faced challenges such as limit of sample size and hours-long image taking speed at small area. The choice of Raman excitation wavelength was also limited. These limits of instruments can leave details of crystal defects undetected in some of the materials. We will continue to track the changes of crystals overtime and characterize the signature wavelength and optimum excitation wavelength for the Raman microscope measurement. Moreover, the success of using crystal damage to measure neutron and neutrino events requires measurement of more than 100-gram scale crystals. Consequently, a more efficient measurement technology, such as selective plane illumination microscopy that illuminates a full plane for microscopic image instead of one focus location, is necessary for further R\&D. The long term goal are to characterize the energy threshold and detector response to nuclear recoils of different energies, which are the key detector 

\subsection*{Acknowledgements}
Part of this work was performed under the auspices of the US Department of Energy by Lawrence Livermore National Laboratory LDRD research program under contract DE-AC52-07NA27344. LLNL-PROC-863705

\FloatBarrier
\newpage

\section{Paleo-Detection of Cosmic Ray Muons}

Authors: {\it L.~Apollonio$^{1,2}$, L.~Caccianiga$^2$, C.~Galelli$^3$, P.~Magnani$^1$, F.M.~Mariani$^{1,2}$, A.~Veutro$^4$} 
\vspace{0.1cm} \\
$^1$University of Milan, $^2$INFN Milano, $^3$Paris Observatory, $^4$Sapienza University 
\vspace{0.3cm} \\
We propose to use the paleo-detectors technique to obtain information of the past flux of cosmic rays. In this contribution, we push forward two ideas. The first one consists in taking advantage of specific geological events to select minerals that have been directly exposed to muons for a specific amount of time. This could give a snapshot of the cosmic ray flux at a certain point of history. The second idea takes advantage of the muon-induced fission tracks on the heavy element inside of the mineral to make sure that the tracks left by the interaction of cosmic rays are indeed measurable.

\subsection{Introduction}
While other works presented in these proceedings plan on retrieving samples from deep underground to avoid cosmic ray contamination, we inversely would like to use this technique to measure the flux of cosmic rays in the past. This technique is, together with the possible detection of atmospheric neutrinos suggested in \cite{Jordan:2020gxx}, one of the few methods proposed to measure the evolution of the cosmic ray flux. In particular, this technique would be sustainable also to detect short-term transients like nearby cataclysmic events such as supernovae.

\subsection{Windows of exposure}
Cosmic rays have the advantage of being shielded by matter. This way, if we know well enough the geological history of a sample, we can identify exposure windows when the sample was exposed to the secondary cosmic ray flux. The cosmic ray-induced tracks in that sample would preserve information on the primary cosmic ray flux in the specific exposure window. An example is the desiccation of the Mediterranean Sea that happened in the Messinian period \cite{Krijgsman:1999}. Evaporites were produced during that event, that were exposed for $500\,\text{kyr}$ and then covered again by a km-scale overburden of water. We investigated the possibility of identifying an excess flux of cosmic rays, due for example to a supernova explosion nearby, by measuring tracks in a $10\,\text{g}$ sample of Halite. From our analysis, available in more detail in \cite{Mariani:2023mwb}, we found that our techniques would be sensitive to the change of flux expected from a supernova exploding at $50\,\text{pc}$ in the exposure time window.

\subsection{Muon-induced fission tracks}
One of the main challenges of the paleo-detectors technique consists in the identification of the tracks formed after the nuclear recoil with astroparticles. However, tracks left by the spontaneous fission of heavy elements (mainly ${}^{238}\text{U}$) are routinely seen in dating minerals such as obsidians and apatites. The fission of heavy nuclei can also be induced by muons stopped inside the minerals, and since muons are directly linked to cosmic rays, we can exploit this feature to obtain useful information on the past flux of cosmic rays. 

Spontaneous fission (SF) and muon-induced fission (MIF) are rare events. To have enough statistics on the number of tracks produced, we need to find minerals rich in Uranium and Thorium. A suitable candidate is therefore Zircon $[\text{ZrSiO}_4]$, which has typical concentrations of both elements around $0.001-0.005$\,g/g \cite{vanSchmus:1995} and can reach a concentration in Thorium of $O(0.01)\,\text{g/g}$~\cite{Sun:2021}.

The rate of spontaneous fissions expected per unit mass of a mineral sample and per time as a function of the atomic numbers of the two daughter nuclei can be defined as 
\begin{equation}  \label{eq:Mil1}
	\text{R}_{\text{X}}^{\text{SF}}(\text{Z}_1,\text{Z}_2)=f_\text{X}\frac{\text{N}_\text{A}}{\text{A}}\frac{\text{BR}_\text{X}^\text{SF}}{\tau_\text{X}^{1/2}}p(\text{Z}_1,\text{Z}_2)\;,
\end{equation}
where $f_\text{X}$ is the concentration of element X inside the sample, $\text{N}_\text{A}$ is the Avogadro number, A is the mass number of element X, $\text{BR}_\text{X}^\text{SF}$ is the branching ratio of spontaneous fission of element X, $\tau_\text{X}^{1/2}$ is its half-life time and $p(\text{Z}_1,\text{Z}_2)$ is the probability of decaying in the couple ($\text{Z}_1,\text{Z}_2$).
\\ The rate for muon-induced fissions is
\begin{equation} \label{eq:Mil2}
	\text{R}^\text{MIF}_\text{X}(\text{Z}_1,\text{Z}_2)=f_\text{X}n^\text{X}_\mu\frac{\text{N}_\mu(\text{T},m)}{\text{T}\times m}p(\text{Z}_1,\text{Z}_2)e^{{-\text{T}}/ \tau_\text{X}}\;,
\end{equation}
where $n_\mu^\text{X}$ is the number of fission induced per muon stopped per mass of material X (the values for Thorium and Uranium were taken from \cite{Measday:2001yr}), $\text{N}_\mu(\text{T},m)$ is the number of muons stopped by a sample of mineral of mass $m$ in time period T and $\tau_\text{X}$ is the decay constant of the element X. It should be noticed that the number of muons stopped depends on the linear size of the sample, the longer the linear size the higher the energy of the muons that can be stopped by the sample.

The spontaneous fission and muon-induced fission are two body decays, therefore once we define the daughter nuclei their recoil energy is fixed by the laws of dynamics. From the rate of tracks as a function of the atomic numbers, we can pass to the rate as a function of the track length by computing the energies of the recoiling nuclei and evaluating the track length using the software \verb+SRIM+.

As can be seen in Eqs.~\eqref{eq:Mil1} and~\eqref{eq:Mil2}, the rate is directly proportional to the concentration of the element inside the mineral ($\text{R}_\text{X}=f_\text{X}C_\text{X}$). The number of tracks expected in a sample of mass $m$ after a time period $t$ can be obtained by multiplying the rate for the mass and the time. However, in the case of equal concentrations of Uranium and Thorium, the number of tracks expected for muon-induced fission is 4 orders of magnitude lower than the number expected for the spontaneous fission of Uranium. In a realistic scenario, the information on the tracks of muon-induced fission would be completely covered by the uncertainties linked to the evaluation of the age of the sample. To avoid the problem we propose to consider two samples produced in the same geological events, but having different concentrations in Uranium and Thorium (e.g.\ samples produced in a volcanic eruption). The number of tracks expected in the two samples are
\begin{equation}
	N_i = (f_\text{U}^i C_\text{U}^\text{SF} + f_\text{U}^i C_\text{U}^\text{MIF} + f_\text{Th}^i C_\text{Th}^\text{SF}  + f_\text{Th}^i C_\text{Th}^\text{MIF}) \times t \times m\,.
\end{equation}
We can define the quantity 
\begin{equation}
	\rho=\frac{\text{N}_1}{\text{N}_2} = \frac{f_\text{U}^1 C_\text{U}^\text{SF} + f_\text{U}^1 C_\text{U}^\text{MIF} + f_\text{Th}^1 C_\text{Th}^\text{SF}  + f_\text{Th}^1 C_\text{Th}^\text{MIF}}{f_\text{U}^2 C_\text{U}^\text{SF} + f_\text{U}^2 C_\text{U}^\text{MIF} + f_\text{Th}^2 C_\text{Th}^\text{SF}  + f_\text{Th}^2 C_\text{Th}^\text{MIF}}\,,
\end{equation}
which is not dependent on the age of the two samples. 

\begin{figure}
   \centering
   \includegraphics[width=0.495\textwidth]{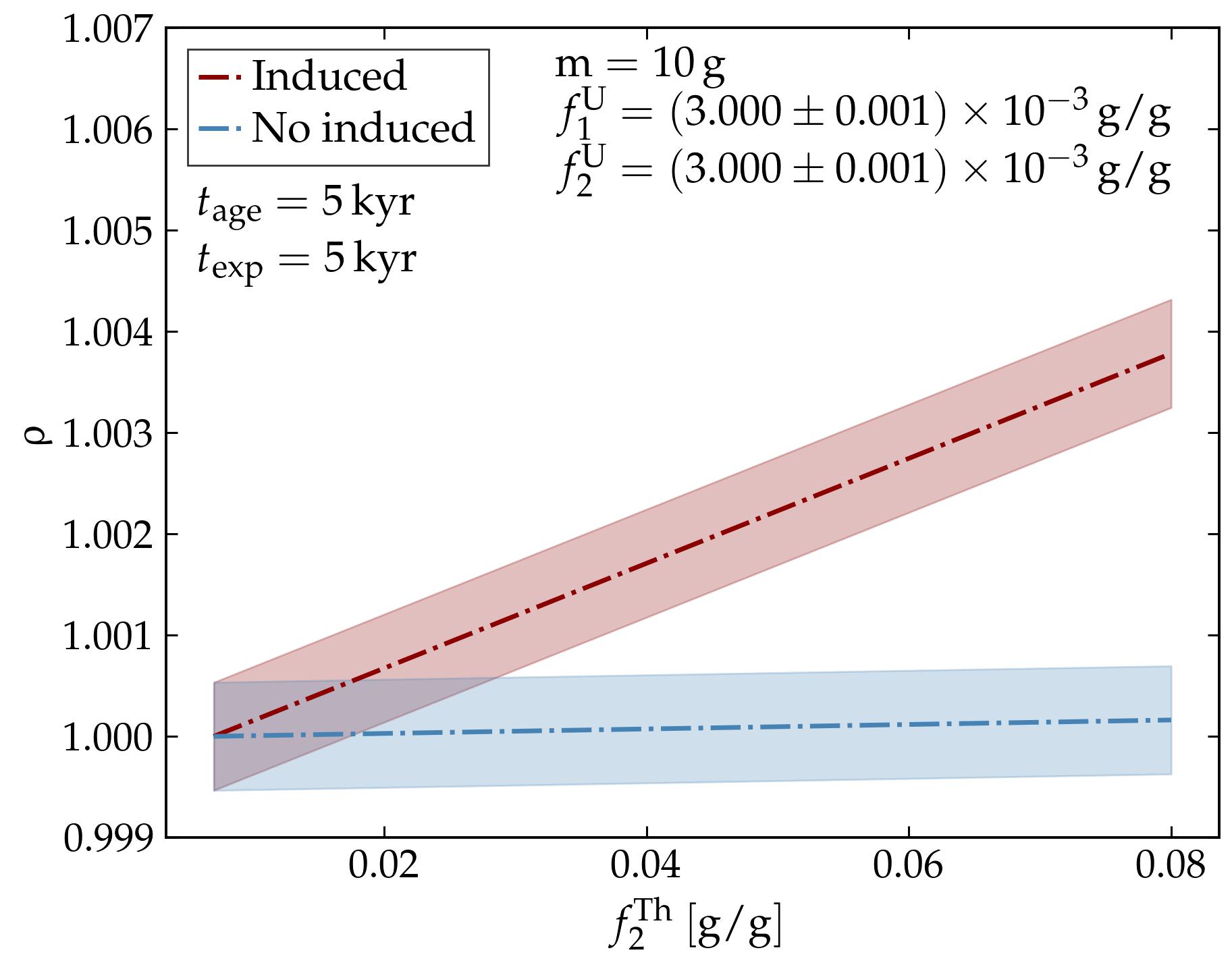}
   \includegraphics[width=0.495\textwidth]{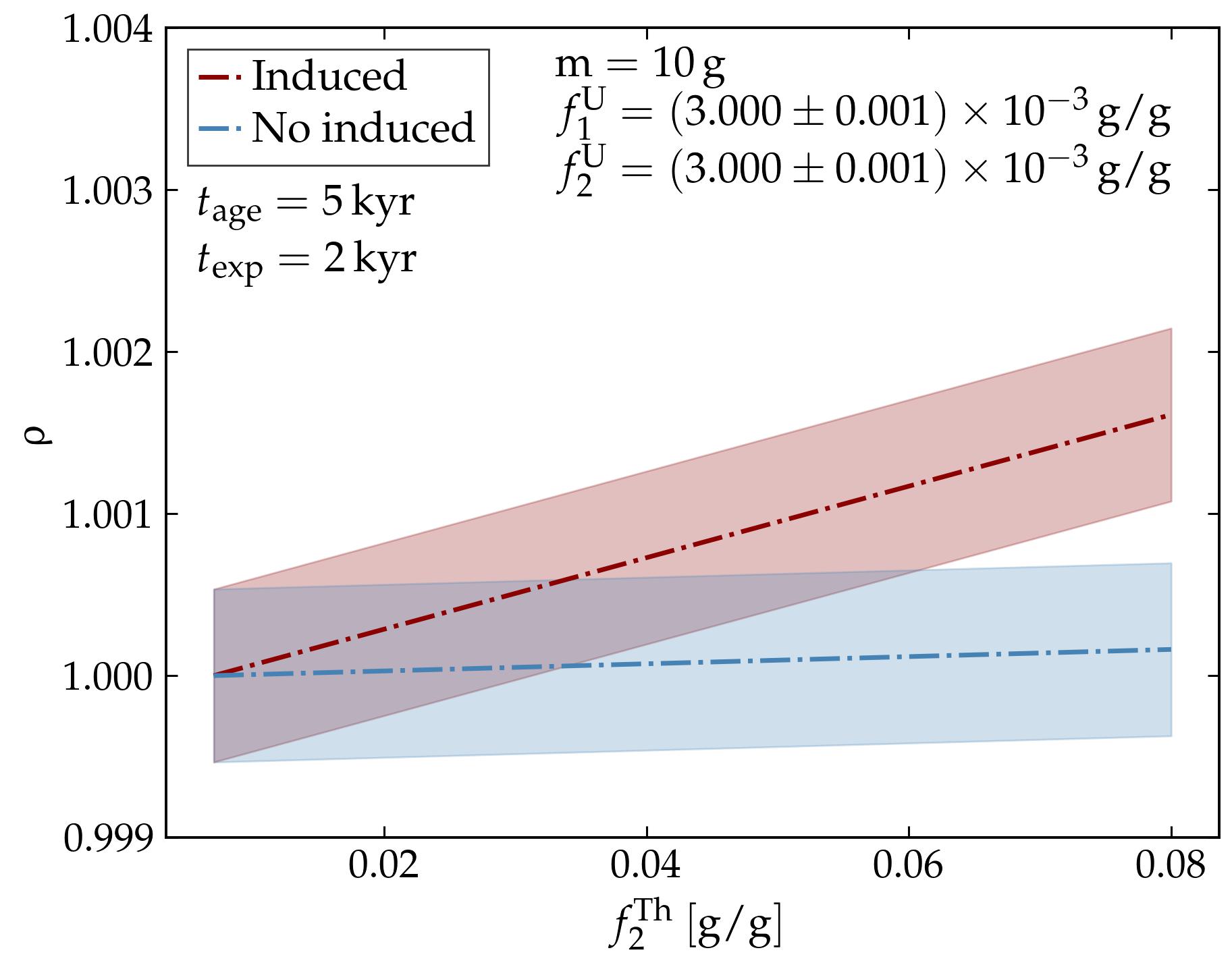}
   \caption{The evolution of the parameter $\rho$ as a function of the Thorium concentration of the second sample, when all the other concentrations have been fixed. The Thorium concentration in the first sample is fixed to be zero. The absolute error on the measure of the concentration has been fixed to $10^{-6}\,\text{g/g}$. The red line indicates the value of $\rho$ considering the formation of the muon-induced fission tracks, the blue line indicates the value of $\rho$ when the muon-induced fission tracks are not formed. On the left, the age of the sample and the exposure time are equal ($t_\text{age}=t_\text{exp}=5\,\text{kyr}$). On the right, the age and the exposure time of the sample are not equal.}
   \label{fig:rhoi}
\end{figure}

To understand if we can distinguish the tracks formed after muon-induced fission we evaluated the evolution of the parameter $\rho$ as a function of the Thorium concentration of the second sample, fixing all the other concentrations. As reported by the left plot of Fig.~\ref{fig:rhoi} in the hypothesis of being able to measure the concentration of Uranium and Thorium with an absolute error of $10^{-6}\,\text{g/g}$, we can obtain information on the muon-induced fission tracks by having concentrations of $O(10^{-3})\,\text{g/g}$. 

After being exposed directly to muons, the mineral sample can be covered by some layers of material. Therefore, the number of muons reaching the sample decreases and the muon-induced fission tracks get suppressed. To understand if we could obtain information on the muon-induced fission tracks even in this hypothesis, we considered a scenario in which the exposure time to muons is not equal to the age of the sample. We selected $t_\text{exp}=2\,\text{kyr}$ and $t_\text{age}=5\,\text{kyr}$. In the right plot of Fig.~\ref{fig:rhoi} we reported the results found, and we can see that even in this scenario we can get a good separation of the parameter $\rho$ with concentrations of $O(10^{-3}\,\text{g/g})$.

In future works, we will consider also possible differences in Thorium and Uranium concentrations at different points of the mineral, as this would remove the issue of finding two samples that are of the same age. 

\subsection*{Acknowledgements}
The authors thank Marco Giulio Giammarchi and Max Stadelmaier for the discussions and feedback received about this work. We also thank Alessandra Guglielmetti, Letizia Bonizzoni, and Cecilia Donini for introducing us to the detection of fission tracks in Obsidian.

\FloatBarrier
\newpage

\bibliographystyle{JHEP.bst}
\bibliography{theBib}

\end{document}